\documentclass[]{jfm} 

\usepackage{graphicx}
\usepackage{comment}
\usepackage{newtxtext}
\usepackage{newtxmath}
\usepackage{natbib}
\usepackage{hyperref}
\usepackage{multirow}
\usepackage{xcolor}
\hypersetup{
    colorlinks = true,
    urlcolor   = blue,
    citecolor  = black,
}

\newcommand{\RomanNumeralCaps}[1]
\linenumbers

\title{Analysis of Resonance in Jet Screech with Large-Eddy Simulations}

\author{Gao Jun Wu\aff{1}
  \corresp{\email{gaojunwu@stanford.edu}},
  Sanjiva K. Lele\aff{2}
 \and Jinah Jeun\aff{3}}

\affiliation{\aff{1}PhD. Candidate, Department of Aeronautics and Astronautics, Stanford University
\aff{2}Professor, Department of Aeronautics and Astronautics and Department of Mechanical Engineering, Stanford University \aff{3} Postdoctoral Scholar, Strömningsmek \& Tekn Akustik, KTH}

\begin{document}
\maketitle

\begin{abstract}
Screech resonance is studied with experimentally validated large-eddy simulation data for a 4:1 rectangular under-expanded jet at three nozzle pressure ratios. The analysis uses spectral proper orthogonal decomposition (SPOD) and spatial cross correlation to characterize the oppositely-traveling waves in the jet at the screech fundamental frequency. The results support recent theoretical framing of screech as absolute instability, and further reveal the spatial separation of individual processes for screech generation. From the leading-order SPOD mode, direct evidence of the guided jet mode being the screech closure mechanism, not the external acoustic feedback, is observed. A match in the spatial wavenumber suggests the guided jet mode is generated via interactions between the Kelvin-Helmholtz wave and the shock cells. The energy of the oppositely-moving waves shows spatially global and non-periodic behavior of the coherent structures in the streamwise direction. The ratio of wave energy identifies regions where distinct processes in screech generation take place by comparing the rate of energy propagation in the downstream direction to that of the upstream direction. The distinct regions correspond to initial shear layer receptivity, sound emission, guided jet mode excitation and decay of coherence. The leading-order SPOD mode also enables the approximation of Lighthill's stress tensor and allows for accurate calculation of the far-field screech tone amplitude with the acoustic analogy formulation. The current findings provide insights on building a physics-based reduced order model for screech amplitude prediction in the future.

\end{abstract}



\section{Introduction}
\label{sec:intro}

Screech is an aero-acoustic resonance phenomenon in imperfectly expanded supersonic jets first discovered by~\citet{powell1953mechanism}. Screech produces high-intensity tones at distinct frequencies, making it a notoriously riling component of supersonic jet noise. In addition, jet screech can induce high-dynamic loading which, if it is close to the structural resonance frequencies, could cause premature fatigue failure of components around the jet engines such as those reported on VC10, F-15, B1-B, and F-35~\citep{raman1997Screech,raman2012aeroacoustics,majumdar2014}. 

Previous research on supersonic jet screech has been comprehensively reviewed by~\citet{raman1999Supersonic} and~\citet{edgington2019Aeroacoustic}, which highlight the experimental observations from 1950s to date and the theories based on these empirical findings. The classical description of screech contains four components: the downstream traveling instability waves originated from the jet initial shear layer, the interactions between instability waves and shock cells in the jet plume, the upstream traveling component that re-excites the initial shear layer, and lastly the receptivity process of the shear layer near nozzle exit. The oppositely moving coherent structures give rise to the presence of spatially modulating standing wave patterns, first documented by~\cite{panda1999AnExp}. More recent work by~\citet{edmitch_2021JFM_waves_in_screech} further investigates the distinct groups of waves in jet screech. Two groups of waves are found to be traveling downstream: the Kelvin-Helmholtz (K-H) instability wave and a duct-like mode. The upstream-traveling component includes external acoustic waves and a guided jet mode, first discovered by~\citet{tam1989onthethree}. 

The role of upstream-traveling waves in screech closure has been a topic of debate until recently. The original depiction of screech by~\citet{powell1953mechanism} considers upstream-traveling acoustic waves outside the jet plume to be the closure mechanism. On the other hand, various experimental and numerical evidence suggests the guided jet mode closes the screech feedback path~\citep{Bogey_Gojon_JM2017_feedback,edgington2018upstream, gojon2019antisymmetric}. This new insight leads to improved model for screech frequency prediction.~\citet{Mancinelli2021} uses a locally parallel linear stability analysis about the fully expanded base state to predict A1 and A2 mode frequencies in under-expanded axisymmetric jets, but the model requires a posteriori calibration of reflection coefficients for the phase resonance condition.~\citet{Nogueira2022_closure_mech_a1_a2,Nogueira2022_absolute_instab} construct a linear stability model by including spatial periodicity in the base state due to shock cells, and finds an absolute instability mechanism for screech generation. This semi-local model is able to accurately predict the staging of screech modes across a range of NPR values. 

Despite the success of these recent models, it is still difficult to predict the screech amplitude due to two major problems. The first problem is the sound generation mechanism. Given the presence of the waves and shock cells, the emission of the highly directive screech tones could be attributed to localized shock leakage~\citep{EdMitch_jfm2022_shockleakage,suzuki_lele_2003,manning2000numerical} or distributed sources by the Mach-wave radiation mechanism~\citep{tam1986proposed}. The second problem is how the gain of screech feedback is determined. In order to answer this, the spatially non-local characteristics of screech feedback need to be considered, including the receptivity close to the nozzle exit, the saturation region where K-H waves and the guided jet mode grow to maximum magnitudes, and the dissipation region where large-scale waves turn into smaller-scale turbulence at the end of the jet potential core. The purpose of the current work is to shed light on the second problem. 

High-fidelity large-eddy simulations (LES) data for a 4:1 rectangular screeching jet with intense anti-symmetric flapping are used to analyze the global behaviors of the screech feedback. Through previous works, the numerical results have been validated against experiments conducted at Florida State University. Excellent agreement between the LES data and the experimental data is found~\citep{wu2022_aiaaj_assessment}. Using the LES data and spectral proper orthogonal decomposition (SPOD)~\citep{lumley1970stochastic, towne_JFM_2018_spectral_relation,Schmidt_Colonius_AIAA2020_Guide}, the dominant coherent structures at the screech fundamental frequency are examined. The magnitude and phase of the external acoustic wave, the guided jet mode and the K-H wave are analyzed. 
This paper is organized as follows: Section~\ref{sec:method} summarizes the numerical and data analysis methods, Section~\ref{sec:dominant coherence} characterizes the dominant SPOD modes at the screech fundamental frequency, Section~\ref{sec:closure_mech} examines the role of upstream-traveling guided jet mode versus the external acoustic wave for screech closure, Section~\ref{sec:wave_energy} quantifies the wave energy and describes the global behavior of these instability waves, Section~\ref{sec:npr_effects} complements the analysis with results from three different NPR conditions, Section~\ref{sec:acoustic_analogy} relates the SPOD modes with the acoustic radiation of the screech tone in the far field, and lastly Section~\ref{sec:conclusions} concludes with the main findings of the study.

\section{Methods}\label{sec:method}
\subsection{LES Simulations}\label{sec:LES_sims}
\begin{figure}[h!] 
 	\centering
 	\begin{tabular}{c} 
 		\includegraphics[width=0.65\linewidth]{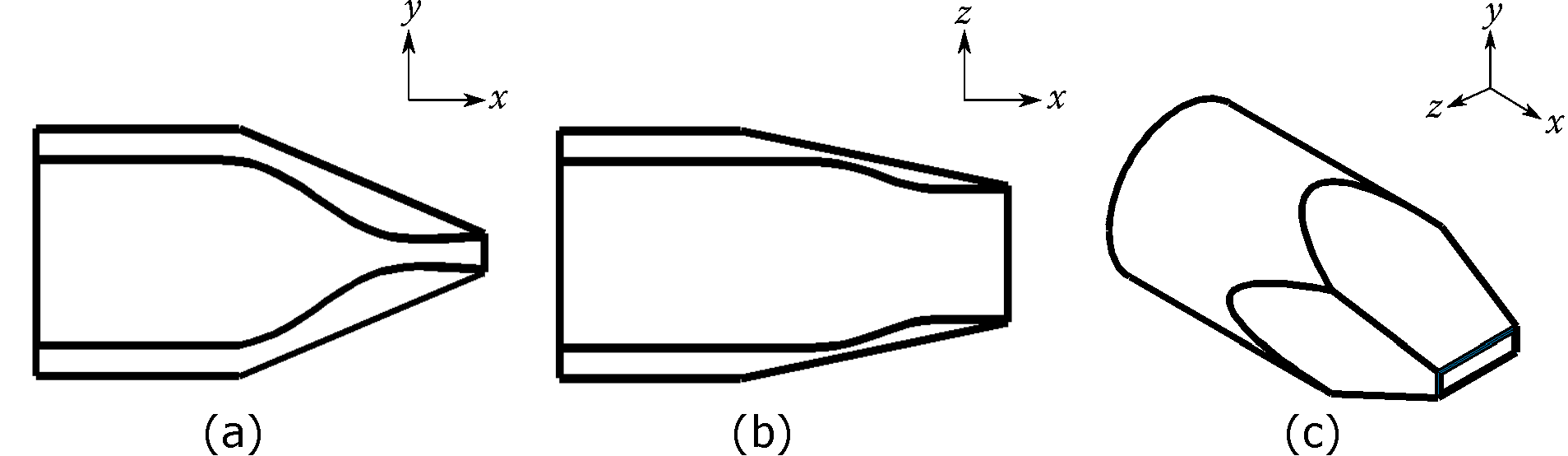} 
 	\end{tabular}
 	\caption{Nozzle geometry shown in the (a) minor-axis plane, (b) major-axis plane, and (c) isometric view.}
 	\label{fig:nozzle_geom}
 \end{figure}

 The rectangular nozzle was designed by the Florida State University (FSU) group for experimental studies of screech in~\citet{alkislar2005structure} and~\citet{valentich2016mixing}. As shown in Figure~\ref{fig:nozzle_geom}, the convergent section is designed using a 5th-order polynomial, and the divergent section in the minor-axis plane is designed by the method of characteristics for an ideal expansion at Mach 1.44. In the major-axis plane, the nozzle surfaces are kept straight downstream of the throat. Starting from a circular cross section of 57.15 mm in diameter, the nozzle smoothly transitions to a rectangular shape at the exit with a 4:1 aspect ratio. The minor dimension of the exit $h$ is 10mm, and the major dimension $b$ is 40mm. The equivalent jet diameter, $D_e = 2\sqrt{hb/\pi}$, is 22.6 mm. In the remainder of this article, $+x$ is used to denote the jet streamwise direction, $y$ is in the minor-axis direction, $z$ is in the major-axis direction, and $\phi$ is the jet polar angle measured with respect to the jet centerline starting from the upstream direction. LES results computed from a mesh containing 140 million grids at three under-expanded screech conditions with $M_j = 1.69$ are used in the analysis, where $M_j$ is the fully expanded jet Mach number under isentropic expansion. The flow parameters are summarized in Table~\ref{tab:flow_param}.

 \begin{table}
 	\centering
 	\begin{tabular}{cccc}
 		
 		$p_{0t}/p_{\infty}$ &  $T_{0t}/T_{\infty}$ & $M_j$ & Simulation Time ($\tilde{t_s}$) \\
 		\hline
 		4.02 & 1 & 1.56 & 1400 \\
 		4.86 & 1 & 1.69 & 1500\\
 		5.57 & 1 & 1.78 & 700\\
 		\hline
 		
 	\end{tabular}
 	\caption{Summary of flow parameters. $p_{0t}$ and $T_{0t}$ are the reservoir stagnation pressure and temperature, $p_{\infty}$ and $T_{\infty}$ are the static pressure and temperature in the ambient. $\tilde{t_s}=t_s c_{\infty}/h$ is the simulation time normalized by the nozzle minor dimension $h$ and ambient speed of sound $c_{\infty}$. The design Mach number of the nozzle is 1.44.}
 	\label{tab:flow_param}
 \end{table}

\begin{figure} 
	\centering
	\begin{tabular}{cc} 
		\includegraphics[width=0.48\linewidth]{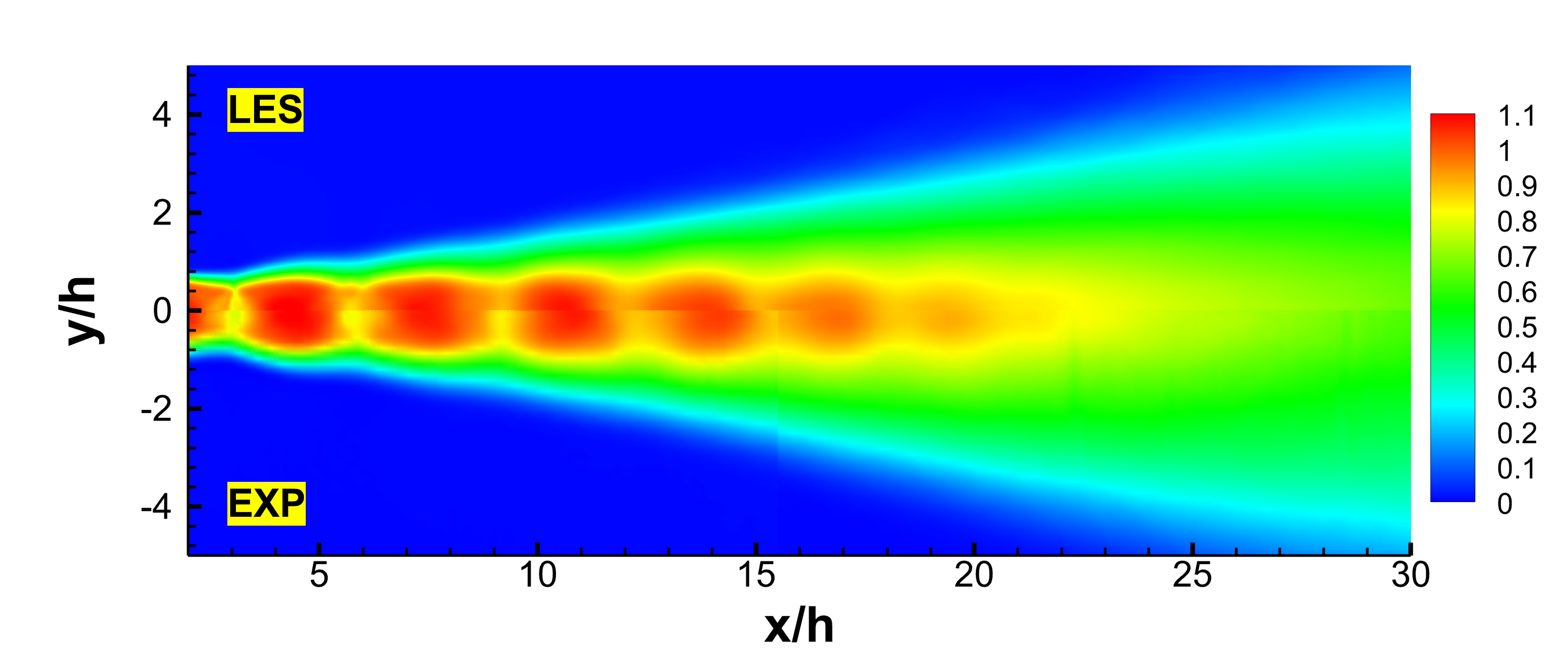} & 
        \includegraphics[width=0.45\linewidth]{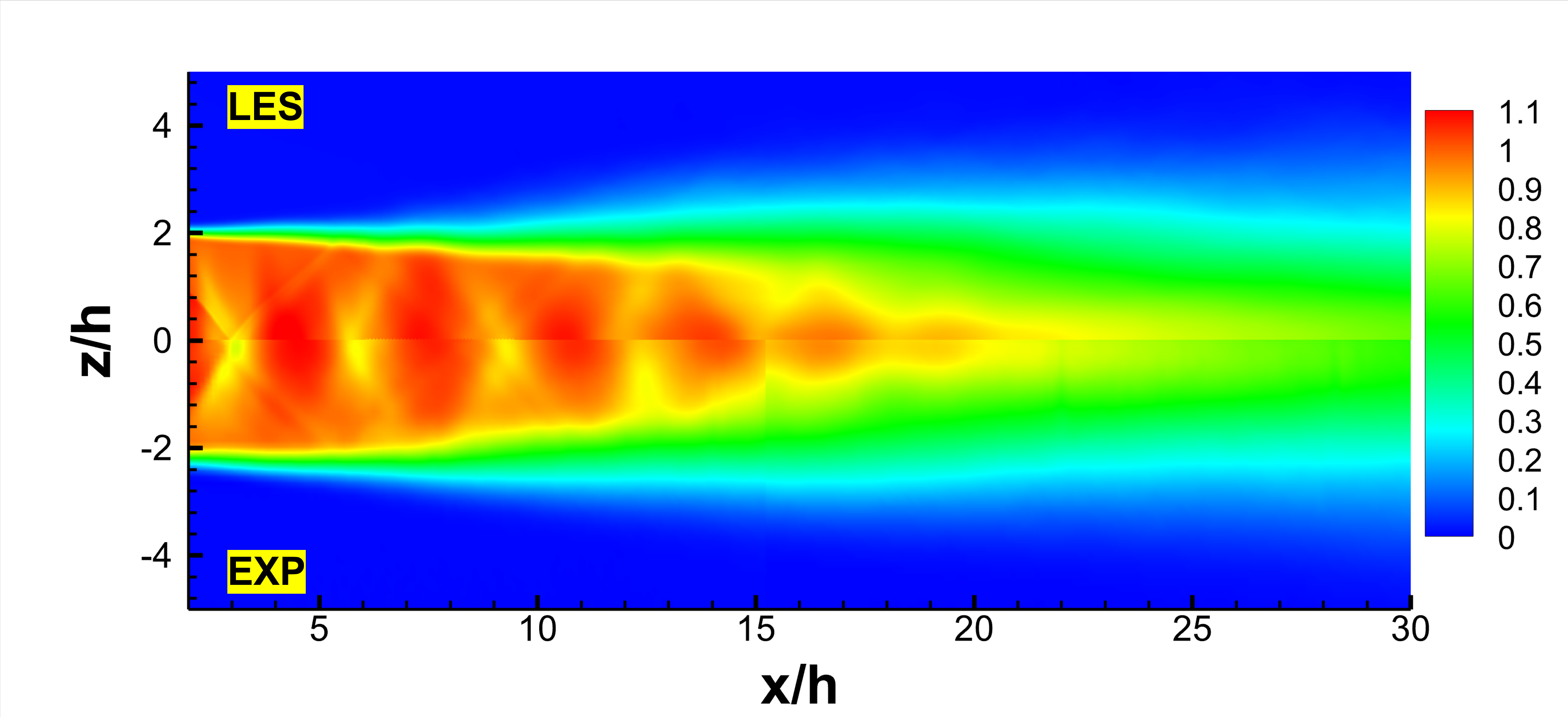}\\
        (a) minor-axis plane & (b) major-axis plane
	\end{tabular}
	\caption{Comparison of mean streamwise velocity contours between LES and PIV measurements for NPR = 4.86}
	\label{fig:mean_velocity_contour}
\end{figure}

\begin{figure} 
	\centering
	\begin{tabular}{cc} 
		\includegraphics[width=0.35\linewidth]{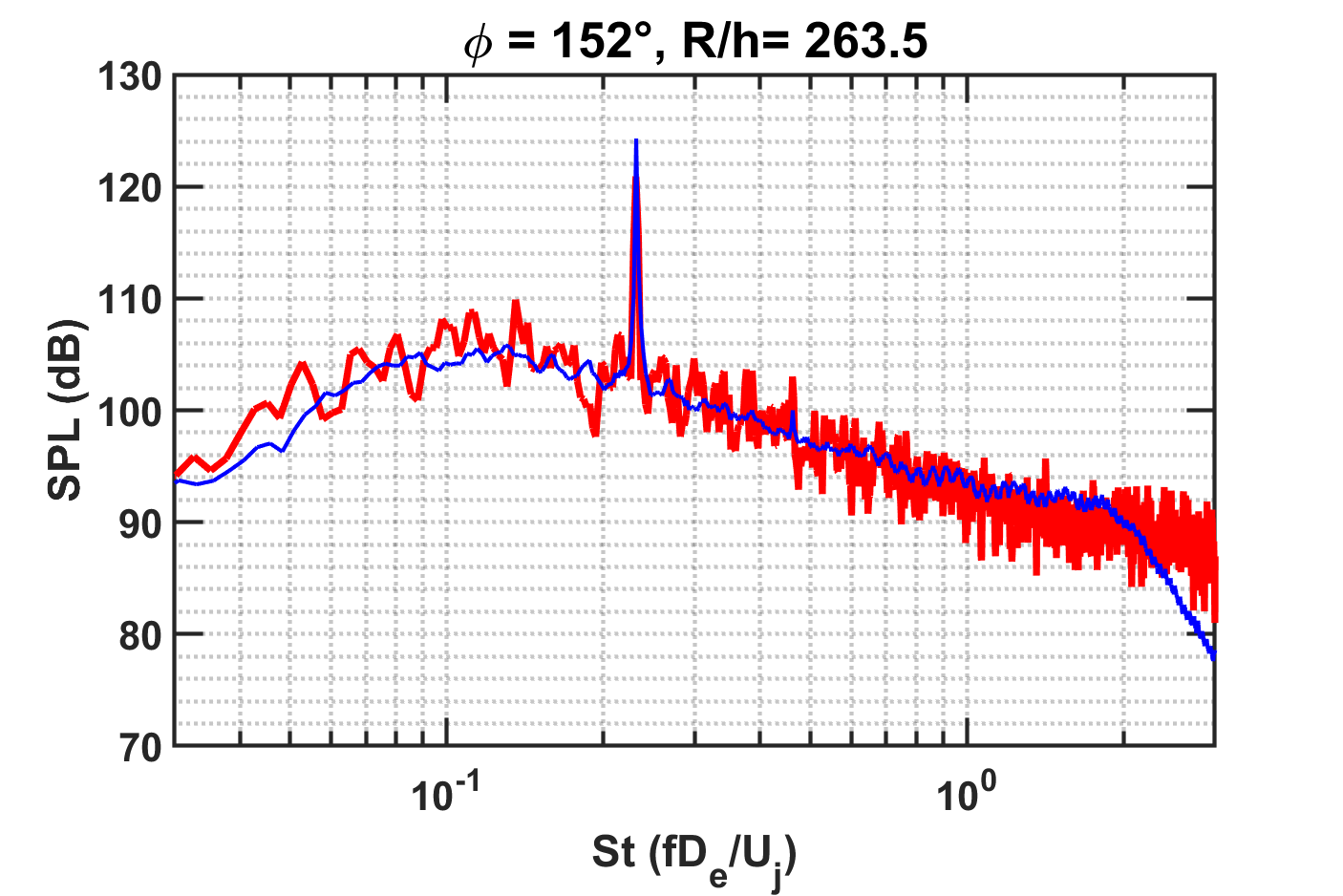} &
        \includegraphics[width=0.35\linewidth]{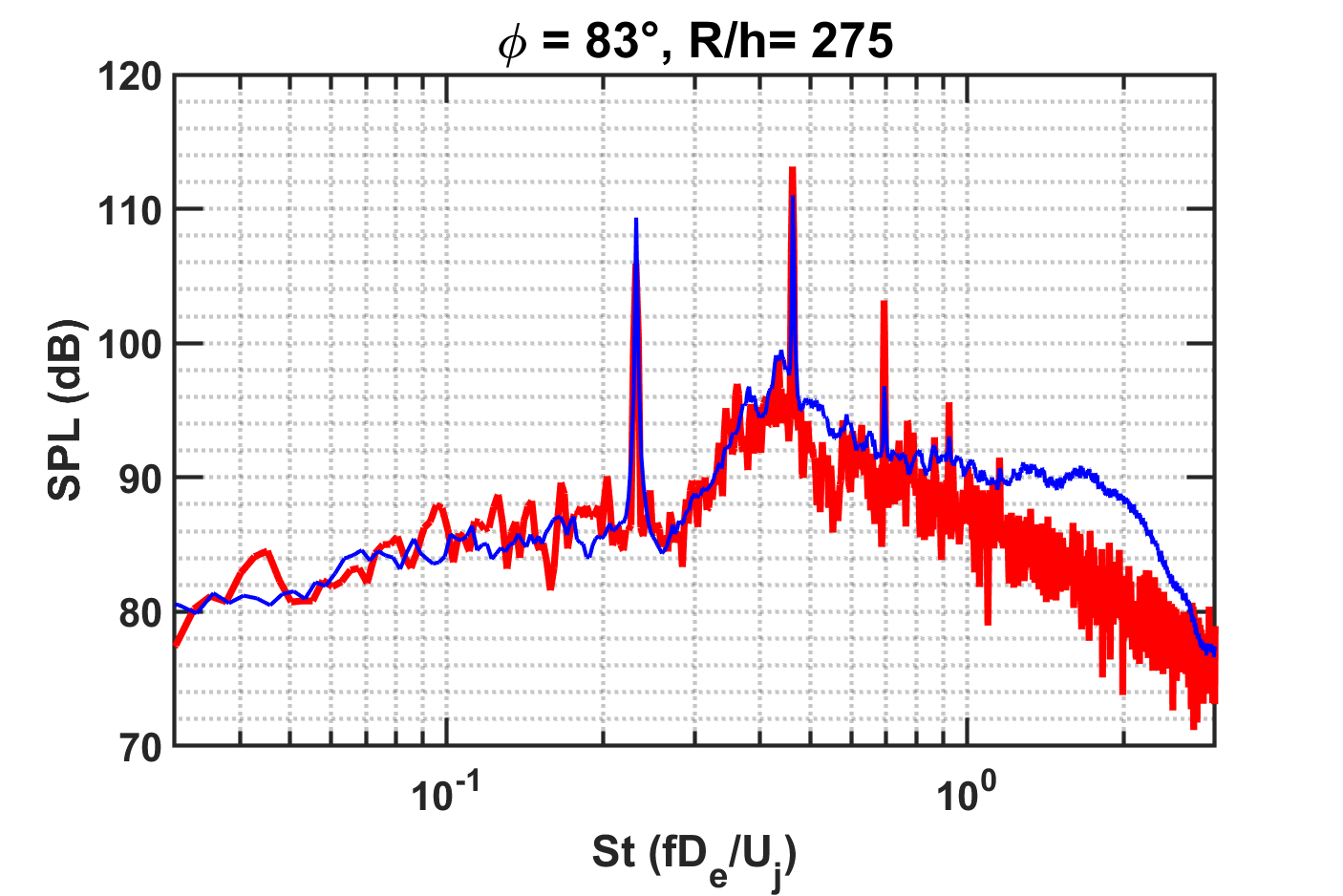}\\
        (a) & (b)\\
	\end{tabular}
	\caption{Comparison of far-field acoustics between LES (red) and Experimental data (blue) in the minor-axis plane.}
	\label{fig:farfield_spl}
\end{figure}

LES are performed using an unstructured compressible flow solver CharLES~\citep{bres2017unstructured} developed at Cascade Technologies, in concert with a Voronoi-based mesh generation paradigm~\citep{bres2018large,bres2019investigating}. CharLES uses a shock-capturing method based on kinetic energy and entropy preserving (KEEP) schemes~\citep{tadmor2003entropy,chandrashekar2013kinetic,fisher2013high}.  The far-field noise of the jet is computed using an efficient permeable formulation~\citep{lockard2000efficient} of the Ffowcs Williams–Hawkings (FW-H) equation~\citep{Ffowcs_etal_1969_Sound} in the frequency domain discussed in details in~\citet{bres2017unstructured}. Figure~\ref{fig:mean_velocity_contour} shows the mean velocity contours in the minor- and major-axis planes at NPR = 4.86. The jet plume is under-expanded and contains a series of shock cells. Figure~\ref{fig:farfield_spl} shows the far-field acoustic spectra at two jet polar angles in the minor-axis plane, with the screech fundamental tone appearing at $St = f D_e/U_j= 0.23$. In the aft-angle direction, the screech fundamental tone is dominant, while near the sideline direction, the second and higher harmonics are present. For a detailed description of the numerical approach for jet screech simulation, and validation of the current LES data with experimental measurements, please see ~\citet{wu2022_aiaaj_assessment}.

\subsection{SPOD Analysis}\label{sec:spod_method}
Spectral proper orthogonal decomposition (SPOD) ~\citep{towne_JFM_2018_spectral_relation,Schmidt_Colonius_AIAA2020_Guide} is used to extract the dominant coherent structures at the screech fundamental frequency from the near-field flow data. Formally, let $\psi(\mathbf{x},t)$ be a zero-mean stochastic variable, and $\Psi$ be the matrix containing the sampled instances of $\psi$,
\begin{equation}
    \Psi = \left[ \psi^{(1)} \psi^{(2)} ... \psi^{(N)} \right] \in\mathbb{R}^{M\times N}
\end{equation}
where $\psi^{(i)},\, i=1,2...,N$ is the $i$th instance of sampled data of dimension $\mathbb{R}^M$. For the current work, $\psi$ is the fluctuations of any state variable from its mean in the turbulent flow field. $M = n_t\times{n_{xyz}}\times{n_{var}}$, where $n_t$ is the number of sampled points in time, $n_{xyz}$ is the number of spatial coordinates, and $n_{var}$ is the number of sampled variables. 
By computing Fourier transform in time on $\Psi$, one could obtain,
\begin{equation}
    \hat{\Psi} =  \left[ \hat{\psi}^{(1)}\hat{\psi}^{(2)}... \hat{\psi}^{(N)} \right] \in\mathbb{C}^{M\times\mathrm N}
\end{equation}
The SPOD analysis solves the following eigenvalue problem,
\begin{equation}
    \left(\frac{1}{N-1} \hat{\Psi} \hat{\Psi}^{\mathrm{H}}\right) \hat{\Phi} = \hat{C} \hat{\Phi} =\hat{\Phi} \Lambda
\end{equation}
where $\hat{\Phi}$ is the matrix containing the right eigenvectors of $\hat{C}$,
\begin{equation}
    \hat{\Phi} =[\hat{\phi}^{(1)} \, \hat{\phi}^{(2)}\, ...  \, \hat{\phi}^{(N)} ], \; \hat{\Phi}\in \mathbb{C}^{M\times N}
\end{equation}
and $\Lambda$ is the diagonal matrix containing the eigenvalues,

$$\Lambda=
\begin{bmatrix}
    \lambda_1 & & &\\
    &  \lambda_2 &  &\\
    &  & \ddots & \\
    & & & \lambda_N
\end{bmatrix}.
$$
In the literature, the eigenvalues are also referred to as the energy spectra of the SPOD modes~\citep{Schmidt_etal_JFM_2018_Spectral_Jet_Turb}. 
By the Karhunen-Loeve theorem, the SPOD mode functions give an optimal representation of $\hat{\psi}$, with $N$ ranked eigenvalues representing the contribution of each mode to the variance of the sampled data. 
\begin{equation}
    \hat{\psi}(\mathbf{x},f) \approx \sum_{j=1}^N \sqrt{\lambda_j (f)} \hat{\phi}_j(\mathbf{x},f), \! \lambda_1 \geq \lambda_2 ... \lambda_N \geq 0
\end{equation}
 For the SPOD analysis used in the current work, data from LES are interpolated onto a 2D structured grid, defined by $x/h \in [0,  150], y/h \in [-20,20], z/h =0$ and $\Delta x/h = \Delta y/h = 0.1$, leading to $n_{var} =5$ and $n_{xyz} = 1501 \times 401 \times 1$. To reduce the spectral noise of the discrete Fourier Transform in time, the Welch's method is used. The whole data sample of size $N_t$ is divided into multiple smaller blocks of size $n_{t}$ with a 50\% overlap. For the case at $\mathrm{NPR} = 4.86$, $N_t$ is $3500$ and $n_{t}$ is $1000$, while for the other two cases, $N_t$ is $2000$ and $n_{t}$ is $800$. As a result, $N=6$ for the case at $\mathrm{NPR}=4.86$ and $N=4$ for the other cases. 

\subsection{Spatial Cross Correlation }\label{sec:spatial_corr_method}
Screech involves interactions between spatially-modulating travelling waves in the streamwise direction. This can be revealed by spatially cross-correlating signals between the nozzle exit and various downstream locations. Following the method outlined in~\cite{wu2020_ctrab}, if a traveling wave is detected by two probes at different streamwise locations, the signals are related as
\begin{equation}
    q(x_2, y_c, t) = \alpha q(x_1, y_c, t-\tau) + n(t) ,
\end{equation}
where $q(x_i, y_c, t)$ is a zero-mean stationary signal of any flow variable at a particular streamwise location $x_i$ along the line $y=y_c$, $\alpha$ is the coefficient of growth or decay in wave magnitude, $\tau$ is the time delay accounting for the traveling time from one location to another, and $n(t)$ is the random noise with a zero mean which is uncorrelated with the signal at station $x_1$.

The cross-correlation function between the stationary signals is 
\begin{align}
    R_{12}(\tau') &= E\left[q(x_1, t) q(x_2, t+\tau') \right] \nonumber\\
    &= E\left[ q(x_1, t) \left(\alpha q(x_1, t+\tau'-\tau) + n(t)  \right)\right] \nonumber \\
    &= \alpha R_{11}(\tau' - \tau) \label{eq:1},
\end{align}
where the autocorrelation function is
\begin{equation}
     R_{11}(\tau') = E\left[q(x_1, t) q(x_1, t+\tau')\right].
\end{equation}
Using the Wiener-Khinchin relations, the cross-correlation function can be calculated by
\begin{equation}
     R_{12}(\tau')= \int_{-\infty}^{\infty} S_{12}(f) e^{i 2\pi f \tau'} \mathrm{d}f.
\end{equation}
$S_{12}$ is the cross-spectral density function between $q(x_1, y_c, t)$ and $q(x_2, y_c, t)$.
Here if we use the signal from the normalized SPOD mode function at the screech fundamental frequency $f_{sc}$, then
\begin{equation}
    q(x,y,t) = \hat{\phi}_1 (x,y,f_{sc})e^{i 2\pi f_{sc} t}
\end{equation}
where $\hat{\phi}_1$ is the leading order SPOD mode. Since $q$ now becomes a harmonic function in time, the cross-spectral density function is simply
\begin{equation}
    S_{12}(f) =\left\{
                \begin{array}{ll}
                   0  \hspace{0.5cm} \textrm{if} f\neq f_{sc} \\
                  \frac{1}{T}\hat{q}^*(x_1,y_c,f_{sc})\hat{q}(x_2, y_c, f_{sc}) \hspace{0.5cm} \textrm{otherwise,}\\
                \end{array}
              \right.
\end{equation}
where $T$ is the screech period. The correlation function becomes 
\begin{align}
    R_{12}(\tau') =  \hat{q}_1^{*}(f_{sc})\hat{q}_2(f_{sc}) e^{i 2\pi f_{sc} \tau'},   \label{eq:2}\\
    R_{11}(\tau'-\tau) = \hat{q}_1^{*}(f_{sc})\hat{q}_1(f_{sc}) e^{i 2\pi f_{sc} (\tau'-\tau)}\label{eq:3}.
\end{align}
Here $\hat{q}_1(f_{sc})$ and $\hat{q}_2(f_{sc})$ are short notations for $\hat{q}(x_1,y_c,f_{sc})$ and $\hat{q}(x_2,y_c,f_{sc})$. By substituting Eqs.~\eqref{eq:2}-\eqref{eq:3} to Eq.~\eqref{eq:1} and solving for $\tau$ and $\alpha$, we get

\begin{gather}
    \tau = \frac{\mathrm{arg}\zeta}{-i2\pi f_{sc}}, \hspace{0.5cm}
    \alpha =|\zeta|, \hspace{0.5cm}  \zeta = \frac{\hat{q}_1^*\hat{q}_2}{\hat{q}_1^{*}\hat{q}_1}.\label{eq:4} 
\end{gather}

If $x_1 = 0$, this analysis can highlight the change in magnitude and phase of a traveling wave relative to the nozzle exit. The leading-order SPOD mode at screech frequency gives the spatial-temporal coherent structure with the largest contribution to the variance of the sampled data. With spatial cross correlation, the fraction of the SPOD mode that is correlated in the streamwise direction is extracted and can be used to interrogate the energy propagation for screech generation.

Equation~\eqref{eq:4} gives $\alpha$, the magnitude variation of the traveling wave relative to the reference point at $x/h=0$. From here, one could also determine the absolute magnitude using the SPOD energy spectra $\lambda$,
\begin{equation}
    \hat{A}(x,y,f_{sc}) =\sqrt{\lambda} \alpha   |\hat{q}_1|
\end{equation}
Finally, we define a quasi-one-dimensional measure of wave energy as
\begin{equation} \label{eq:wave_energy}
    \mathrm{E}(x,f_{sc}) = \int_{y_1}^{y_2} |\hat{A}|^2 \mathrm{d}y
\end{equation}


\section{Dominant Coherent Structures in Screech Generation} \label{sec:dominant coherence}
\begin{figure}
	\centering
	\begin{tabular}{cc}
		\includegraphics[width=0.45\textwidth]{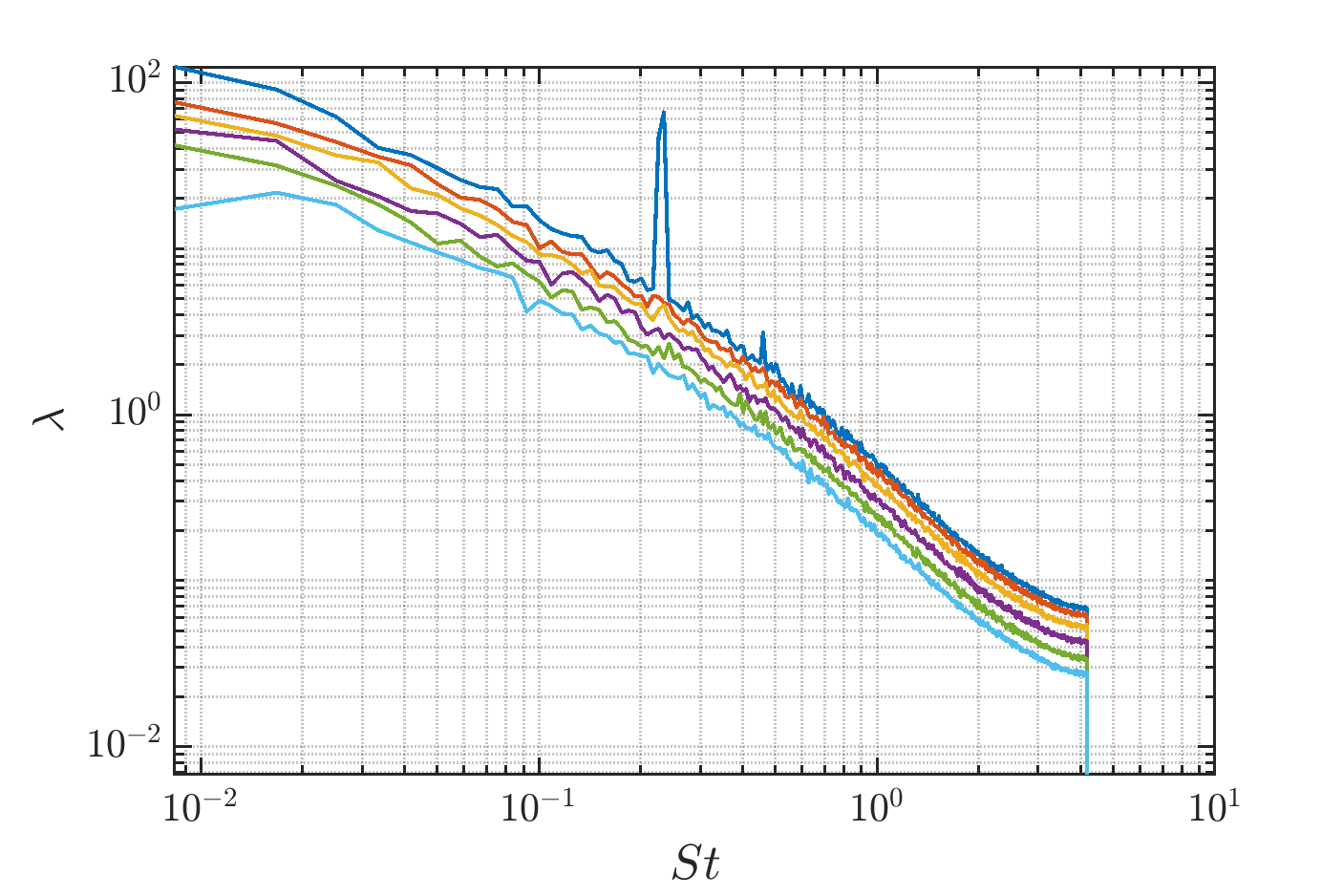} &
		\includegraphics[width=0.45\textwidth]{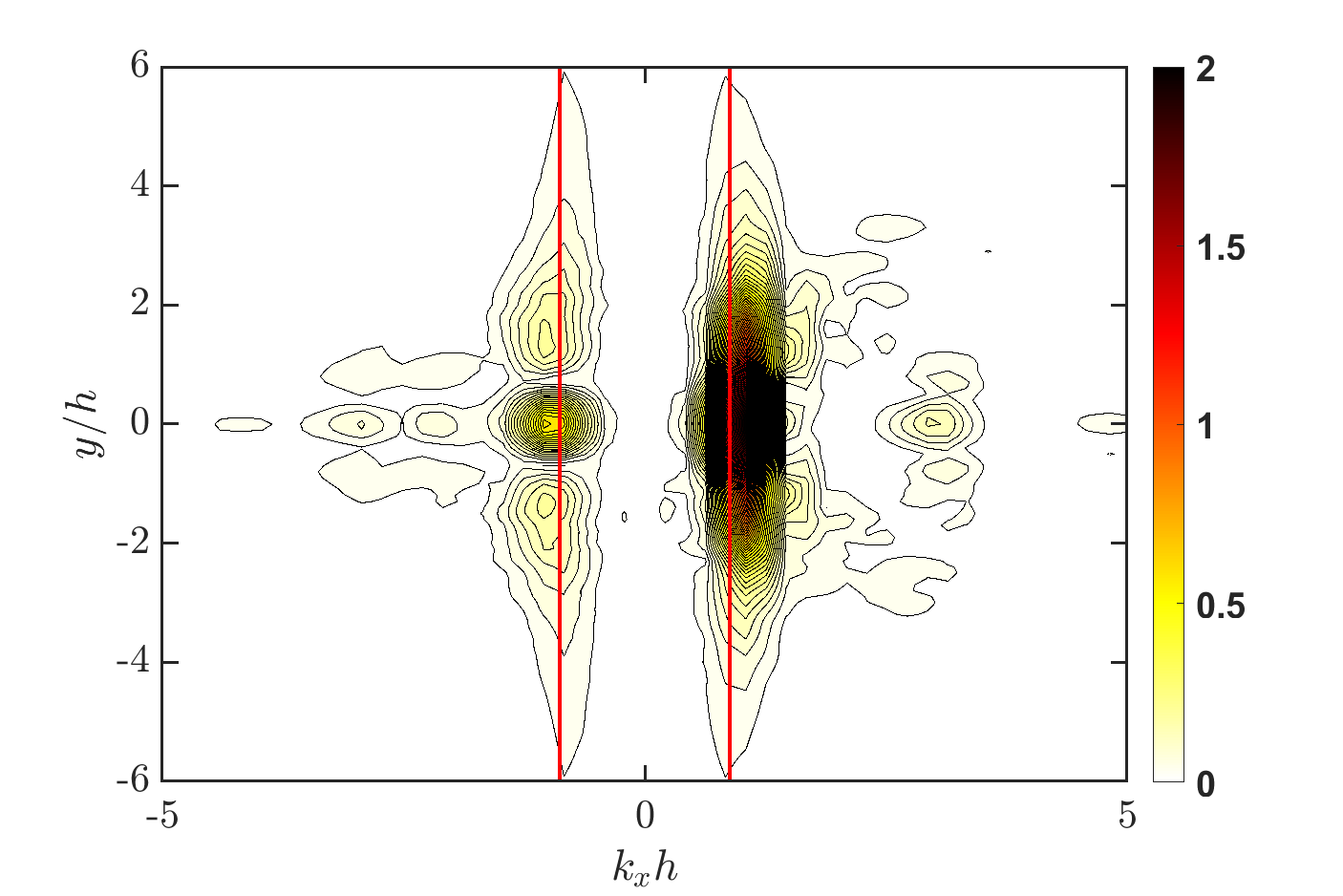} \\
		(a) & (b) \\
	\end{tabular}
	\caption{(a) SPOD energy spectra for $v'$ and (b) modulus of the first SPOD mode shape function with respect to the streamwise wavenumber $k_x$ and $y$. }
	\label{fig:Spod_energy_v}
\end{figure}

With the procedure outlined in Section~\ref{sec:spod_method}, the leading-order SPOD mode for the case at $M_j=1.69$ is computed using $700$ acoustic time units of data. The acoustic time unit is defined by the time it takes for sound in the ambient to travel a distance of $h$, the minor dimension of the nozzle exit. Figure~\ref{fig:Spod_energy_v} (a) shows the energy spectra of the SPOD modes for transverse velocity fluctuations $v'$. A high-energy peak is seen at the screech fundamental frequency, $St =0.23$, with clear rank separation between the first mode and the others. Figure~\ref{fig:Spod_energy_v}(b) shows the modulus of the first SPOD mode function at $St = 0.23$ with respect to the wavenumber in the streamwise direction $k_x$ and the transverse coordinate $y$. The contour plot indicates the presence of waves with opposite signs of wavenumbers. Here, the sign of the wavenumber can be used as a proxy for the direction of wave propagation. The interference of these oppositely-traveling waves give rise to spatially modulating standing waves in the jet plume, which have been observed and studied in detail in previous screech experiments~\citep{panda1999AnExp}.

\begin{figure}
	\centering
	\begin{tabular}{cc}
		\includegraphics[width=0.45\textwidth]{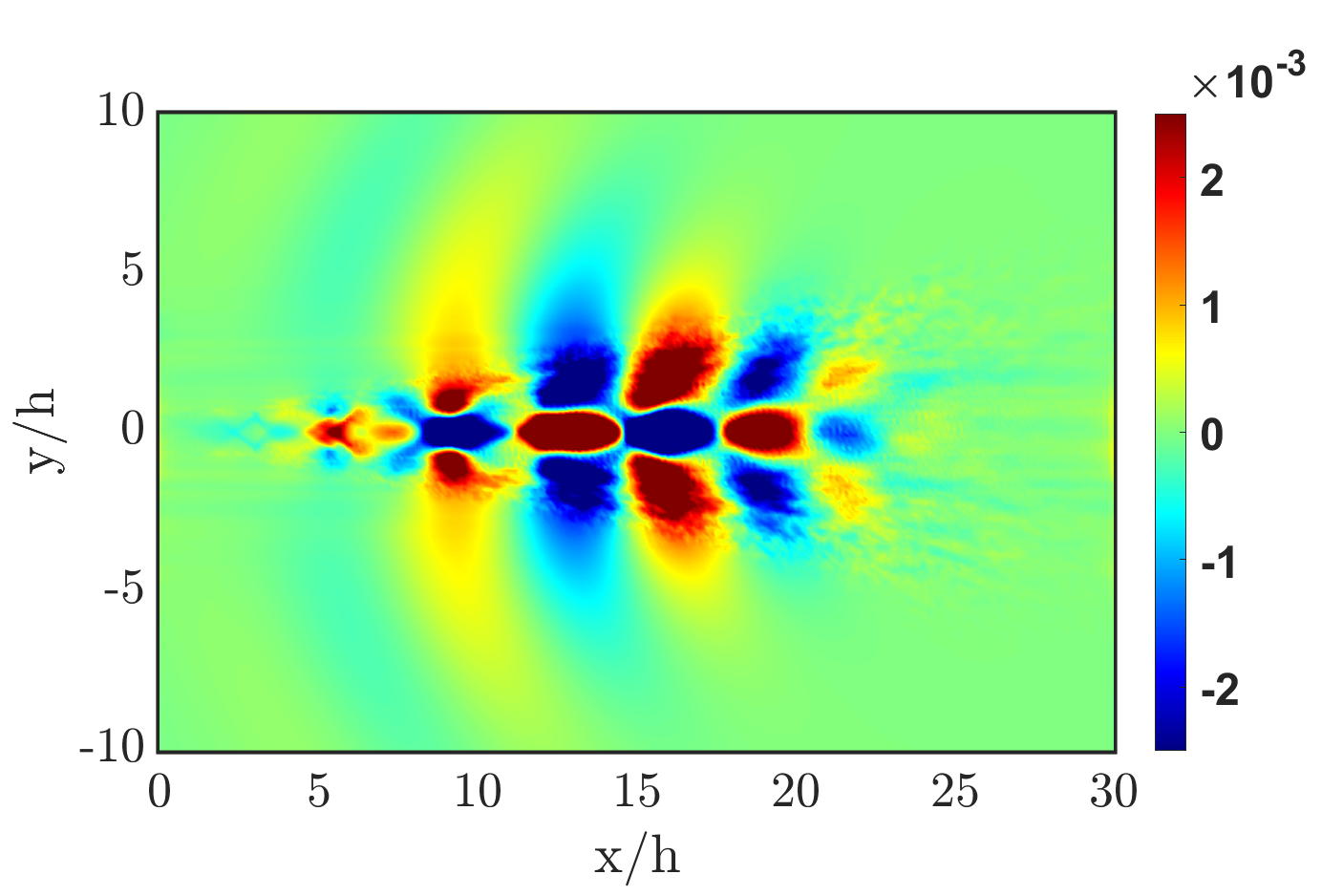} &
		\includegraphics[width=0.45\textwidth]{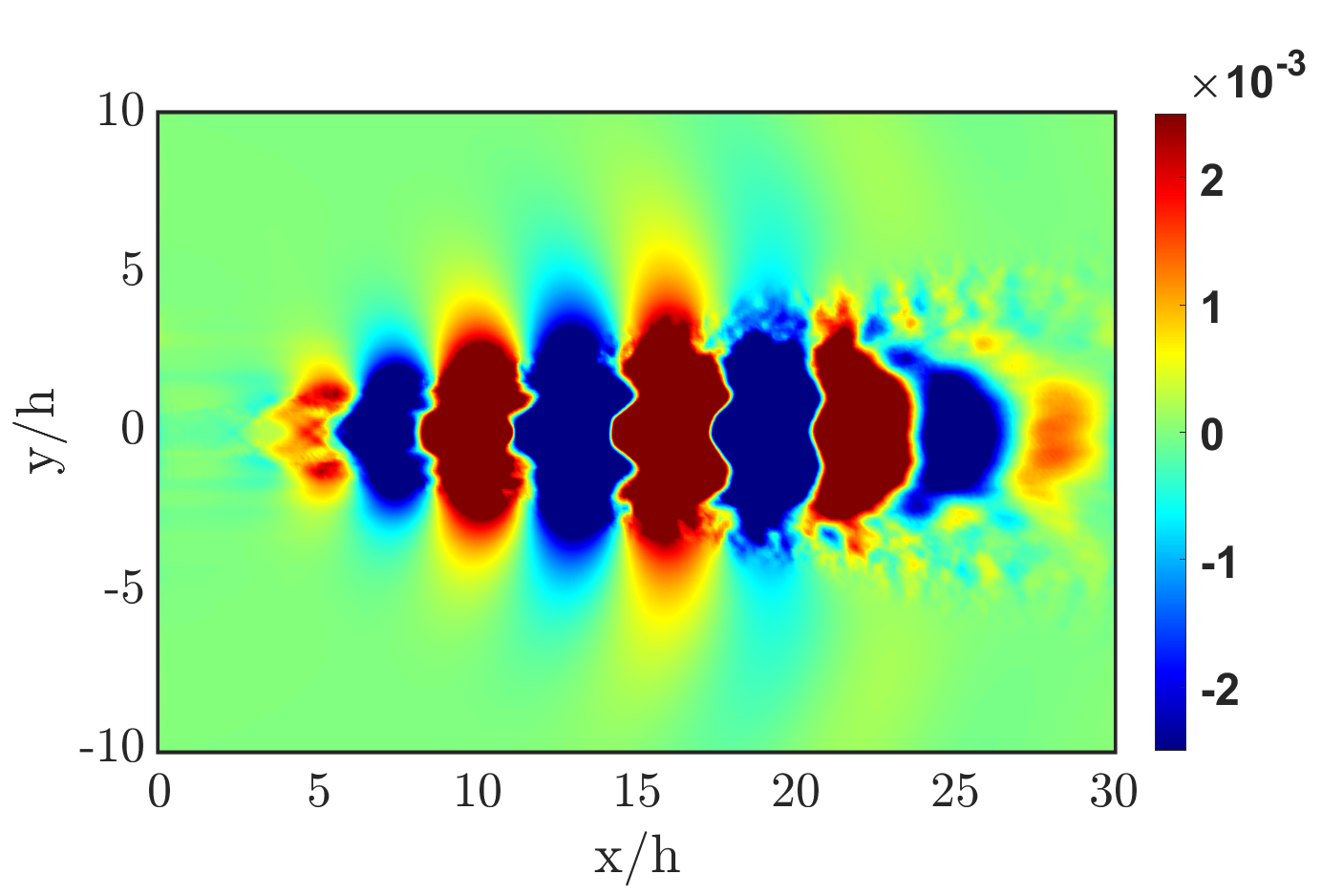} \\
		(a) & (b) 
	\end{tabular}
	\caption{(a) Upstream-traveling waves in the $v'$ SPOD mode; (b) downstream-traveling waves in the $v'$ SPOD mode.}
	\label{fig:Spod_decomp_v}
\end{figure}

By only using the positive or negative $k_x$ for spatial reconstruction, figure~\ref{fig:Spod_decomp_v} separates the downstream-traveling and upstream-traveling components from the SPOD mode. The downstream-traveling waves are the K-H instability waves commonly seen in subsonic and supersonic jets \citep{Schmidt_etal_JFM_2018_Spectral_Jet_Turb,Nekkanti2020}. The upstream-traveling waves contain both external acoustic waves and the guided jet mode~\citep{tam1989onthethree}. To facilitate the remainder of the discussion, we use $c_-$ to denote the upstream-traveling acoustic wave, $k_-$ to denote the upstream-propagating guided mode and $k_+$ to denote the downstream-propagating K-H wave.

Figure~\ref{fig:Spod_energy_v}(b) shows both the upstream- and the downstream-traveling waves are composed of multiple Fourier modes, with one dominant group close to the acoustic wavenumber and additional subdominant groups at higher wavenumbers. The peak wavenumbers for $k_-$ and $k_+$ are summarized in Table~\ref{tab:kx_summary}, alongside with the peak wavenumber of the shock cells $k_s$ from the mean flow data. Figure~\ref{fig:Spod_energy_v}(b) indicate the peak wavenumbers for $k_-$ and $k_+$ have a phase velocity slightly less than the ambient speed of sound. This feature agrees with other experimental and theoretical works~\citep{Tam_Ahuja_JFM1990,edmitch_2021JFM_waves_in_screech}. Values of $k_-$ and $k_+ - |k_s|$ indicate the guided jet mode, the K-H wave and the shock cell structures could exchange energy via triadic interactions. Strong interactions between the K-H wave and the shock cells could excite the guided jet mode, and the growth of the guided jet mode could further reinforce the K-H and shock interactions. This has also been suggested by recent experimental and theoretical works~\citep{edgington2018upstream,edmitch_2021JFM_waves_in_screech,edmitch_jfm2022_unifying}. The SPOD modes for other variables, including fluctuations of density, fluctuations of velocity in $x$ and $z$ direction as well as pressure, have also been computed. The results are presented in Appendix~\ref{appA}.

\begin{table}
	\centering
	\begin{tabular}{ccccccc}
         NPR & $\mathrm{St_{sc}}$ & Range of $x/h$ & $k_-$ & $k_+$ & $k_s$ & $k_+ - |k_s|$ \\
		\hline 
        4.02 & 0.31 & $[0, 20]$ & $ -1.26$ & $+1.57$ & $\pm 2.83 $ & $ -1.26 $ \\
		\textbf{4.86} & \textbf{0.23}& $\mathbf{[0,30]}$  &$\mathbf{-1.05}$ & $\mathbf{+1.05}$ & $\mathbf{\pm 2.09}$ & $\mathbf{-1.04}$\\
        5.57 & 0.20 &  $[0, 35]$ & $ -0.90$ & $+0.90$ & $\pm 1.80 $ & $ -0.90 $ \\
	\end{tabular}
	\caption{Summary of the peak wave numbers from SPOD at the screech frequency and the mean shock cell structures. Case in bold is analyzed in detail here and the other two are shown in Section \ref{sec:npr_effects}.}
	\label{tab:kx_summary}
\end{table}

\section{Screech Closure Mechanism} \label{sec:closure_mech}
\begin{figure}
	\centering
	\begin{tabular}{cc}
		\includegraphics[width=0.45\textwidth]{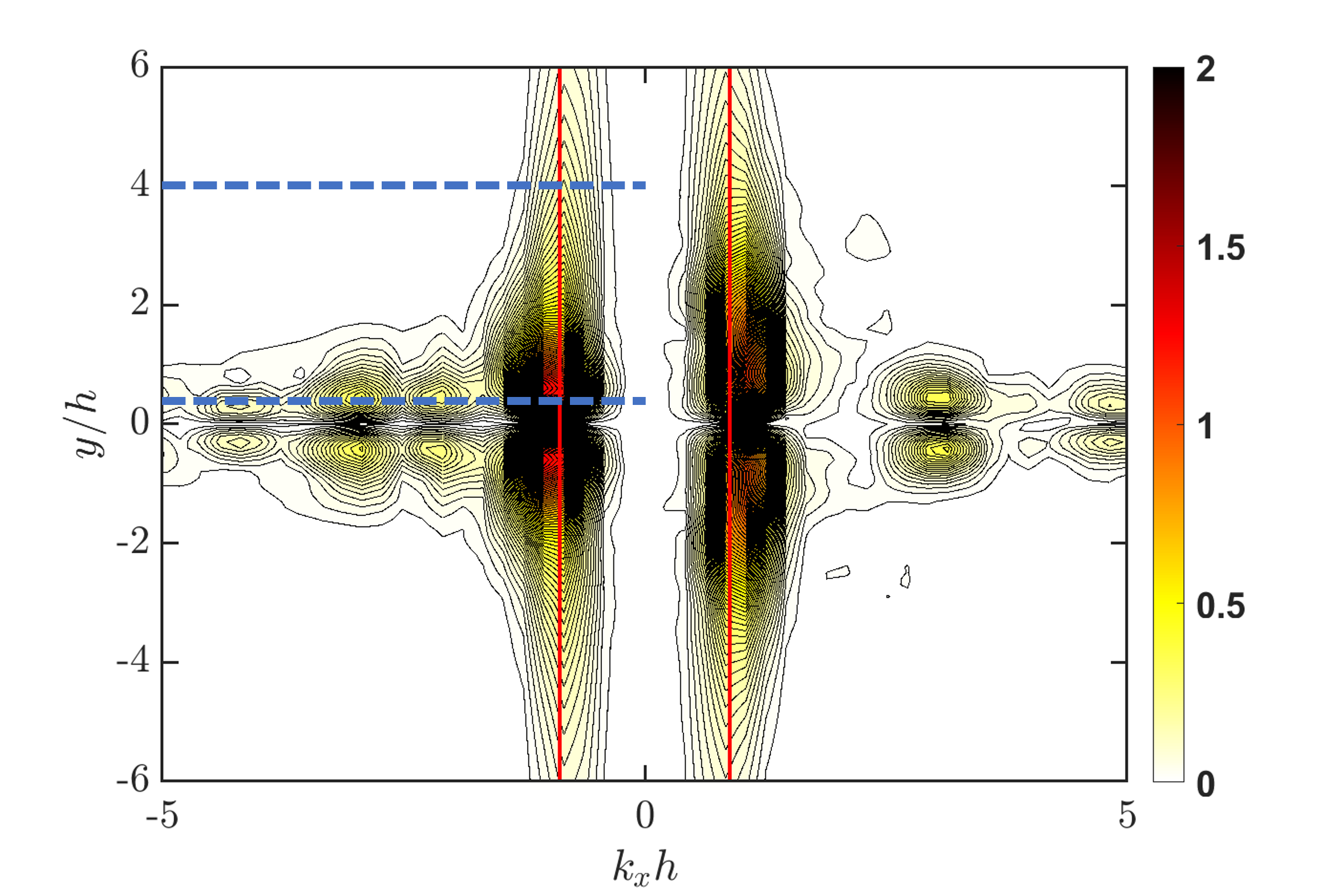} &
        \includegraphics[width=0.45\textwidth]{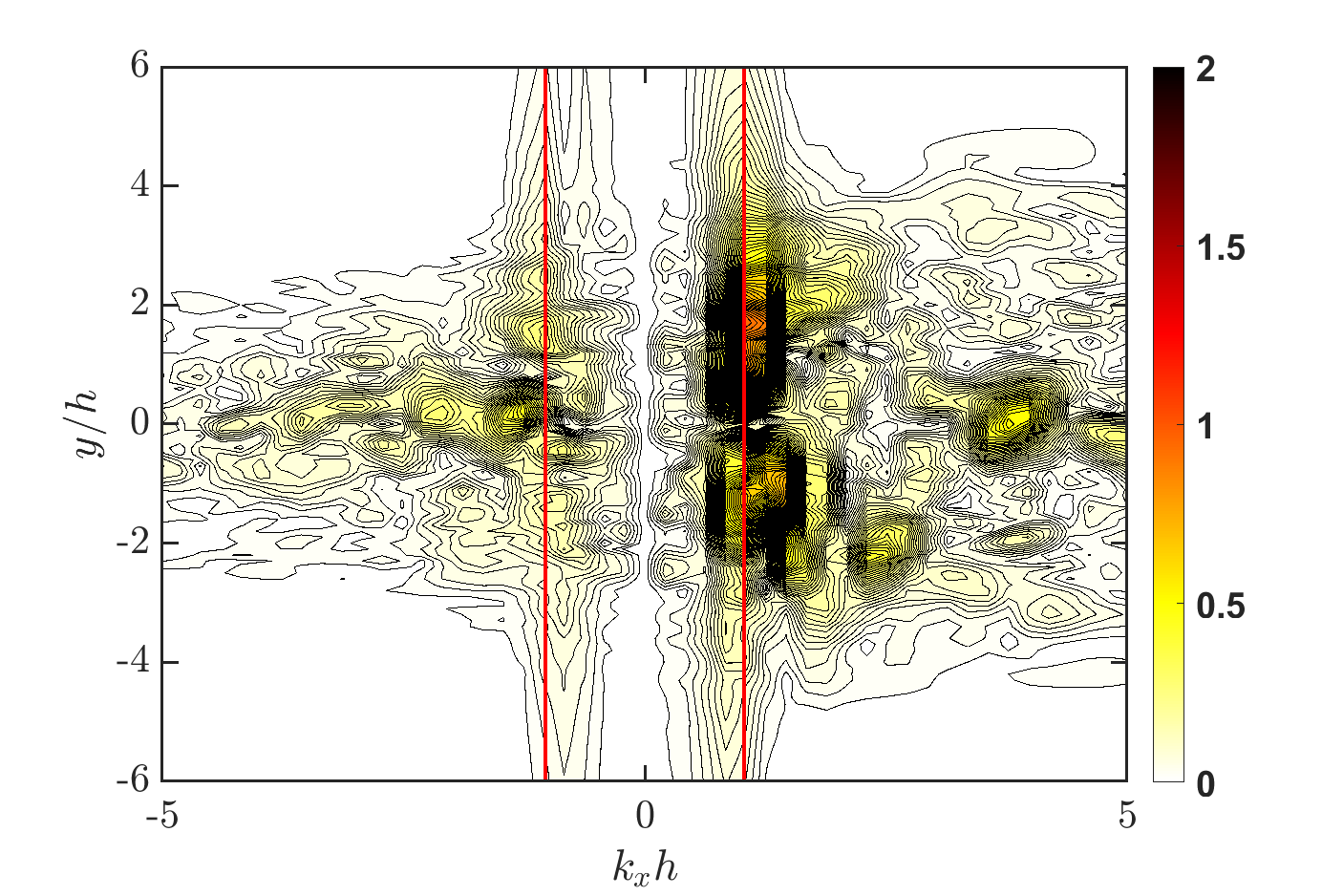} \\
		(a) & (b)\\
	\end{tabular}
	\caption{The modulus of the $p'$ SPOD mode shape function at (a) the screech frequency $St_1=0.234$ and (b) one neighboring frequency.$St_1 + \Delta St = 0.276$. Solid red lines mark the acoustic wavenumber and dashed lines indicate the locations where the spatial cross correlation for the $c_-$ and $k_-$ waves are computed.}
	\label{fig:spod_amp_p_freqs}
\end{figure}

Having obtained the dominant coherent structures at the screech frequency using SPOD, one can further examine the upstream-traveling waves to investigate the screech closure mechanism, the pathway that redirects energy from downstream back to the nozzle exit. To do this, the spatial cross correlation method outlined in Section~\ref{sec:spatial_corr_method} is used to analyze the leading-order SPOD mode for pressure fluctuations $p'$. Using Eq.~\eqref{eq:4}, the time delay $\tau$ and relative magnitude variation $\alpha$ for $c_-$ and $k_-$ with respect to the nozzle exit are computed. To differentiate between acoustic and hydrodynamic fluctuations, the spatial cross-correlation for $c_-$ is computed along $y/h = 4$, and that for $k_-$ is done along $y/h =0.2$. These locations are marked by the dashed blue lines in Figure~\ref{fig:spod_amp_p_freqs}(a).  Figure~\ref{fig:spod_amp_p_freqs}(a) also marks the acoustic wavenumber with the red solid lines. The acoustic waves outside the jet plume center around the acoustic wave number. The $k_-$ wave for $p'$ appears as a pair of symmetric lobes with respect to the jet centerline, and the peak wavenumber lies to the left of the acoustic wavenumber. Figure~\ref{fig:spod_amp_p_freqs}(b) shows the modulus of the SPOD shape function at a slightly higher frequency than the screech frequency. At this non-resonant frequency, there are still upstream-traveling acoustic waves outside the jet plume, but there is no $k-$ wave inside the jet plume, judging by the lack of symmetry with respect to the jet centerline. In addition, the $k+$ wave can be found at both frequencies, so energy still propagates in the downstream direction.  Therefore, the $k-$ wave seems to be the differentiating factor between the screech and the non-screech conditions.

\begin{figure}
	\centering
	\begin{tabular}{cc}
		\includegraphics[width=0.45\textwidth]{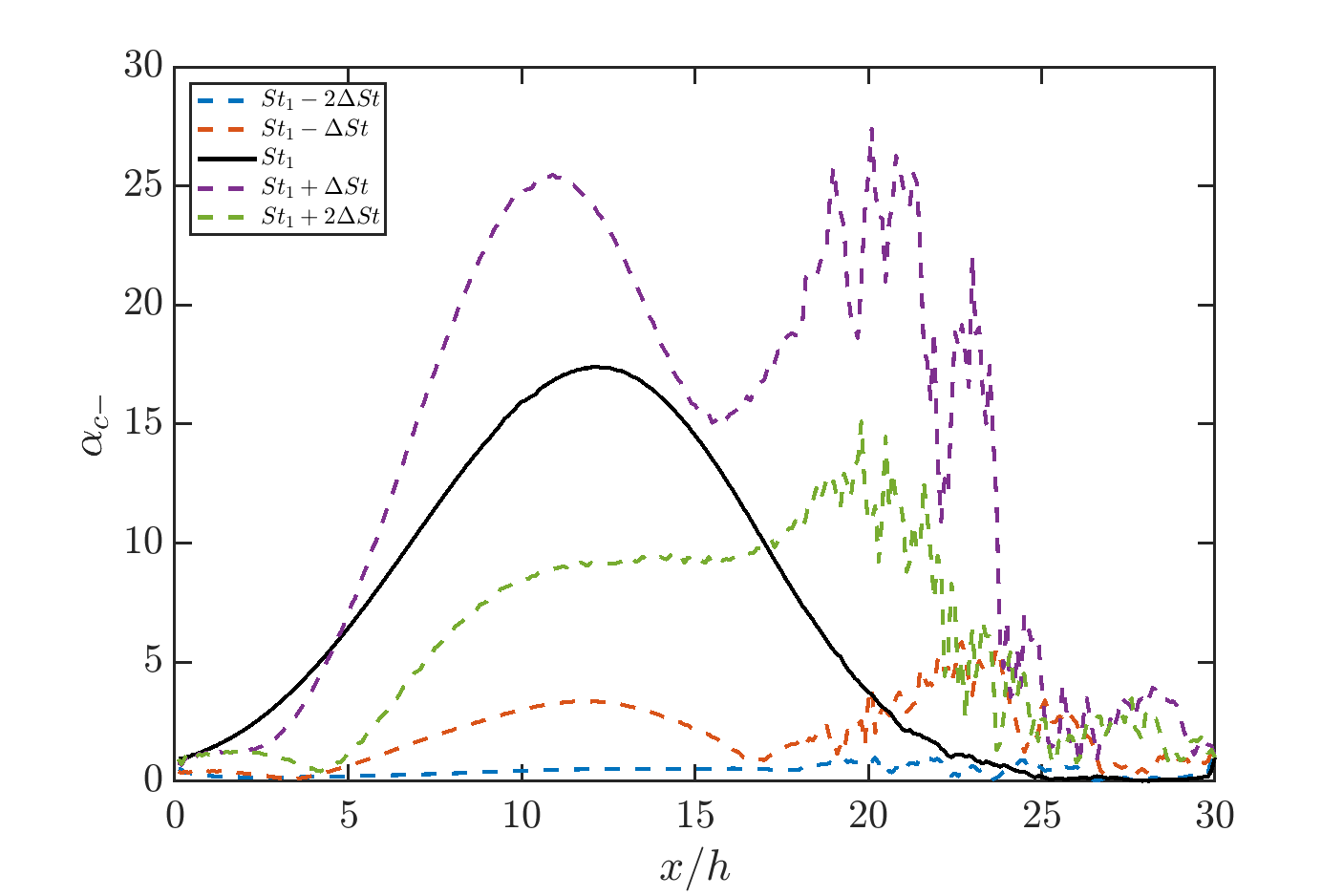} &
        \includegraphics[width=0.45\textwidth]{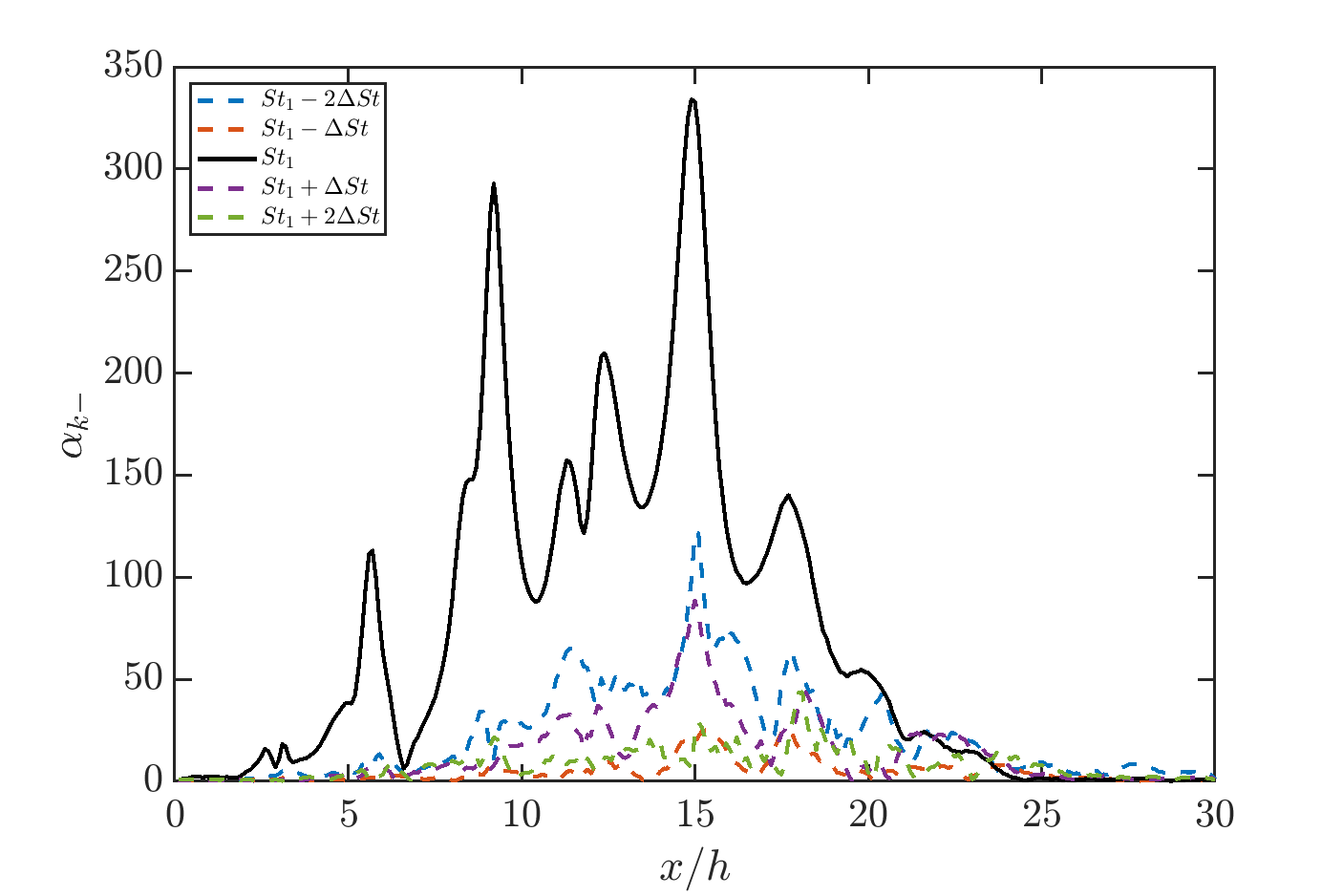} \\
		(a) & (b)\\
	\end{tabular}
	\caption{(a) relative magnitude for the acoustic wave $c_-$, tracked along $y/h=4$ in reference to the point $x/h=0, y/h=4$ and (b) relative magnitude for the guided jet mode $k_-$, tracked along $y/h=0.2$ in reference to the point $x/h=0, y/h=0.2$. Solid lines are for the screech frequency and dashed lines for the neighboring frequencies.$St_1 = 0.234$, $\Delta St = 0.042$.}
	\label{fig:cross_corr_fundamental}
\end{figure}

Results from the cross-correlation further confirms the growth of the $k-$ wave is the key for screech to take place. Figure~\ref{fig:cross_corr_fundamental}(a) shows the relative magnitude $\alpha$ for the wave signal  tracked along $y/h=5$. Because of the relatively large distance away from the jet boundary, this signal would mostly be associated with the acoustic wave $c-$. Figure~\ref{fig:cross_corr_fundamental}(b) cross correlates the wave signal along $y/h=0.2$, indicating the growth of the $k-$ wave inside the jet plume. The spatial growth pattern between $c_-$ and $k_-$ differs in two ways. Firstly, the maximum of $\alpha_{k_-}$ is larger than $\alpha_{c_-}$ by one order of magnitude at the screech fundamental frequency. This shows the spatial growth of $k_-$ wave relative to the nozzle exit is much more significant than that of $c_-$. Secondly, $\alpha_{k_-}$ is highly sensitive to frequency; the spatial growth is the largest at the screech frequency, $St_1$. In contrast, the main peak of $\alpha_{c_-}$ does not vary significantly between $St_1$ and a few of its neighboring frequencies. At the two higher non-resonant frequencies, there are secondary peaks in the ${c_-}$ wave between $x/h=20$ and $x/h=25$ that reach similar or slightly higher values of $\alpha$ than the screech acoustic wave, and these may be the upstream component of the super-directive broadband acoustic radiation, due to the K-H wave near the end of the jet potential core, as discussed by~\citet{Nekkanti2021}. Overall, unlike $k_-$, the growth of $c_-$ is non-sensitive to the range of frequencies close to the neighborhood of the screech frequency. These differences are direct evidence showing the dominance of the $k_-$ wave over the $c-$ wave in closing the screech feedback loop. 

\section{Estimation of Wave Energy} \label{sec:wave_energy}
Section~\ref{sec:dominant coherence} shows energy transfer is possible via triadic interactions among the shock cells $k_s$, the $k_+$ and the $k_-$ waves. Section~\ref{sec:closure_mech} then shows the $k_-$ wave is the deciding factor for propagating energy in the upstream direction. These arguments can be further supported by quantifying the wave energy carried by these oppositely-traveling waves using Eq.~\eqref{eq:wave_energy}. 

\begin{figure}
	\centering
	\begin{tabular}{ccc}
		\includegraphics[width=0.3\textwidth]{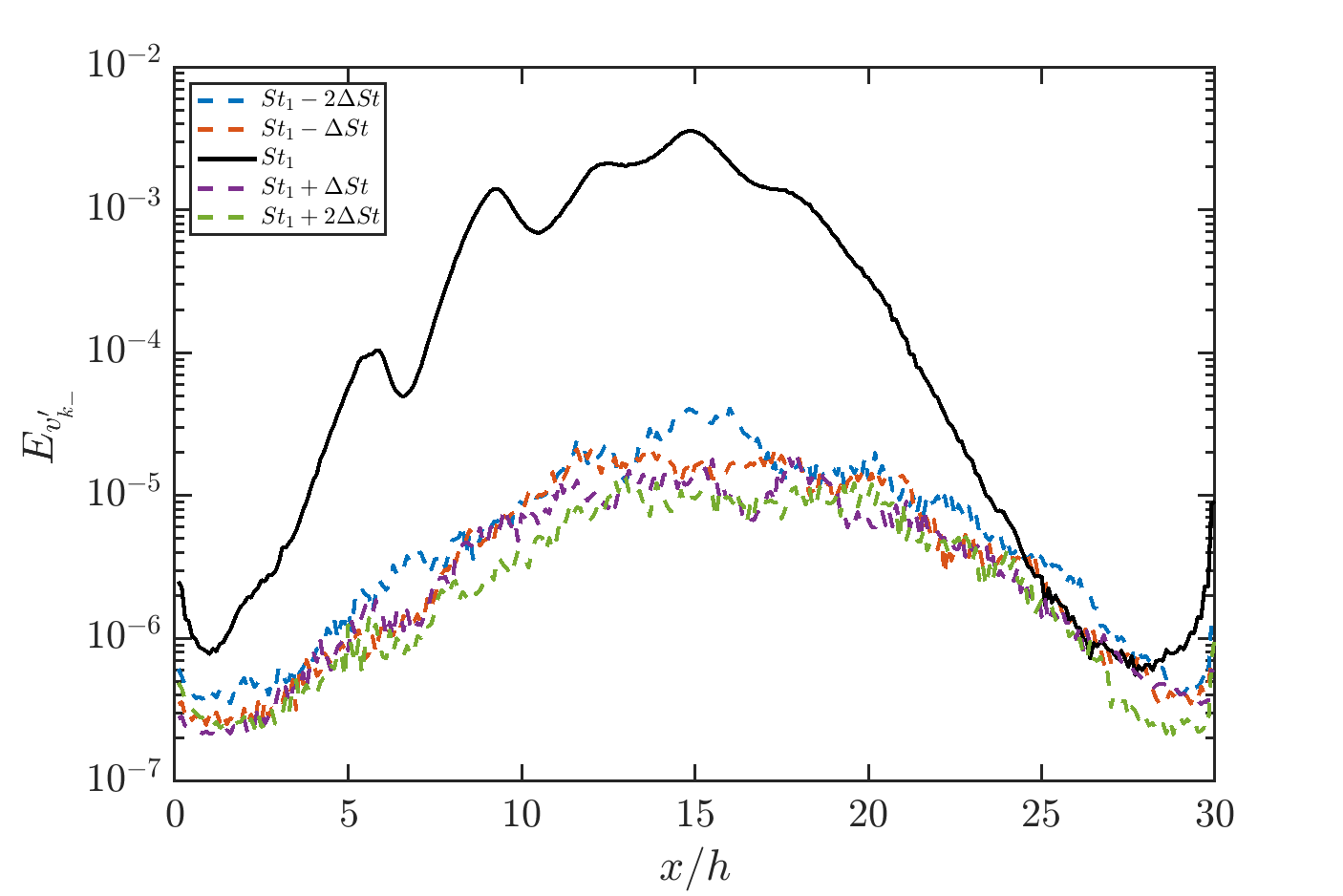} &
		\includegraphics[width=0.3\textwidth]{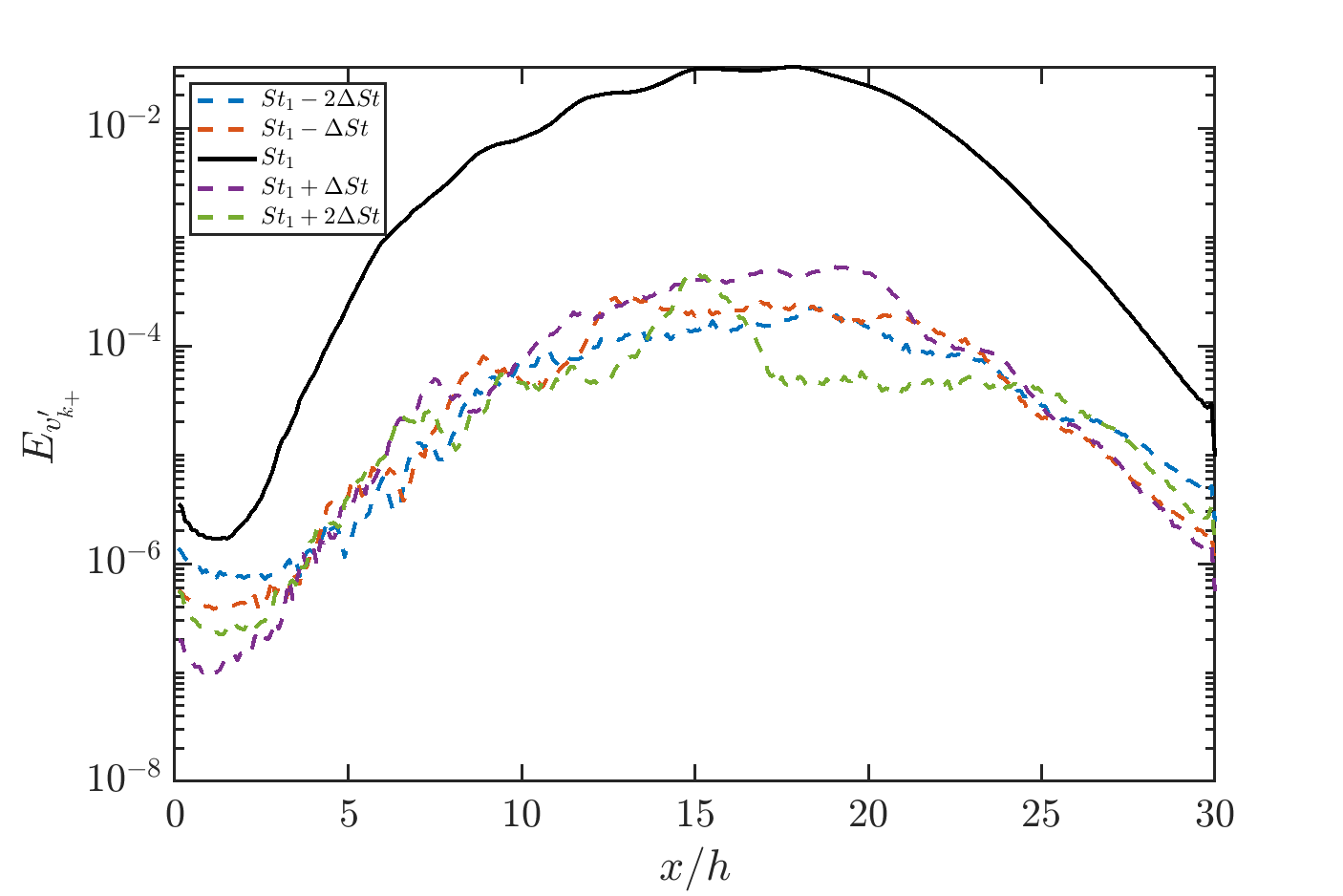} & 
		\includegraphics[width=0.3\textwidth]{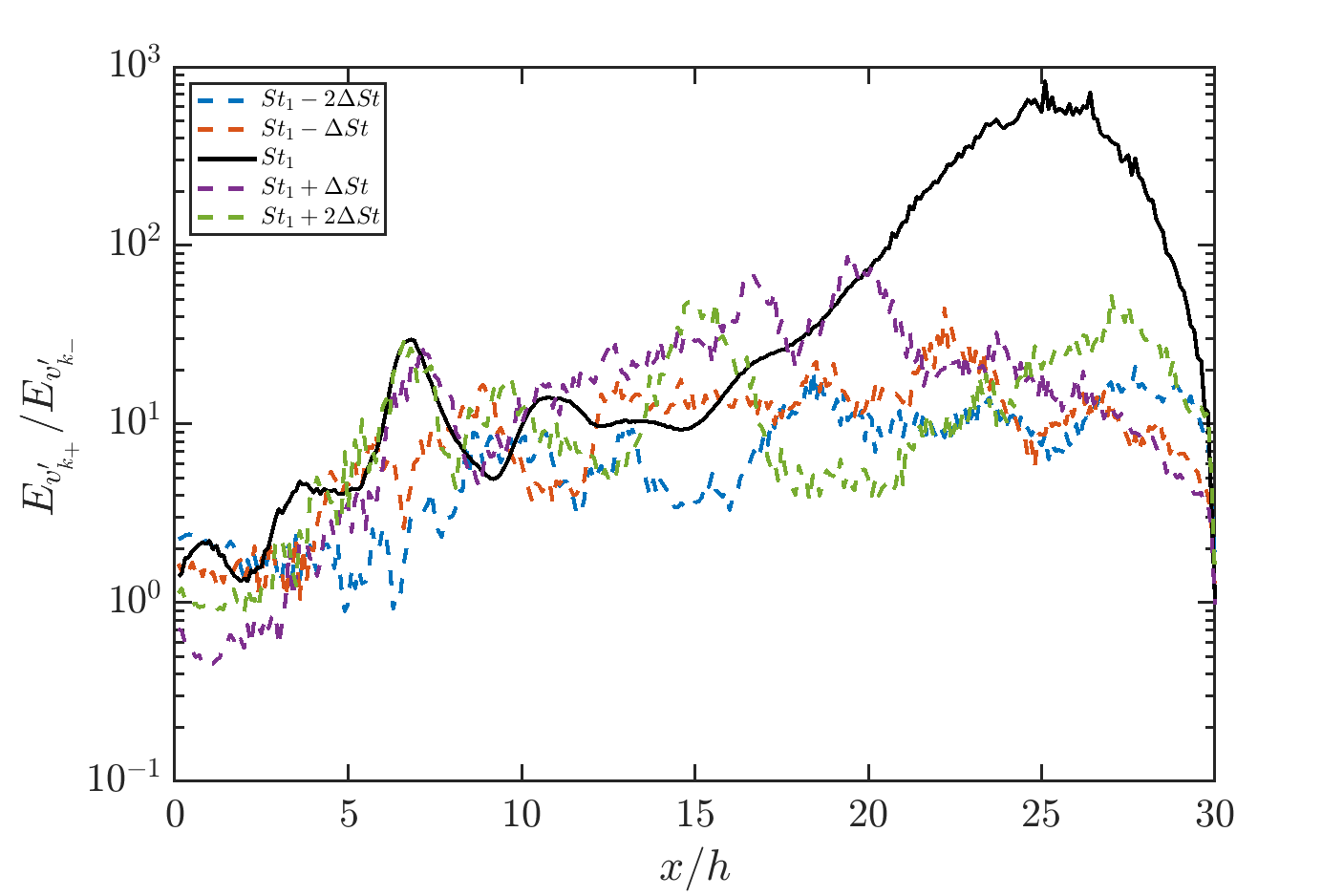} \\
		(a)  & (b) & (c)
	\end{tabular}
        \caption{Wave energy for the $k-$, $k+$ waves and their ratio at various frequencies. Solid lines are for the screech frequency and dashed lines for the neighboring frequencies. }
	\label{fig:E_mult_freq}
\end{figure}

 \begin{figure} 
 	\centering
 	\begin{tabular}{c} 
 		\includegraphics[width=0.5\linewidth]{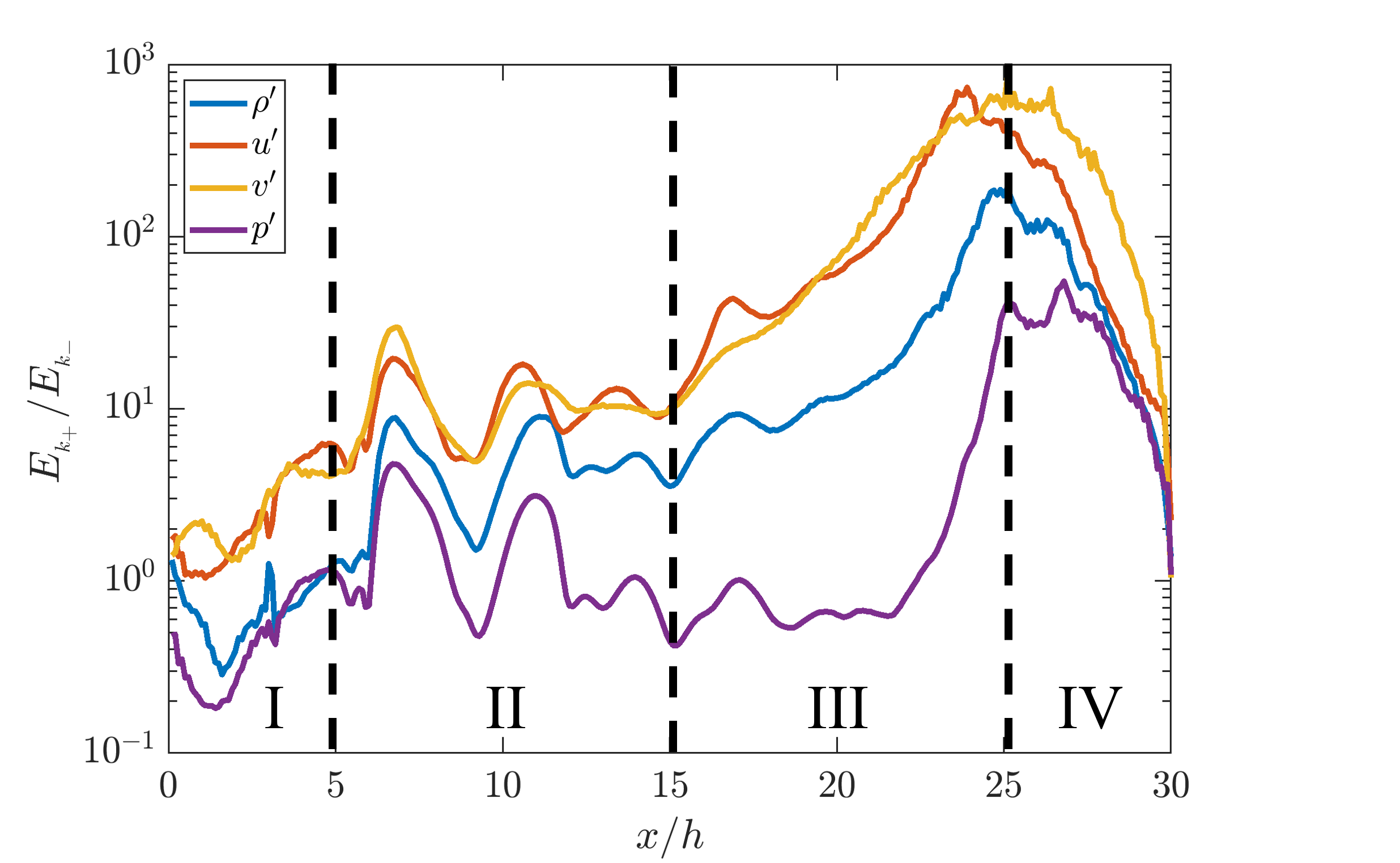} 
 	\end{tabular}
 	\caption{Ratio of downstream-traveling to upstream-traveling energy propagation from waves of $\rho'$,$u'$,$v'$ and $p'$. The four distinct regions corresponding to different processes of screech are marked by dashed lines and labelled in Roman numerals.}
 	\label{fig:E_all_vars}
 \end{figure}

Figure~\ref{fig:E_mult_freq} shows $E_{k_-}$, $E_{k_+}$, and $E_{k_+}/E_{k_-}$ with respect to $x/h$ using the SPOD data for $v'$. These results can be best interpreted together with figures~\ref{fig:mean_velocity_contour} and~\ref{fig:Spod_decomp_v}. $E_{k_-}$ and $E_{k_+}$ are reflective of the nonlinear growth and decay of the $k_-$ and $k_+$ wave over the streamwise direction. They are also functions dependent on frequency. At the screech fundamental frequency, the maxima of $E_{k_-}$ and $E_{k_+}$ are found to be at $x/h=14.4$ and $x/h= 16.9$, respectively. Interestingly, the value of $E_{k_+}/E_{k_-}$, with focus on the screech tone $St_1$, indicates four distinct regions in the jet plume: $x/h$:$[0,5]$, $[5, 15]$, $[15, 25]$, $[25, 30]$. The first region ($\mathrm{I}$) is between $x/h =0$ and $x/h = 5$, where $E_{k_+}/E_{k_-}$ undergoes growth from a low initial value. This likely corresponds to the receptivity process where energy of the $k_{-}$ excites the jet initial shear layers to form embryonic $k_+$ wave. The second region ($\mathrm{II}$) is at $5 \leq x/h \leq 15 $, where $E_{k_+}/E_{k_-}$ reaches a plateau. In this region, both $E_{k_-}$ and $E_{k_+}$ grow with respect to the downstream direction at a similar rate. The first two regions also coincide with the jet supersonic core where effects of the shock cells are visible. The third region ($\mathrm{III}$) is located between $x/h=15$ and $x/h=25$, where $E_{k_+}/E_{k_-}$ increases rapidly due to the fast fall-off of $E_{k_-}$. At $x/h=25$, $E_{k_+}/E_{k_-}$ has a maximum value. The maximum of $E_{k_+}/E_{k_-}$ indicates the endpoint where the $k_-$ wave reaches a significant magnitude with respect to the $k_+$ wave. Recall that the $k-$ wave propagates in the $-x$ direction, so the profile between $x/h=15$ and $x/h=25$ indicates that $E_{k_-}$ grows, saturates and decays in the $-x$ direction. Therefore, the third region corresponds to the excitation of the guided jet mode. Considering the mean velocity contours in figure~\ref{fig:mean_velocity_contour}, the third region extends to the end of the jet supersonic core. Additionally, both the second and third regions align with locations where acoustic waves are emitted as seen in the SPOD structures in figure~\ref{fig:Spod_decomp_v}. This is further verified in Section~\ref{sec:acoustic_analogy}. Lastly, the fourth region ($\mathrm{IV}$) is located downstream of $x/h=25$, where $E_{k_+}/E_{k_-}$ drops rapidly as the $k_{+}$ wave breaks down at the end of the jet potential core. 

The four distinct regions are also observed across other flow variables, as shown by figure~\ref{fig:E_all_vars} using the SPOD data for $\rho'$, $u'$, $v'$ and $p'$. In summary, the values of $E_{k_-}$, $E_{k_+}$, and $E_{k+}/E_{k-}$ highlight the spatially non-local and non-periodic characteristics of various processes involved in screech resonance. $E_{k+}/E_{k-}$ can be used as a metric to identify distinct regions in the jet plume that correspond to different processes of screech. For the K-H wave: region $\mathrm{I}$ corresponds to the receptivity/excitation of the K-H wave ($k_+$) in the initial shear layer, region $\mathrm{II}$ corresponds to the growth of the $k_+$ wave and interaction with shock-cells, and regions $\mathrm{III}$ and $\mathrm{IV}$ are related to the saturation and decay of the k+ wave. For the guided jet mode $k_-$, region $\mathrm{III}$ corresponds to the excitation and growth in the $-x$ direction, and region $\mathrm{II}$ is related to the saturation and shock associated modulation of the $k_-$ wave.

\section{Effects of NPR Variation} \label{sec:npr_effects}
\begin{figure}
	\centering
	\begin{tabular}{cc}
		\includegraphics[width=0.3\textwidth]{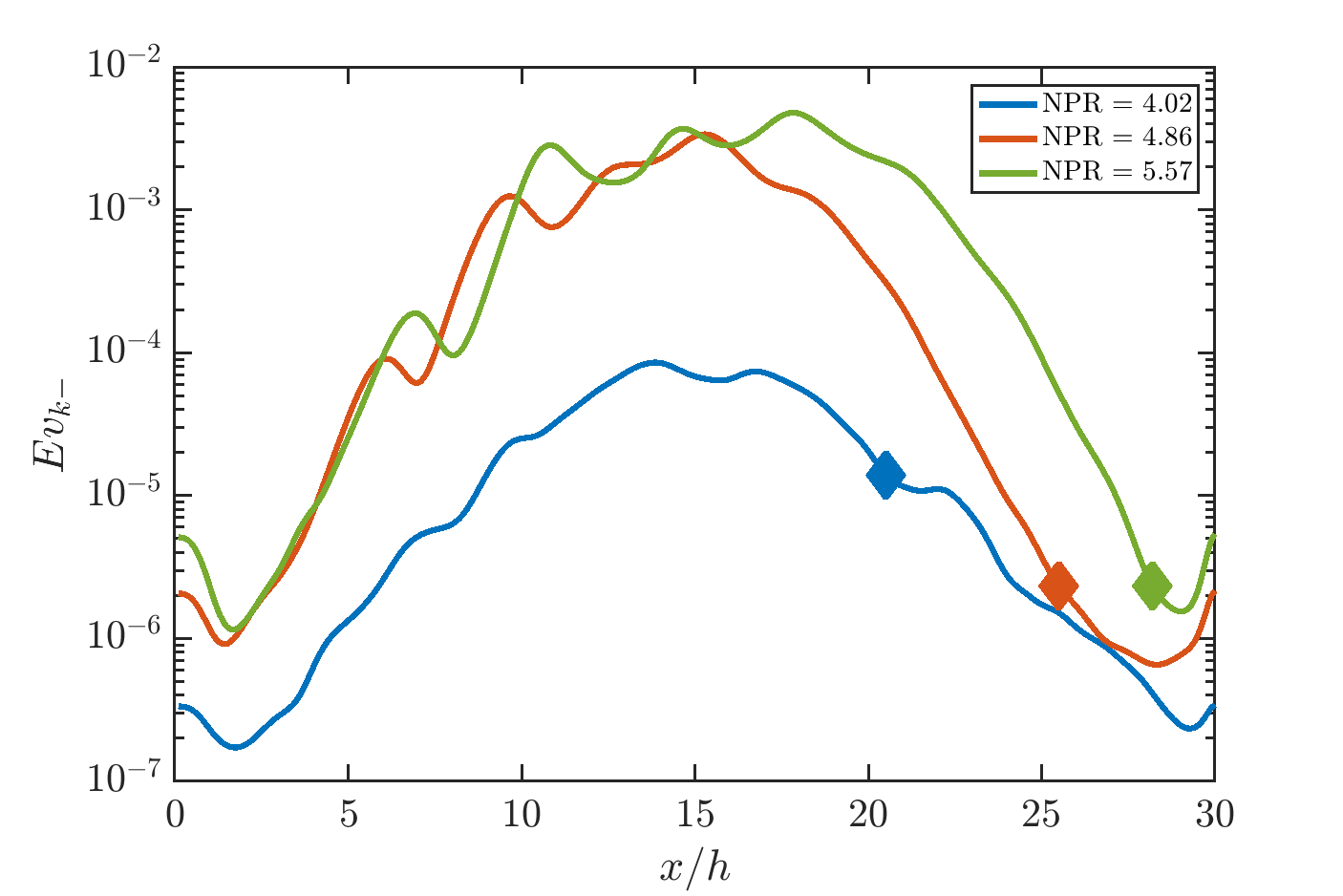} &
		\includegraphics[width=0.3\textwidth]{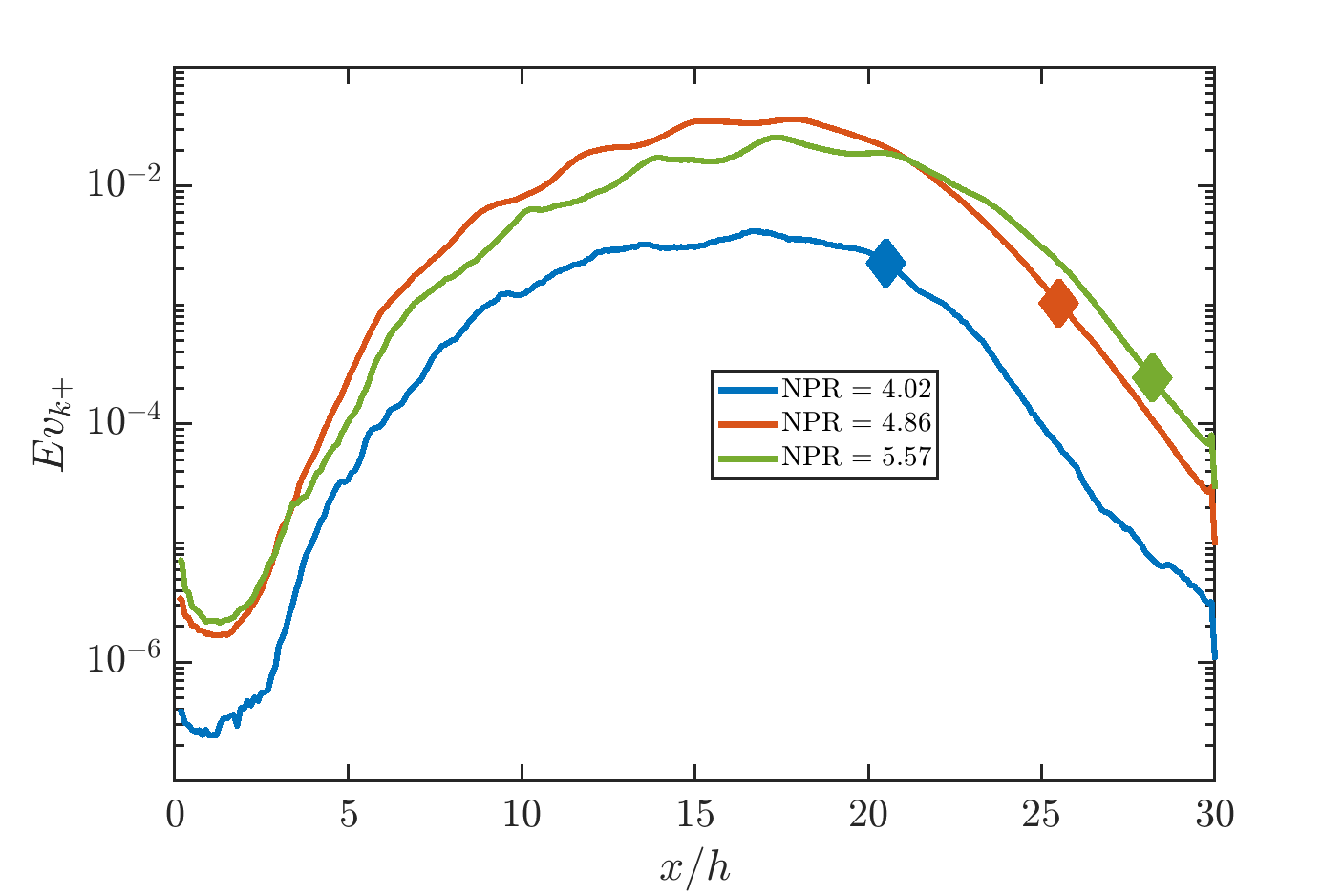}\\ 
            (a) $E_{k_-}$ & (b) $E_{k_+}$\\
		\includegraphics[width=0.3\textwidth]{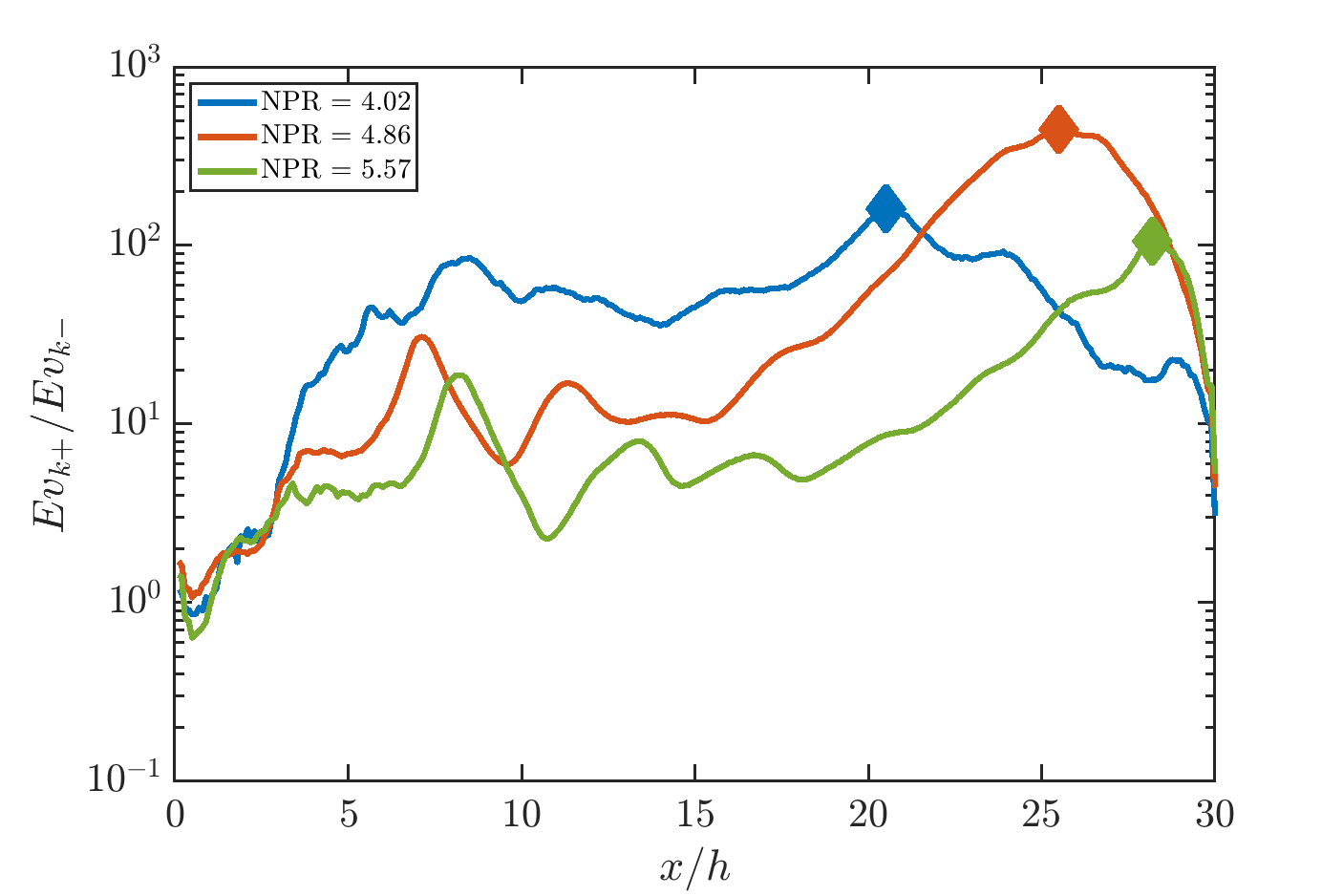} &
            \includegraphics[width=0.3\textwidth]{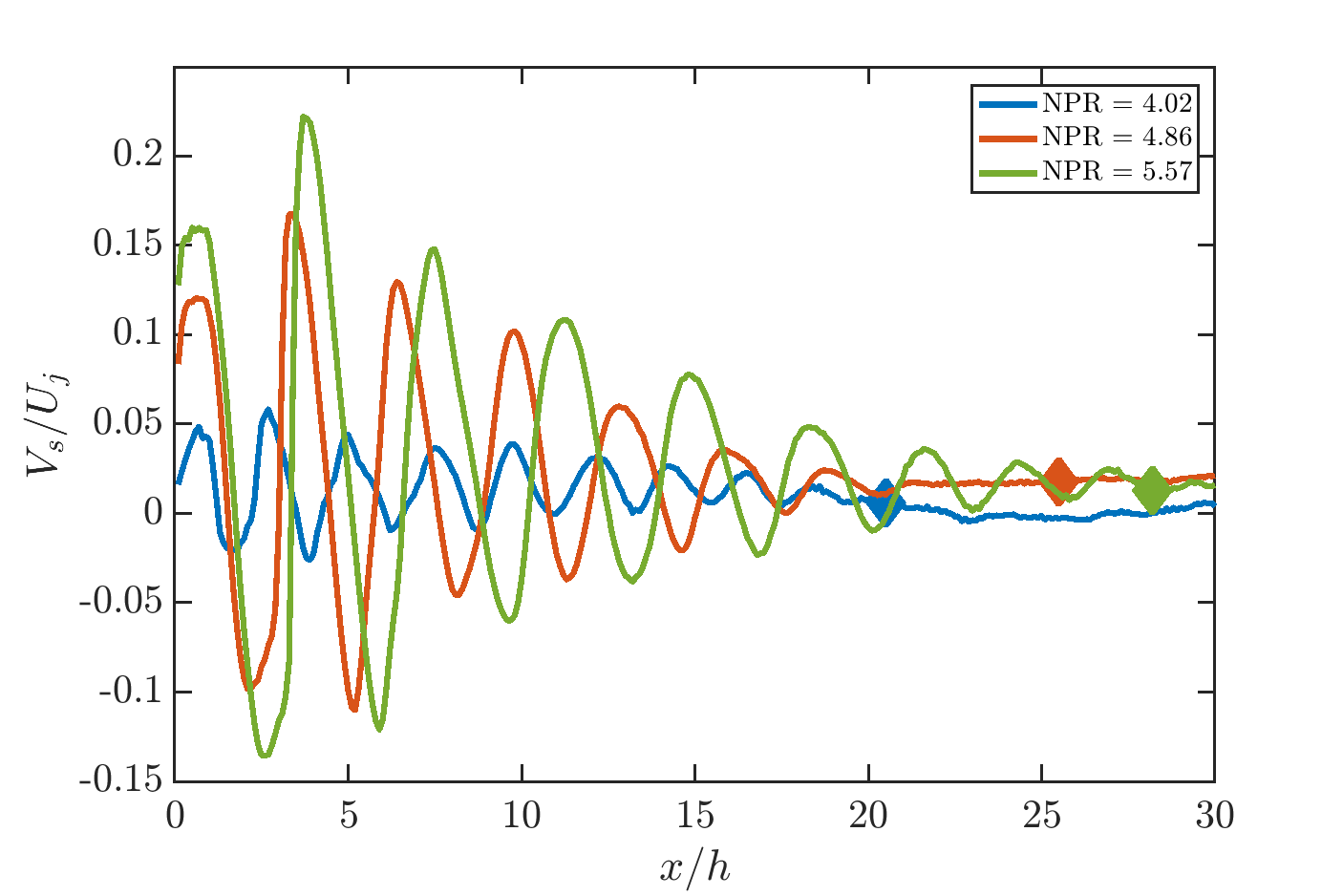} \\
		  (c) $E_{k_+}/E_{k_-}$& (d) $V_{lip}/U_j$
	\end{tabular}
    \caption{Comparison of wave energy and shock between three NPR cases. Solid diamonds indicate the locations of the maximum point of $E_{k_+}/E_{k-}$ for each case.}
	\label{fig:int_amp_wave_mult_NPR}
\end{figure}

This section repeats the same analysis carried out in Section~\ref{sec:wave_energy} for two more NPR conditions. For all three screech cases considered, similar spatial variation regarding $E_{k_-}$, $E_{k_+}$, and $E_{k+}/E_{k-}$ are observed in Figure~\ref{fig:int_amp_wave_mult_NPR}. Figure~\ref{fig:int_amp_wave_mult_NPR}(c) shows as NPR increases, region I shortens while region II and III lengthens. At higher NPR values, the increased velocity difference between the jet plume and the ambient makes the initial shear layer more unstable and more receptive to Kelvin-Helmholtz instability. The stronger shock cells and longer supersonic core also enable stronger excitation of the $k-$ wave. The maxima of $E_{k+}/E_{k_-}$ are marked by the solid circles in Figure~\ref{fig:int_amp_wave_mult_NPR}(c). Their streamwise locations are also indicated in the other subfigures. Figure~\ref{fig:int_amp_wave_mult_NPR} (d) indicates the endpoint can be taken to be the end of the jet supersonic core which moves further downstream with increasing NPR. Table~\ref{tab:screech_NPR_summary} summarizes the screech frequency and the screech amplitude measured at $100h$ away from the nozzle exit from two jet polar angles. As NPR increases, the screech amplitude increases, which is consistent with the observed increase in length for regions II and III. As verified later in Section~\ref{sec:acoustic_analogy}, these two regions are indeed associated with the distributed acoustic source for the screech tone.

\begin{table}
	\centering
	\begin{tabular}{cccc}
         NPR & $\mathrm{St_{sc}}$ & SPL(dB) at $\phi = 30\deg $  & SPL(dB) at $\phi = 150\deg$\\
		\hline 
        4.02 & 0.31 & 119 & 116 \\
	4.86 & 0.23 & 131 & 124 \\ 
        5.57 & 0.20 & 136 & 134 \\
	\end{tabular}
	\caption{Summary of screech frequency and amplitude measured $100h$ away from the nozzle exit for different NPR conditions}
	\label{tab:screech_NPR_summary}
\end{table}

\section{SPOD Modes and Sound}\label{sec:acoustic_analogy}
At the screech frequency, the near-field hydrodynamic fluctuations are dominated by spatial-temporal coherent structures, represented by the leading order SPOD mode, as shown in previous sections. In this section, we present direct evidence linking the acoustics of screech with the SPOD structures using Lighthill's equation~\citep{Lighthill1952}. By an exact rearrangement from mass and momentum balance, Lighthill shows the pressure fluctuations, $p' = p-p_0$ in the far field from any flow with arbitrary motion are analogous to the acoustic response in an ideal medium at rest due to quadrupole acoustic sources. 
\begin{equation}
    \frac{1}{c_0^2} \frac{\partial ^2 p'}{\partial t^2} -\nabla^2 p' = \frac{\partial^2 T_{ij}}{\partial x_i \partial x_j}
\end{equation}
and
\begin{equation}
    T_{ij} = \rho u_i u_j - \tau_{ij} + (p' - c_0^2\rho')\delta_{ij} 
\end{equation}
The term $T_{ij}$ is Lighthill's stress tensor, which includes all the actual flow effects on noise generation in the acoustic analogy, where $\rho u_i u_j$ is the momentum component, $\tau_{ij}$ is the viscous stress, and $p' - c_0^2 \rho'\delta_{ij} $ is due to entropy changes. In the following analysis, we show that the SPOD modes can be directly used to approximate the $T_{ij}$ associated with screech tone.

\begin{figure}
	\centering
	\begin{tabular}{cc}
		\includegraphics[width=0.4\textwidth]{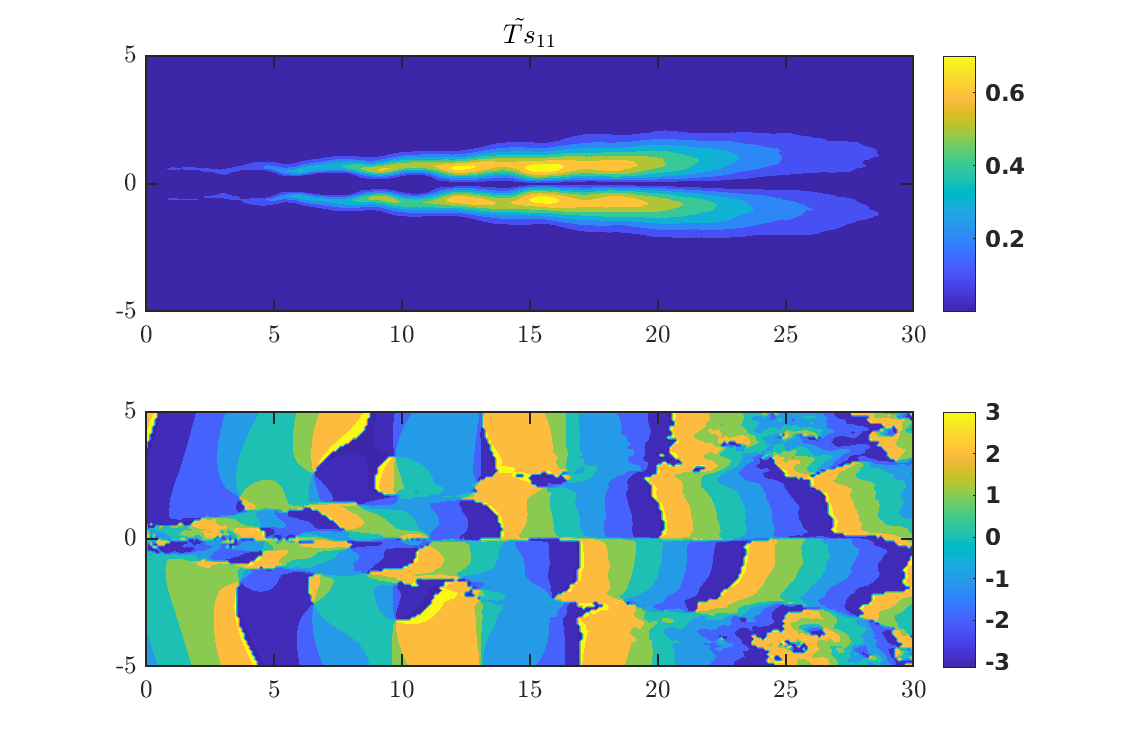} &
		\includegraphics[width=0.4\textwidth]{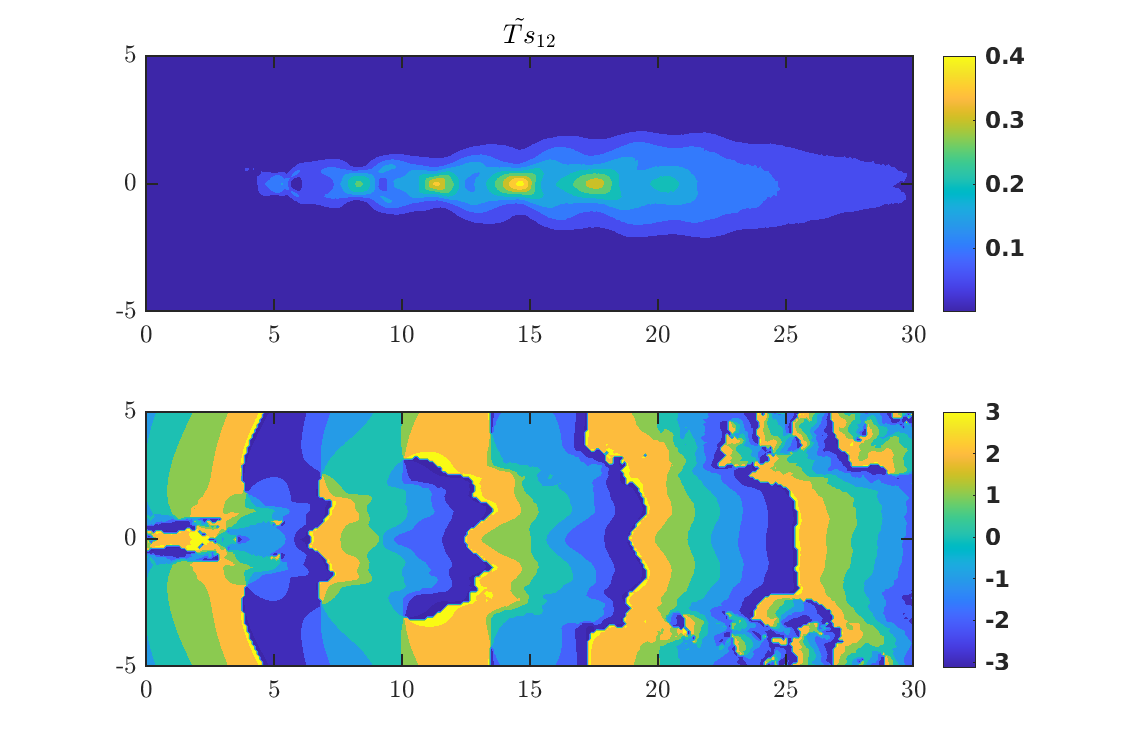} \\
  	(a) & (b)  \\
        \includegraphics[width=0.4\textwidth]{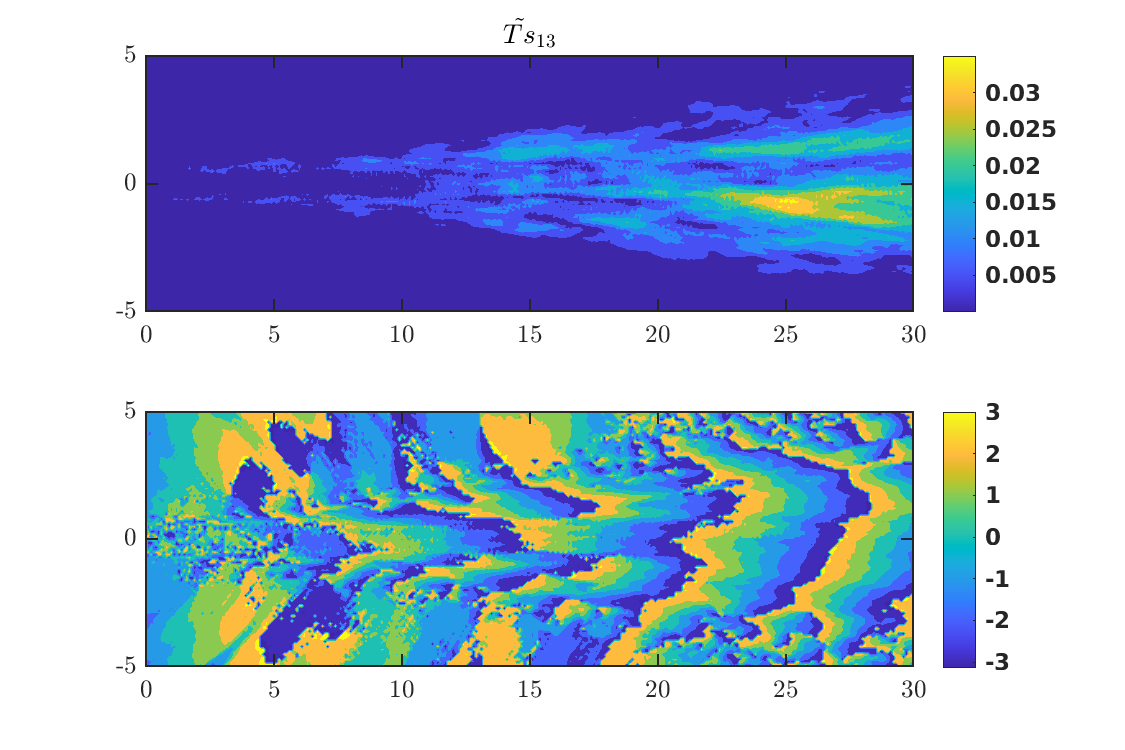} &
        \includegraphics[width=0.4\textwidth]{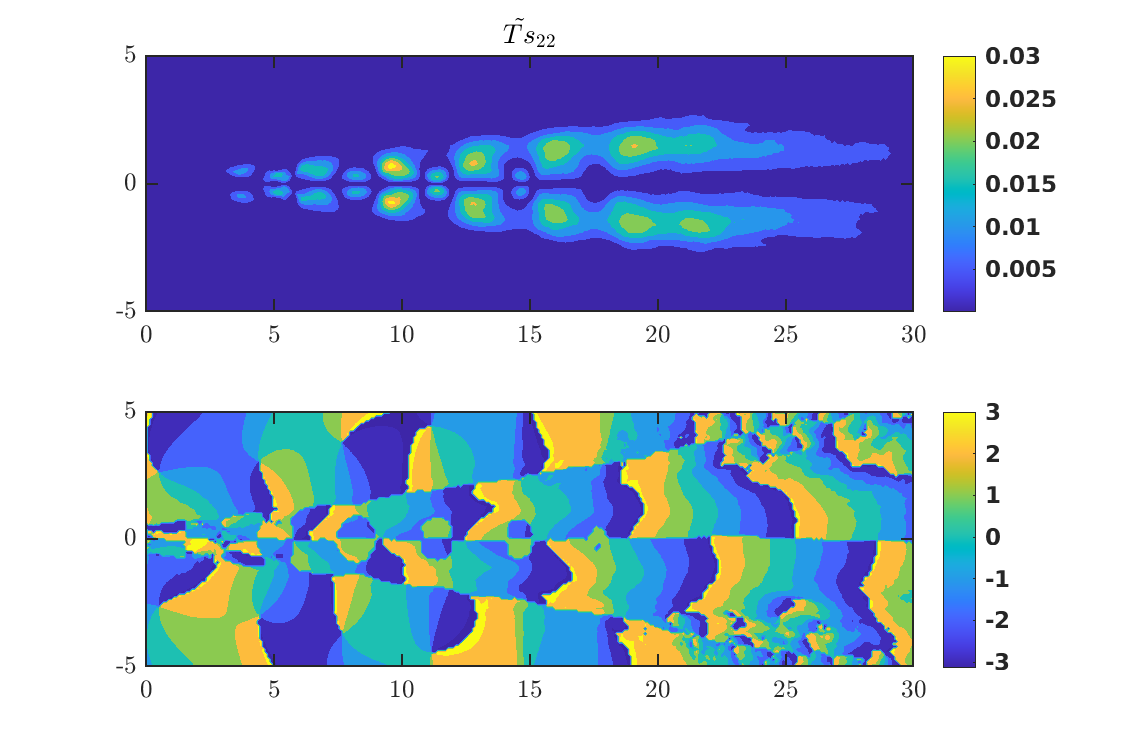} \\
	(c) & (d) \\
		\includegraphics[width=0.4\textwidth]{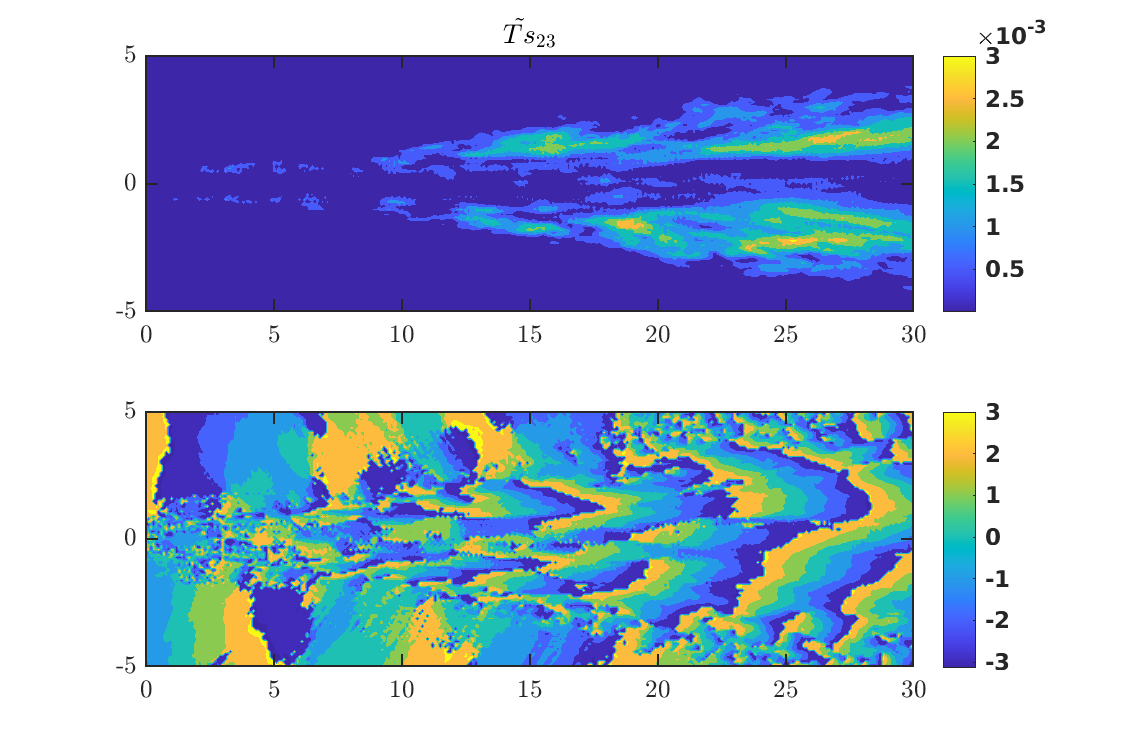} &
        \includegraphics[width=0.4\textwidth]{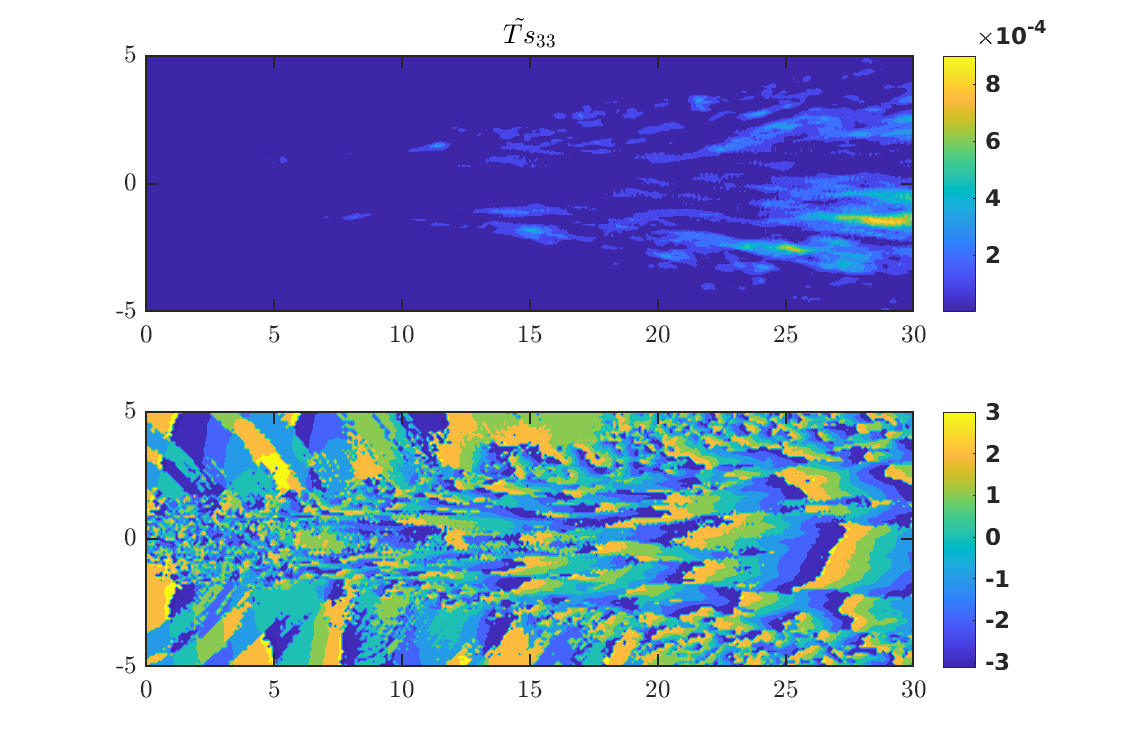} \\
		(e) & (f) 
	\end{tabular}
	\caption{Estimated $T_{ij}$ with SPOD data. }
	\label{fig:Tij_approx_spod}
\end{figure}

In the frequency domain, the solution for $p'$ far away from the acoustic source region at a location $\mathbf{y}$ can be written as
\begin{equation}
    \hat{p}' (\mathbf{y},\omega) = \int_{V} \frac{\partial^2\hat{T}_{ij}(\mathbf{x}, \omega)}{\partial x_i \partial x_j} G_0(\mathbf{x},\mathbf{y};\omega) \mathrm{d}x
\end{equation}
where $\omega$ is the angular frequency, $V$ is the control volume enclosing the acoustic source, $\hat{T}_{ij}(\mathbf{x}, \omega)$ is the Fourier component of $T_{ij}$, and $G_0$ is the free-space Green's function. 
Applying integration by part and the divergence theorem, and taking the surface integrals to be zero, we can shift the spatial derivatives from $\hat{T}_{ij}$ to the Green's function.
\begin{equation}\label{eq:acoustic_analogy}
     \hat{p}' (\mathbf{y},\omega) = \int_{V} \hat{T}_{ij}(\mathbf{x}, \omega) \frac{\partial^2 G_0(\mathbf{x},\mathbf{y},\omega)}{\partial x_i \partial x_j}  \mathrm{d}x
\end{equation}
In non-heated high-Reynolds-number jets, effects of viscous stress and entropic inhomogeneity are small compared to momentum, so
\begin{equation}
T_{ij} \approx \rho u_i u_j. 
\end{equation}
Using Reynolds decomposition, $\rho u_i u_j$ can be expressed as
\begin{equation}
    \rho u_i u_j = (\bar{\rho} + \rho^t)  (\bar{u}_i+ u_i^t) (\bar{u}_j+ u_j^t)
\end{equation}
where $\bar{()}$ denotes the mean quantity in time and $()^t$ denotes the unsteady component. For the purpose of calculating the far-field acoustic pressure, we ignore the steady term in time and only keep the first-order unsteady terms.
\begin{equation}
   T_{ij} \approx  \bar{\rho}  \bar{u}_i  u_j^t + \bar{\rho}  u_i^t \bar{u}_j + \rho^t \bar{u}_i\bar{u}_j 
\end{equation}
Then the time-Fourier component of $T_{ij}$ simply becomes
\begin{equation}\label{eq:T_ij}
    \hat{T}_{ij}\approx  \bar{\rho}  \bar{u}_i  \hat{u}_j^t + \bar{\rho}  \hat{u}_i^t \bar{u}_j + \hat{\rho}^t \bar{u}_i\bar{u}_j 
\end{equation}
At the screech frequency, the near-field hydrodynamic fluctuations are dominated by spatial-temporal coherent structures as shown in previous sections. This suggests the generation of screech noise could be well approximated by the waves from the leading-order SPOD mode. To verify this hypothesis, we substitute the SPOD data in place for $\hat{\rho}^t$, $\hat{u}_i^t$ and $\hat{u}_j^t$ to calculate $\hat{T}_{ij}$. Figure~\ref{fig:Tij_approx_spod} shows the amplitude and phase for the complex valued $\hat{T}_{ij}$ at the screech frequency for the case at $\mathrm{NPR}=4.86$. The $\hat{T}_{11}$ and $\hat{T}_{12}$ are the most dominant terms because of the large streamwise velocity fluctuations, and the spatial distribution spans between $x/h=5$ and $x/h=25$, which aligns with regions $\mathrm{II}$ and $\mathrm{III}$ identified by $E_{k+}/E_{k_-}$ in Section~\ref{sec:wave_energy}. 

The term involving the free-space Green's function in the integral solution for $\hat{p}'$ is
\begin{equation}\label{eq:green function}
    \frac{\partial^2 }{\partial x_i \partial x_j} G_0 (\mathbf{x},\mathbf{y}, \omega) = \frac{r_i r_j}{4\pi |\mathbf{r}|^5} \left(3- 3\mathrm{i} k_c |\mathbf{r}| -  k_c^2 |\mathbf{r}|^2 \right) e^{\mathrm{i} k_c |\mathbf{r}|}
\end{equation}
where $\mathbf{r} = \mathbf{y} - \mathbf{x}$ is the displacement vector from the source to the observer location, and $k_c = \omega /c_0$ is the acoustic wavenumber. For $|\mathbf{r}|/L_s >> 1$, where $L_s$ is the length scale associated with the distributed acoustic source, the term involving the Green's function can be approximated as
\begin{equation}\label{eq:far green}
    \frac{\partial^2 }{\partial x_i \partial x_j} G_0 (\mathbf{x},\mathbf{y}, \omega) \approx \left(\frac{-k_c^2 r_{0i}r_{0j}}{4\pi |\mathbf{r_0}|^3}\right) e^{\mathrm{i} k_c \left(|\mathbf{r_0}| +  \frac{\mathbf{r_0}}{|\mathbf{r_0}|} \cdot \mathbf{\xi} \right)}
\end{equation}
where $\mathbf{r_0} = \mathbf{y} - \mathbf{x_c}$ is the displacement from the center of the acoustic source to the observer, and $\mathbf{\xi} = \mathbf{x_c} - \mathbf{x}$. For the case analyzed, $x_c$ is chosen to be $15h$. 

\begin{figure}
	\centering
	\begin{tabular}{c}
		\includegraphics[width=0.8\textwidth]{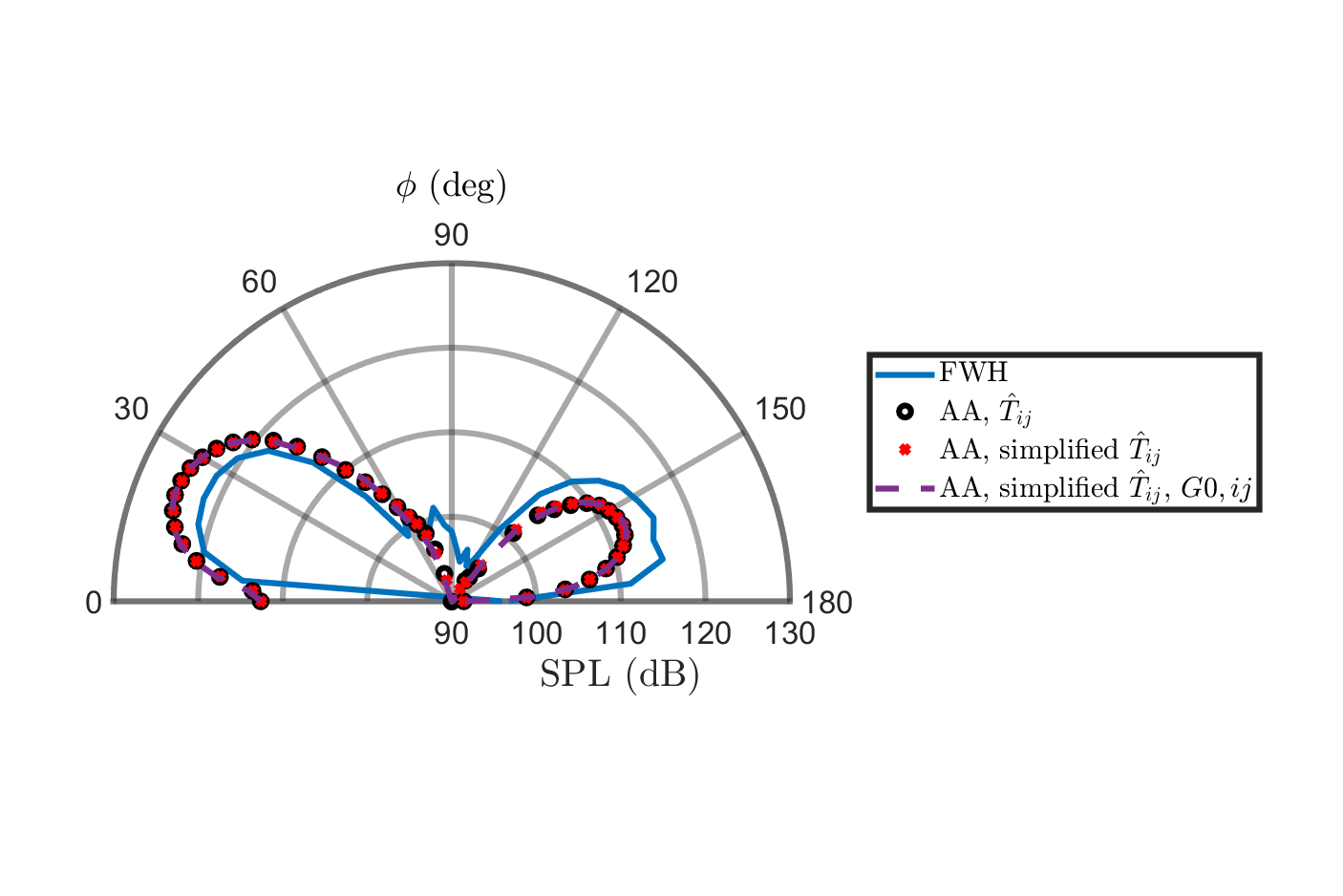} \\
	\end{tabular}
	\caption{Far-field acoustic pressure evaluated with Lighthill's acoustic analogy at $|\mathbf{r}|/h=270$, in comparison with the LES-FWH calculation.}
	\label{fig:acoustic_p_compare}
\end{figure}

The far-field acoustic pressure at $270h$ away from the nozzle exit in the minor-axis plane is calculated using Eq.~\ref{eq:acoustic_analogy}. $T_{ij}$ is assumed to have a uniform distribution in the major-axis direction between $z/h = -5$ and $z/h = 5$ and zero elsewhere. This is a crude simplification of the acoustic source, which is not homogeneous in the $z$ direction due to the rectangular nozzle geometry. Nevertheless, the result of the screech tone amplitude compares favorably with the LES-FWH calculation, as shown in Figure~\ref{fig:acoustic_p_compare}. In particular, values marked by the black circles are calculated with Eq.~\ref{eq:acoustic_analogy}, Eq.~\ref{eq:T_ij} and Eq.~\ref{eq:green function}. Since $T_{11}$ and $T_{12}$ are much larger compared to the other components of $T_{ij}$, values shown by the red crosses are computed only using these two as the source terms. Those indicated by the purple line are obtained with further simplification on $\frac{\partial^2 }{\partial x_i \partial x_j} G_0$ using Eq.~\ref{eq:far green}. The various forms of simplification do not yield significant variation in the calculated screech tone amplitude. With varying jet polar angles in the minor-axis plane, the acoustic analogy solutions also capture the screech tone directivity very well.

The good agreement between the acoustic analogy and the FW-H results have a few important implications. First, this verifies the coherent wave structures from SPOD are good representation of the noise generation mechanism that produces far-field screech tones. The SPOD data include the 
$k_-$ and $k_+$ waves, which propagate energy in the upstream and downstream directions inside the jet plume and establish the feedback resonance. Simultaneously, areas where the growth and interaction of these waves with the shock cells
take place (regions $\mathrm{II}\&\mathrm{III}$) also correspond to the region of a spatially distributed acoustic source for screech tones. The outcome of the
interaction of the $k_+$ wave with the mean flow, including the mean shock cell structures, 
includes
the $k-$ and the acoustic waves. The acoustic waves are also subject to refraction in propagation across the mean jet shear layer. These processes are reflected in 
the 
rapid phase variation of $\hat{T}_{ij}$ across the jet in Figure~\ref{fig:Tij_approx_spod}.
The far-field Green's function~\ref{eq:far green} contains the geometric angular factors as well as phase variation associated with retarded time variation over the source region. When this is combined with the spatially distributed source in Equation~\ref{eq:acoustic_analogy} 
the far-field acoustic radiation embodies the
constructive and destructive interference of the
effective acoustic sources,
resulting in the screech tone directivity pattern. In future, these results can be used
to improve 
screech amplitude prediction models, by constructing the acoustic sources with a functional form suggested by the SPOD mode and the mean jet state.

\section{Conclusions}\label{sec:conclusions}
Using high-fidelity large-eddy simulation data, this work studies the screech generation mechanism in a 4:1 rectangular jet under three under-expanded NPR conditions. The analysis combines spectral proper orthogonal decomposition (SPOD), Fourier decomposition in the streamwise direction and spatial cross correlation using the LES data. For the case at NPR = 4.86, a detailed assessment of the dominant coherent structures and their relevance in screech generation is conducted. The effects of NPR variation are evaluated by repeating the same analysis for the other two NPR conditions. The first SPOD mode at the screech fundamental frequency is composed of acoustic waves, the upstream-traveling guided jet mode and the downstream-traveling Kelvin-Helmholtz wave. Estimates of the wave energy indicate global nonlinear characteristics in the streamwise direction.

The main findings of the analysis agree with recent theoretical understanding of screech instability and shed light on the spatial separation of individual processes for screech generation. Direct evidence of the guided jet mode being the screech closure mechanism, not the external acoustic wave, is observed by considering the growth of waves across screech frequency and its neighboring frequencies. Quantification of the wave energies ($E_{k_+}$, $E_{k_-}$, and their ratio $E_{k_+}/E_{k_-}$) identifies regions where distinct processes in screech generation take place, including initial shear layer receptivity, sound generation, guided jet mode excitation and coherence decay due to turbulence. The initial shear layer receptivity region is marked by a rapid spatial growth of $E_{k_+}/E_{k_-}$ near the nozzle. From there, $E_{k_+}/E_{k_-}$ plateaus then sharply increases in value, corresponding to the origin, growth and saturation of the guided jet mode in the $-x$ direction. This region also aligns with the location where Lighthll's stress tensor $T_{ij}$, approximated by the SPOD mode and the mean flow (including the shock cells), is large in amplitude. The acoustic analogy formulation with $T_{ij}$ and the free-space Green's function provides a good estimate of the far-field screech tone noise and matched well with the FW-H calulations using the original LES data. This further verifies the coherent structures highlighted by the leading-order SPOD mode capture both the dominant hydrodynamic and acoustic processes involved in screech generation. The current findings can be extended 
construct amplitude prediction model for jet screech in future.

\section*{Acknowledgment}
 We would like to thank Cascade Technologies for their LES solver and Dr. Guillaume A. Brès for his substantial support and guidance on our work. We thank Professor Rajan Kumar from Florida State University for sharing with us the experimental data. 
\section*{Funding}
This work is supported by the ONR grant N00014-18-1-2391 with Dr. Steve Martens as the project manager and the XSEDE (TG-CTS190021) program.
\section*{Declaration of Interests}
The authors report no conflict of interest.

\appendix

\section{SPOD modes for $p'$, $u'$, $w'$, and $\rho'$}\label{appA}
\begin{figure}
	\centering
	\begin{tabular}{cc}
		\includegraphics[width=0.325\textwidth]{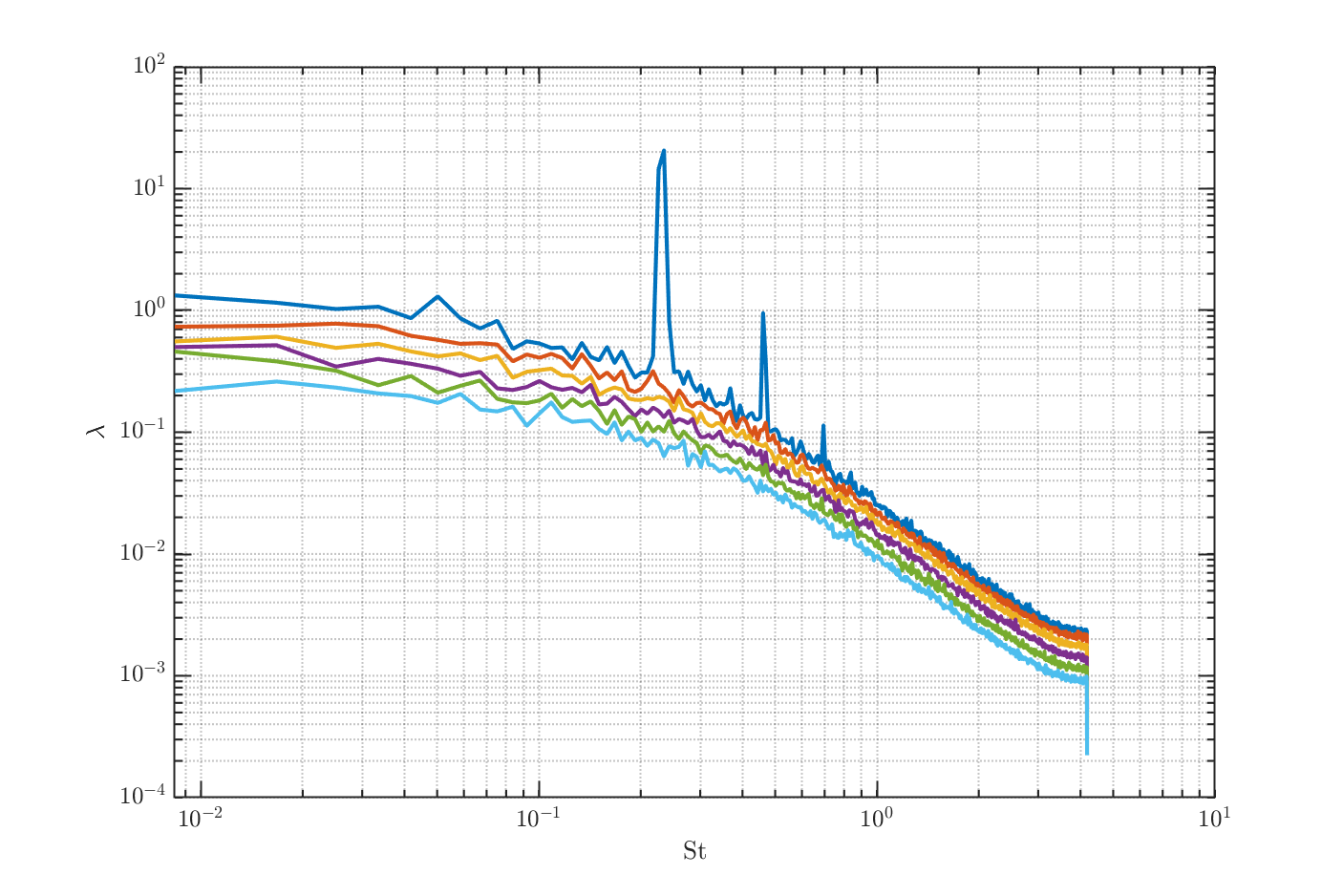} &
		\includegraphics[width=0.35\textwidth]{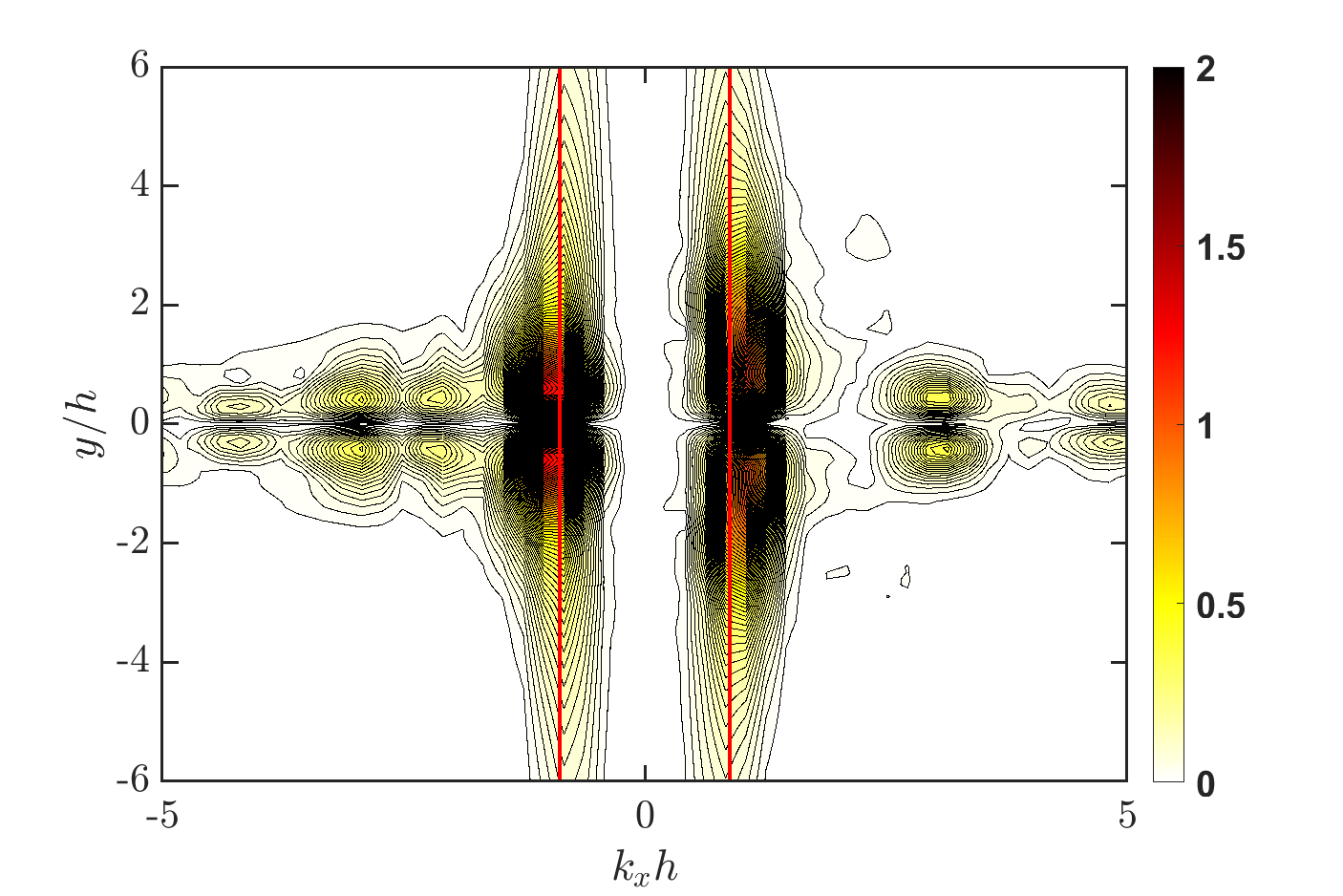} \\
        (a) & (b) \\
        \includegraphics[width=0.35\textwidth]{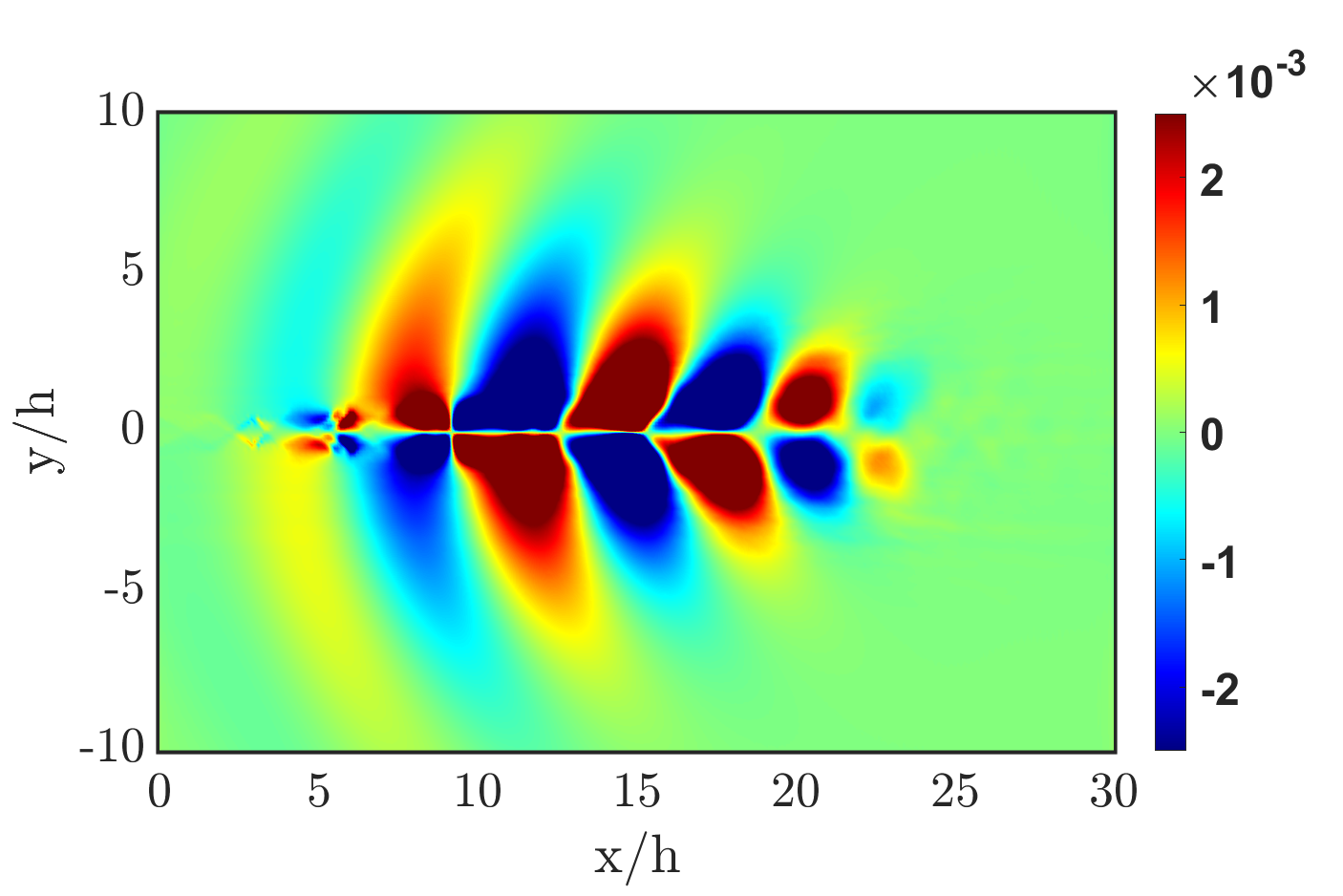} &
		\includegraphics[width=0.35\textwidth]{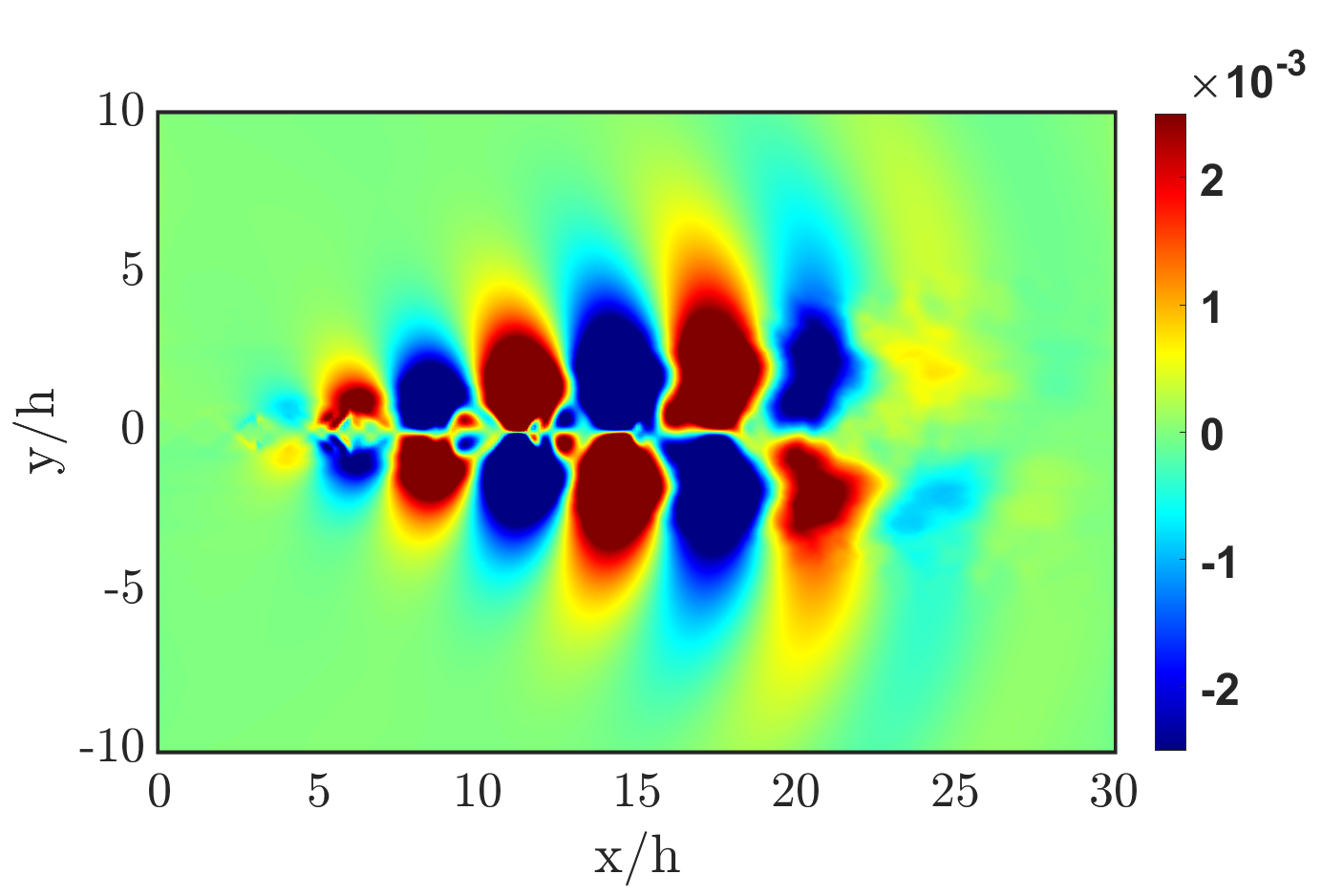} \\
		(c) & (d) \\
	\end{tabular}
	\caption{(a) Energy spectra of $p'$ SPOD modes, (b) modulus of the leading-order SPOD mode shape function for $p'$ at the screech fundamental frequency, (c) upstream-traveling waves in the leading-order $p'$ SPOD mode, (d) downstream-traveling waves in the leading-order $p'$ SPOD mode. }
	\label{fig:Spod_p}
\end{figure}

\begin{figure}
	\centering
	\begin{tabular}{cc}
		\includegraphics[width=0.325\textwidth]{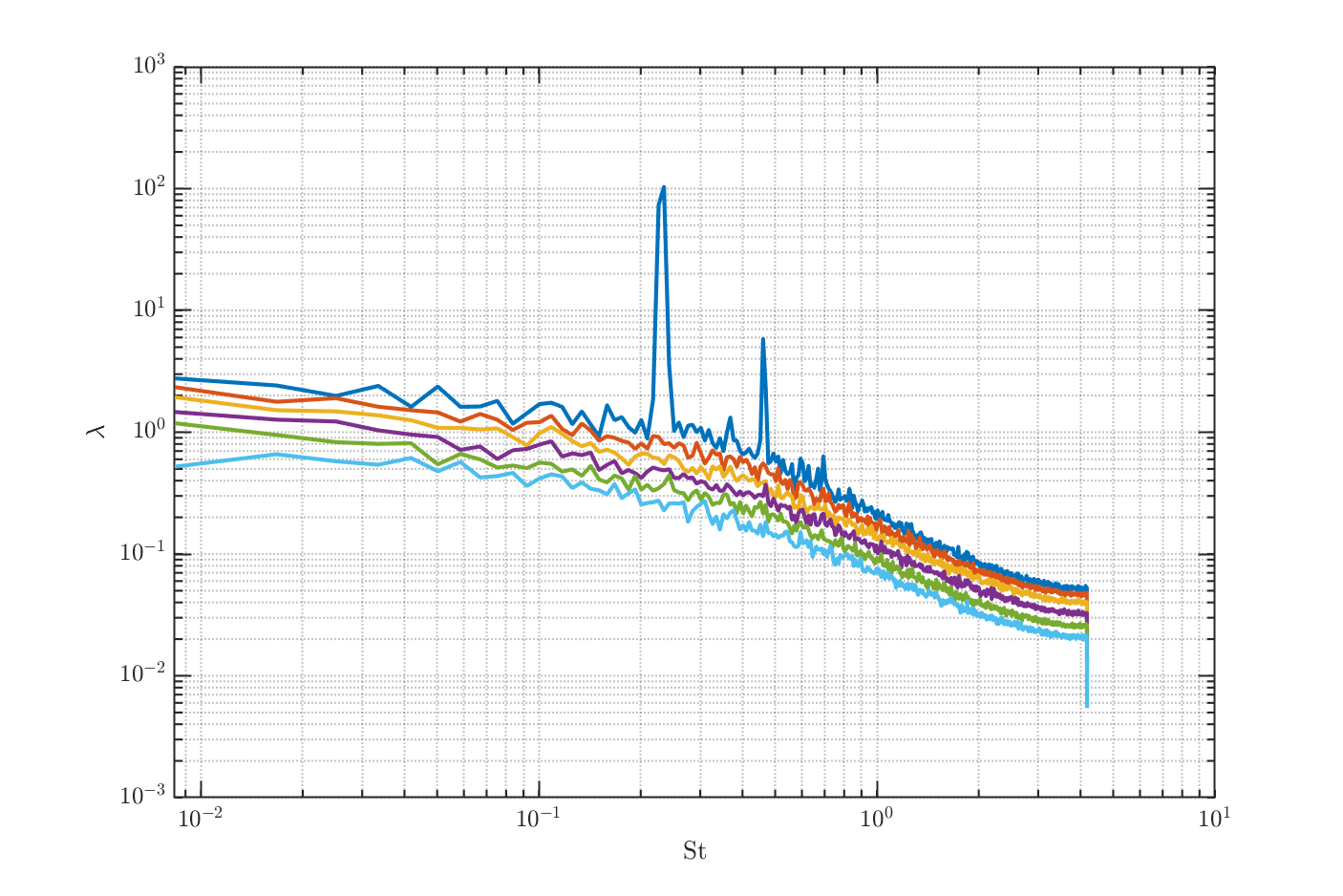} &
		\includegraphics[width=0.35\textwidth]{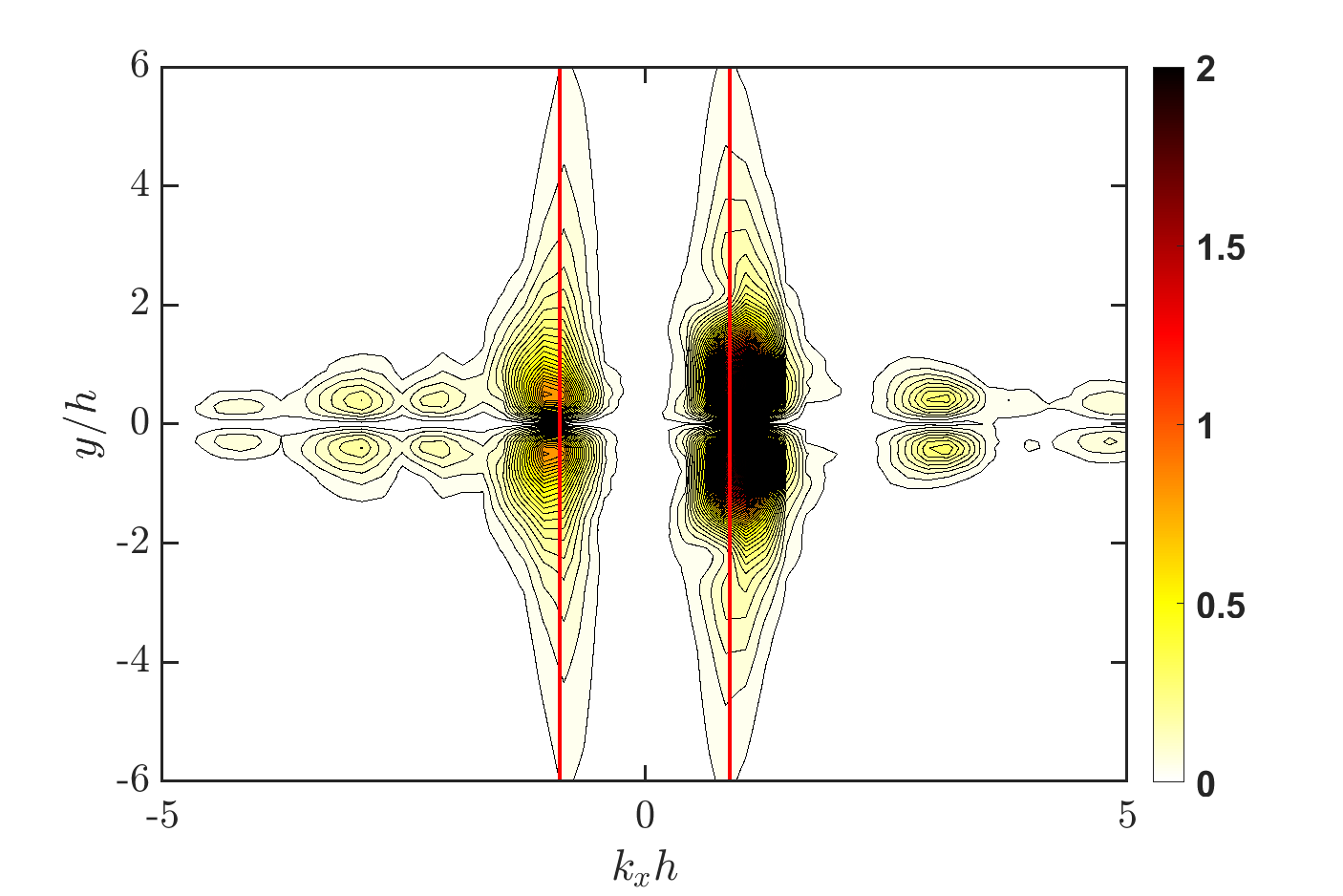} \\
        (a) & (b) \\
        \includegraphics[width=0.35\textwidth]{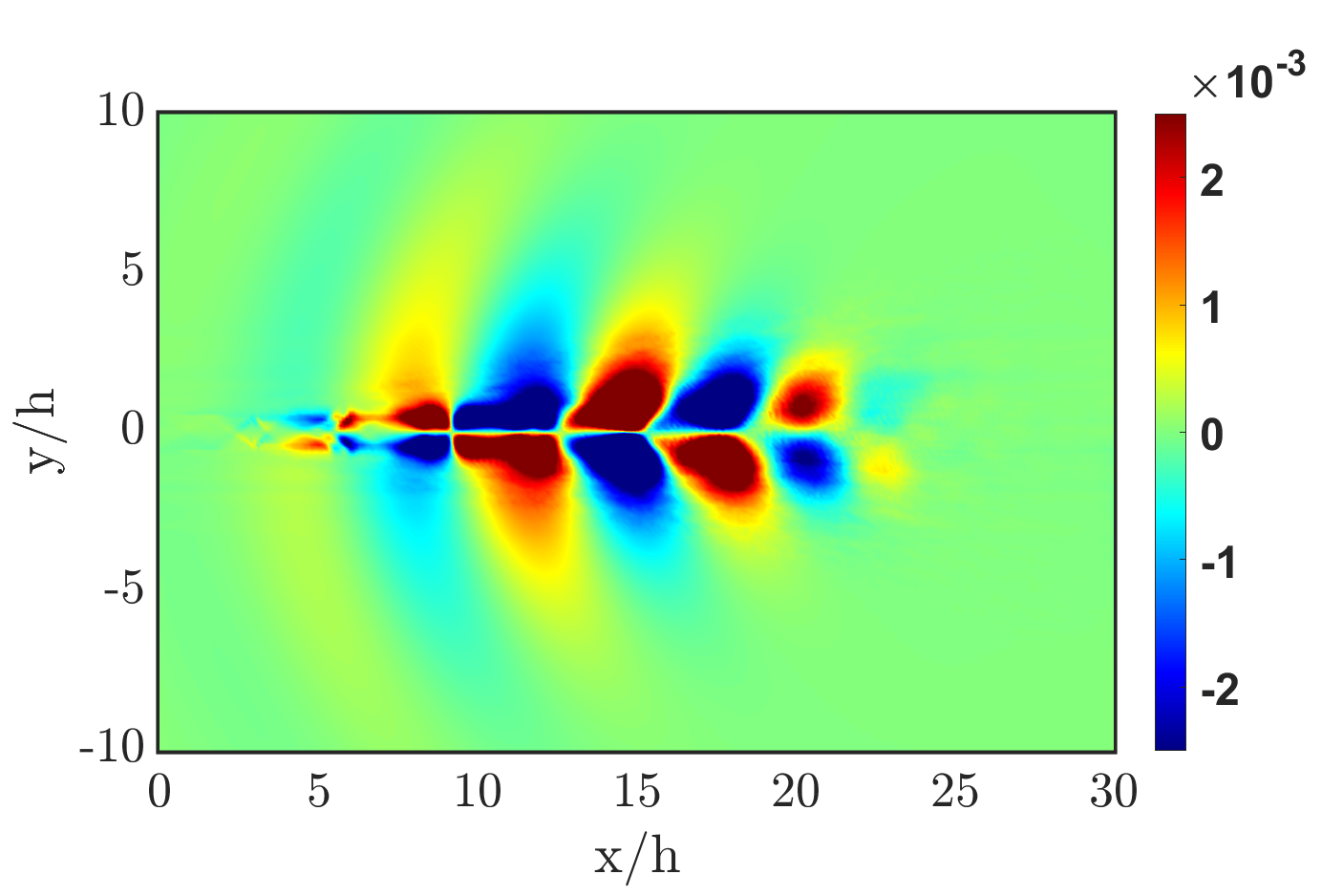} &
		\includegraphics[width=0.35\textwidth]{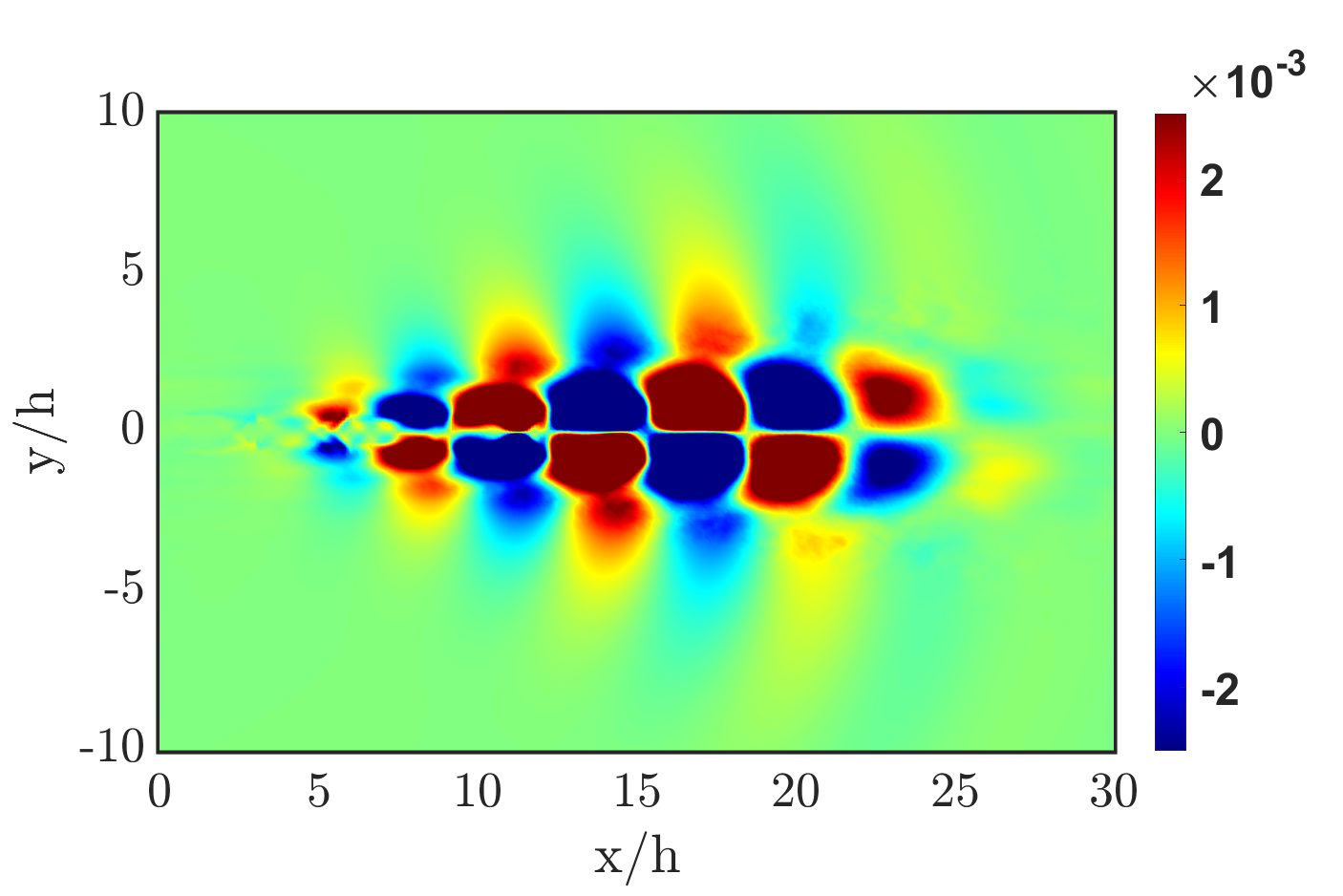} \\
		(c) & (d) \\
	\end{tabular}
	\caption{(a) Energy spectra of $\rho'$ SPOD modes, (b) modulus of the leading-order SPOD mode shape function for $\rho'$ at the screech fundamental frequency, (c) upstream-traveling waves in the leading-order $\rho'$ SPOD mode, (d) downstream-traveling waves in the leading-order $\rho'$ SPOD mode. }
	\label{fig:Spod_rho}
\end{figure}

\begin{figure}
	\centering
	\begin{tabular}{cc}
		\includegraphics[width=0.325\textwidth]{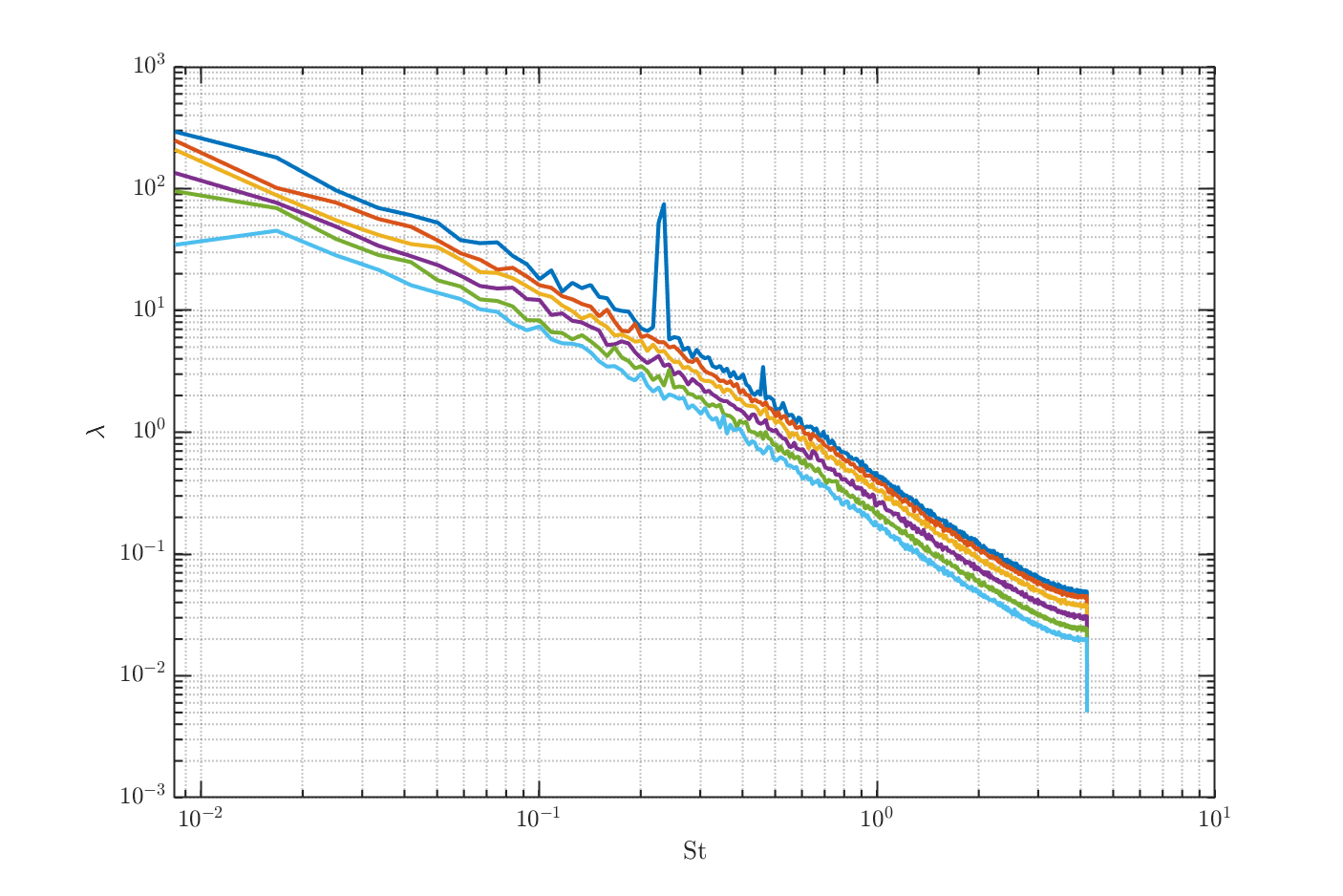} &
		\includegraphics[width=0.35\textwidth]{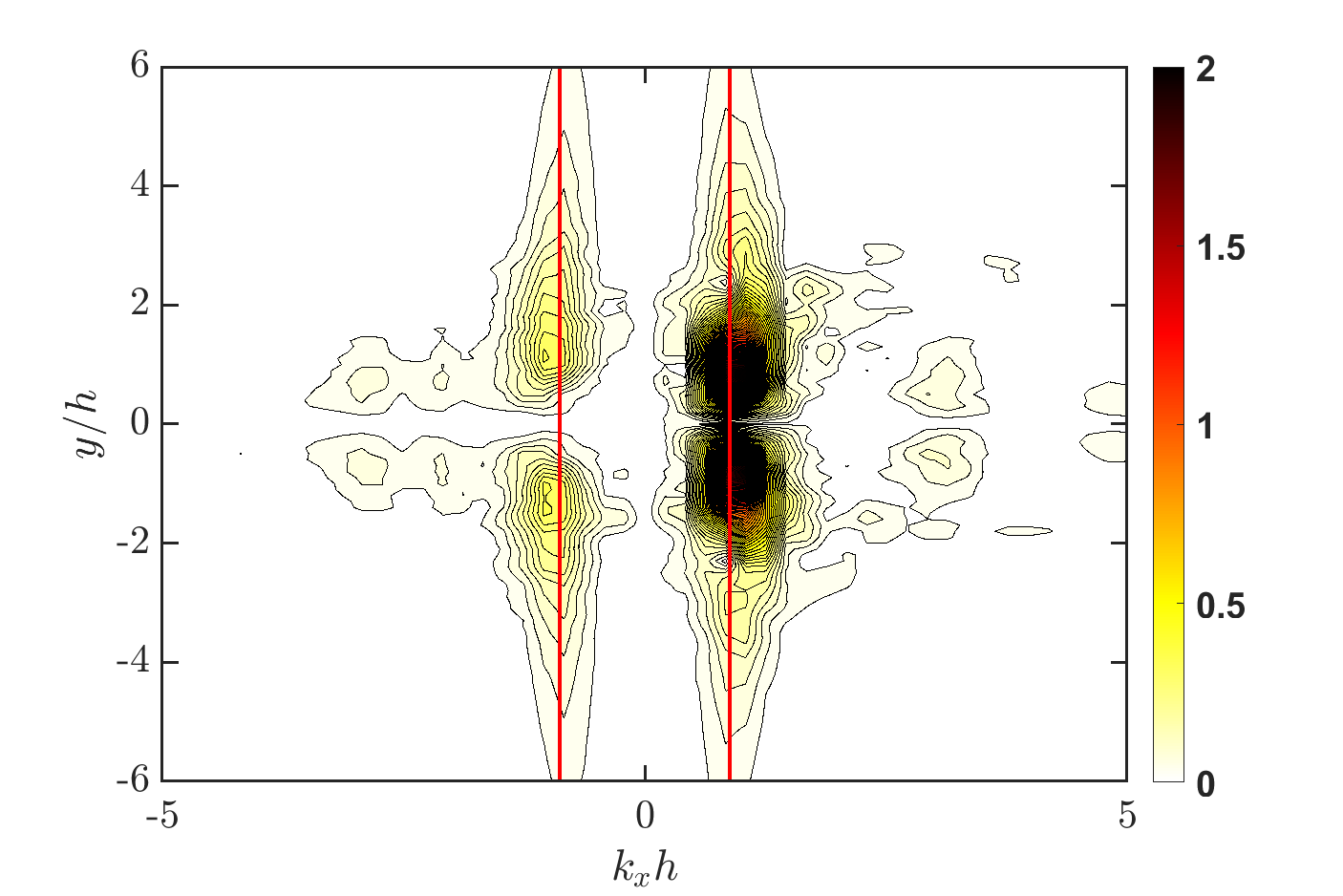} \\
		(a) & (b) \\
        \includegraphics[width=0.35\textwidth]{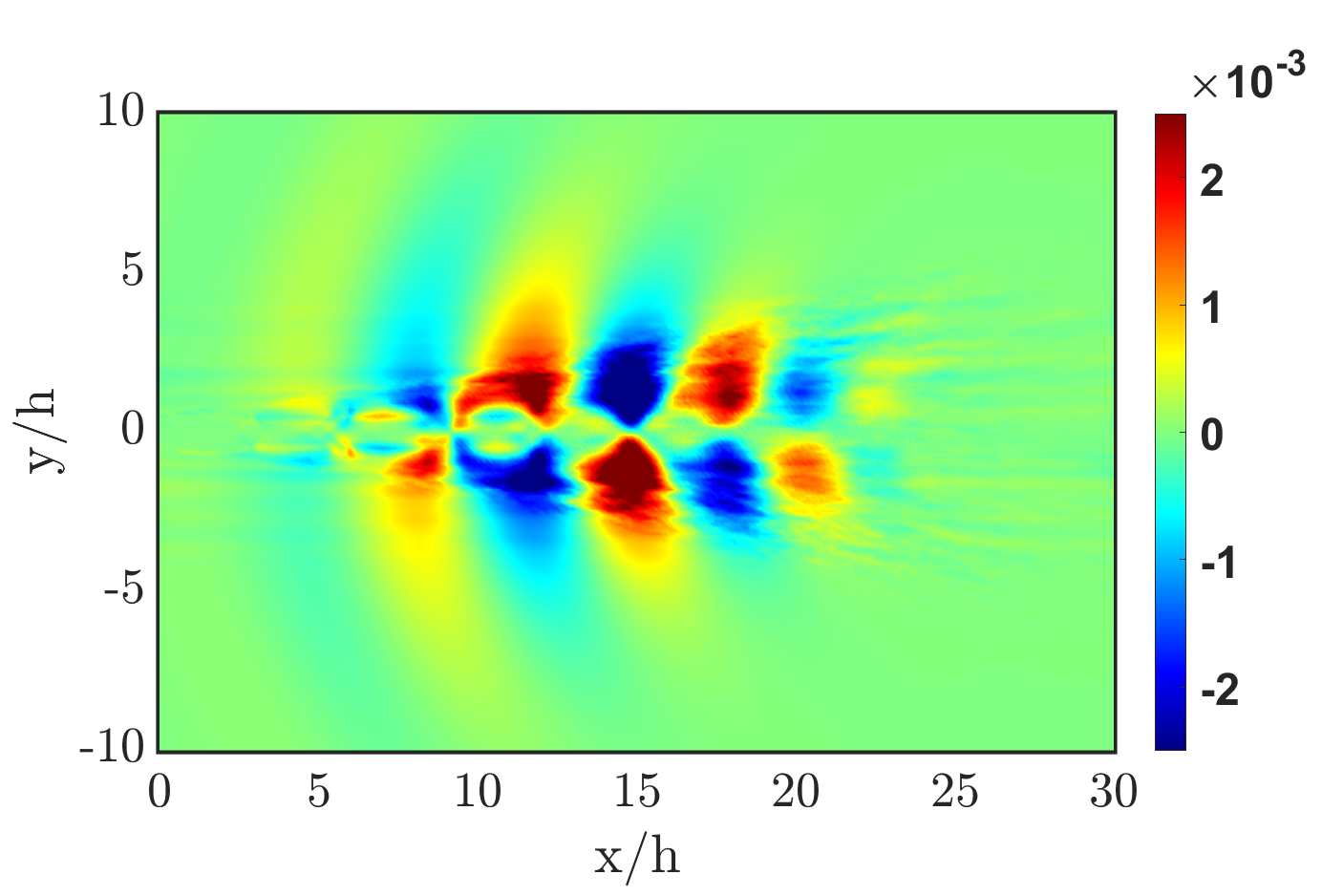} &
		\includegraphics[width=0.35\textwidth]{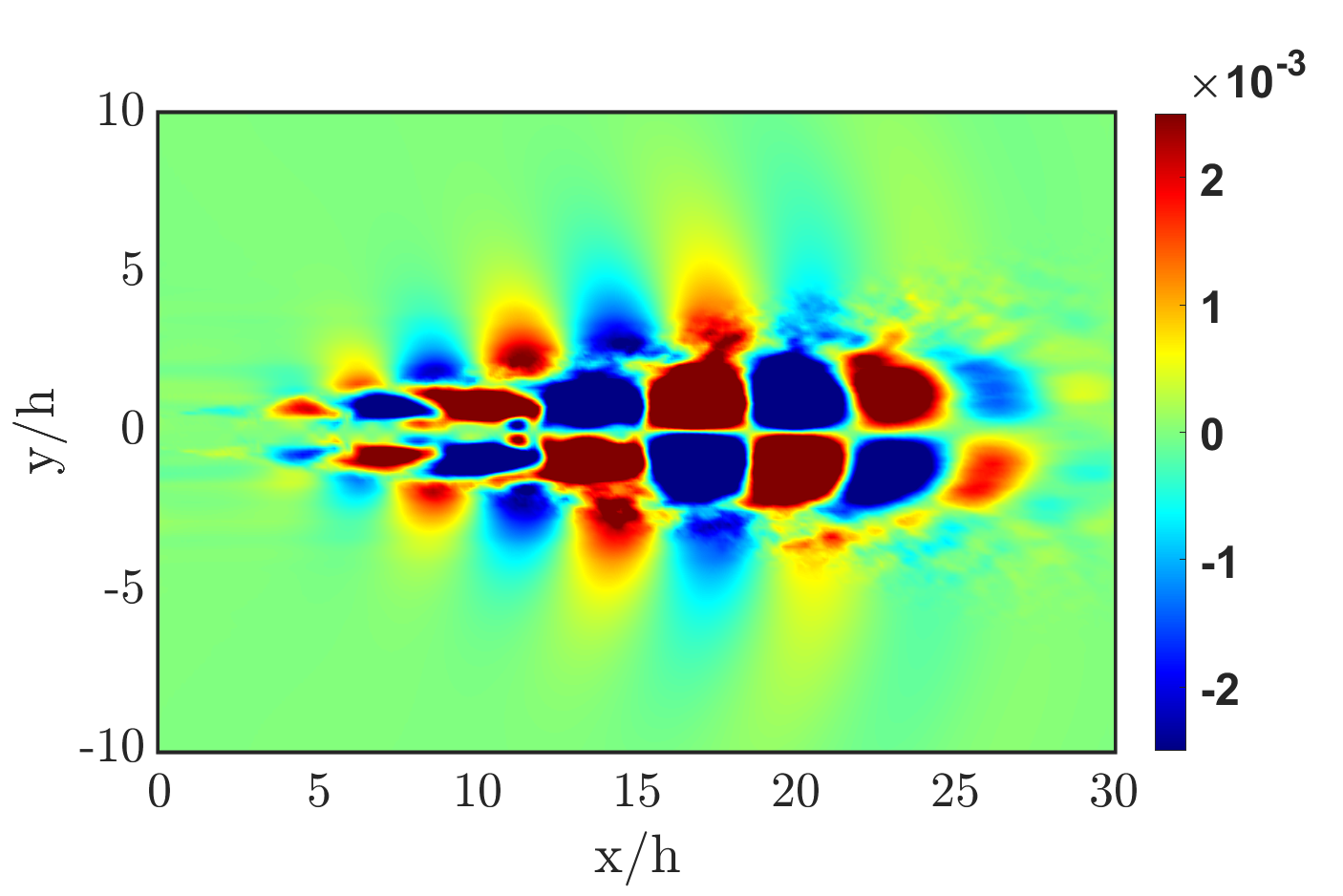} \\
		(c) & (d)
	\end{tabular}
	\caption{(a) Energy spectra of $u'$ SPOD modes, (b) modulus of the leading-order SPOD mode shape function for $u'$ at the screech fundamental frequency, (c) upstream-traveling waves in the leading-order $u'$ SPOD mode, (d) downstream-traveling waves in the leading-order $u'$ SPOD mode.}
	\label{fig:Spod_u}
\end{figure}

\begin{figure}
	\centering
	\begin{tabular}{cc}
		\includegraphics[width=0.325\textwidth]{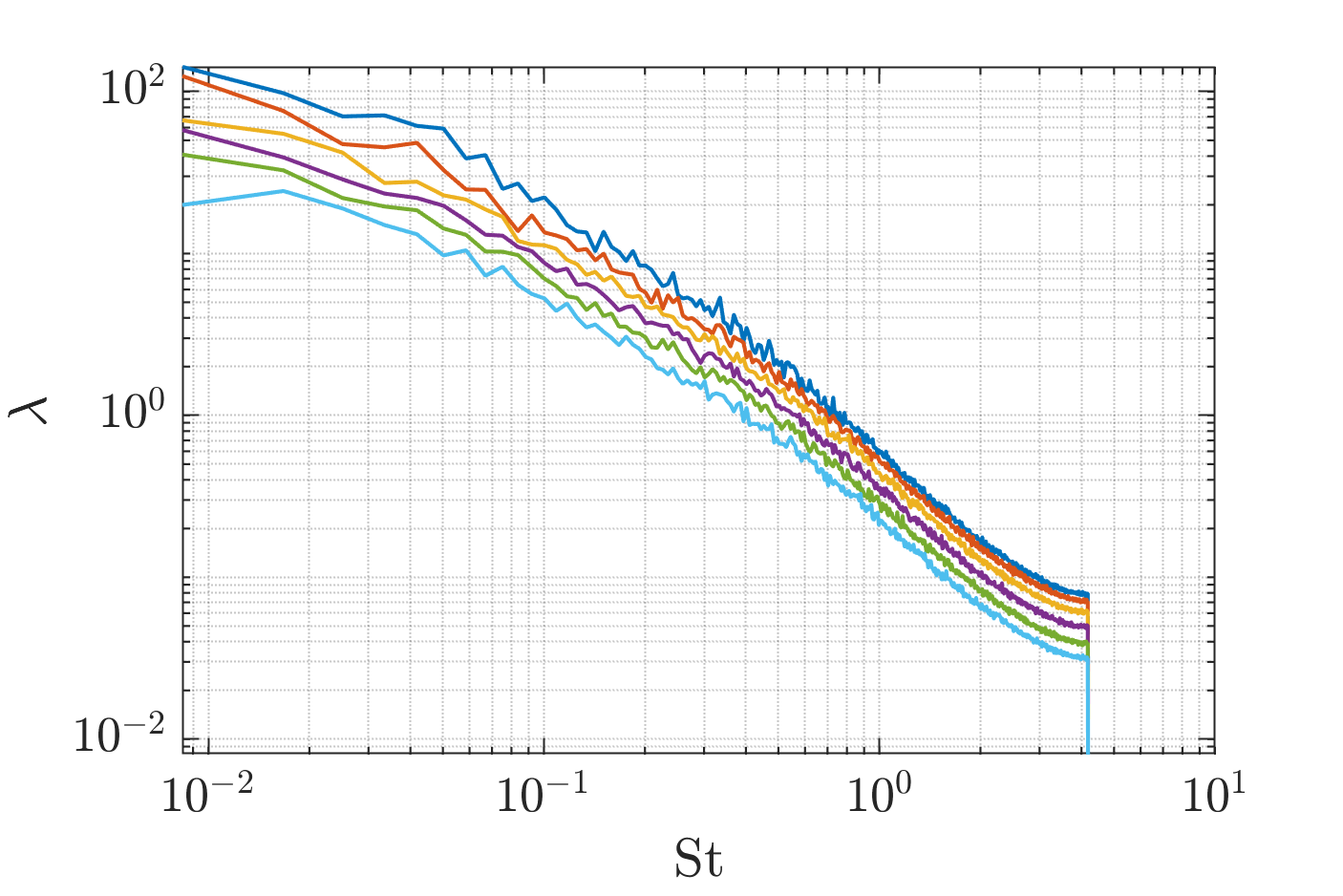} &
		\includegraphics[width=0.35\textwidth]{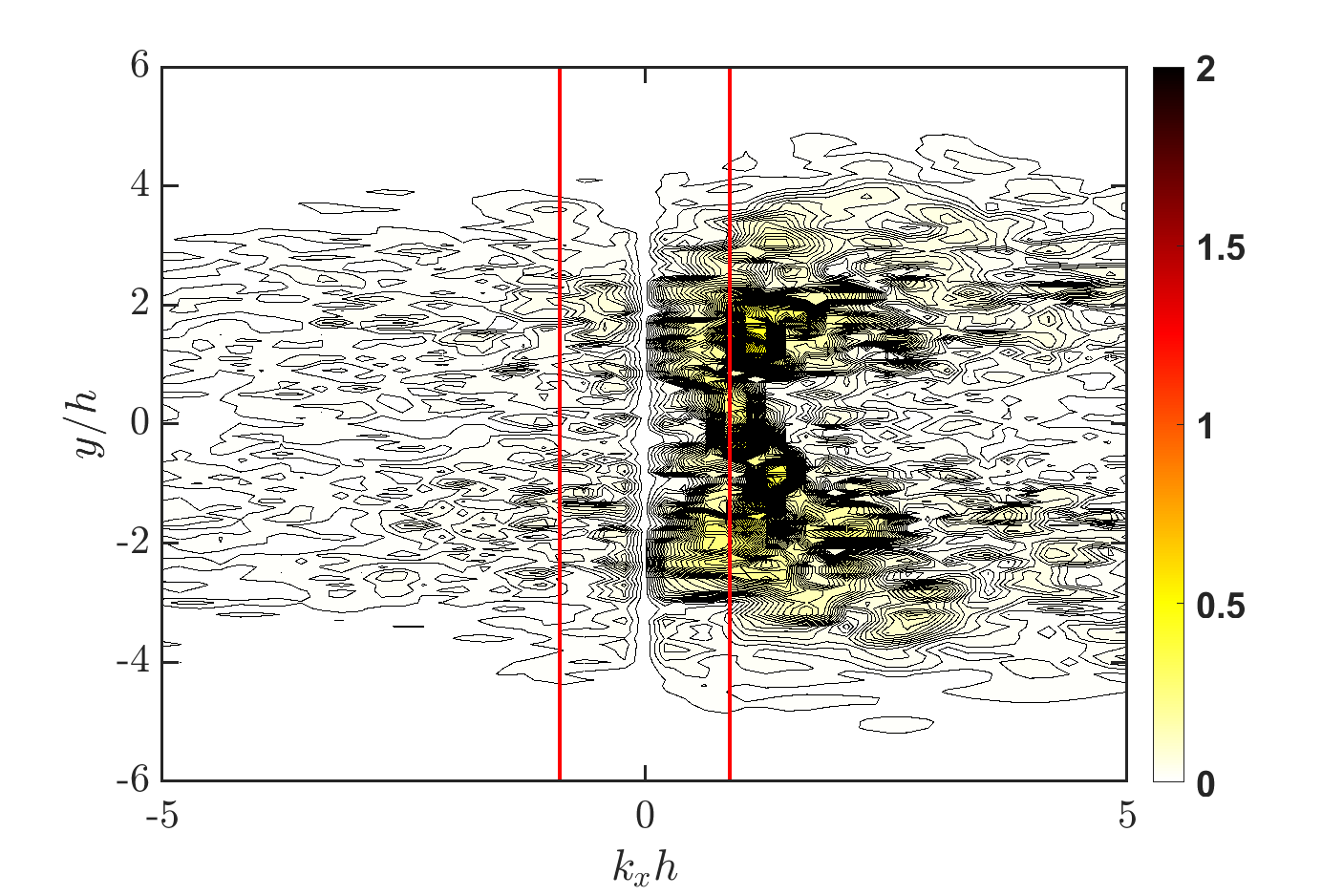} \\
		(a) & (b) \\
        \includegraphics[width=0.35\textwidth]{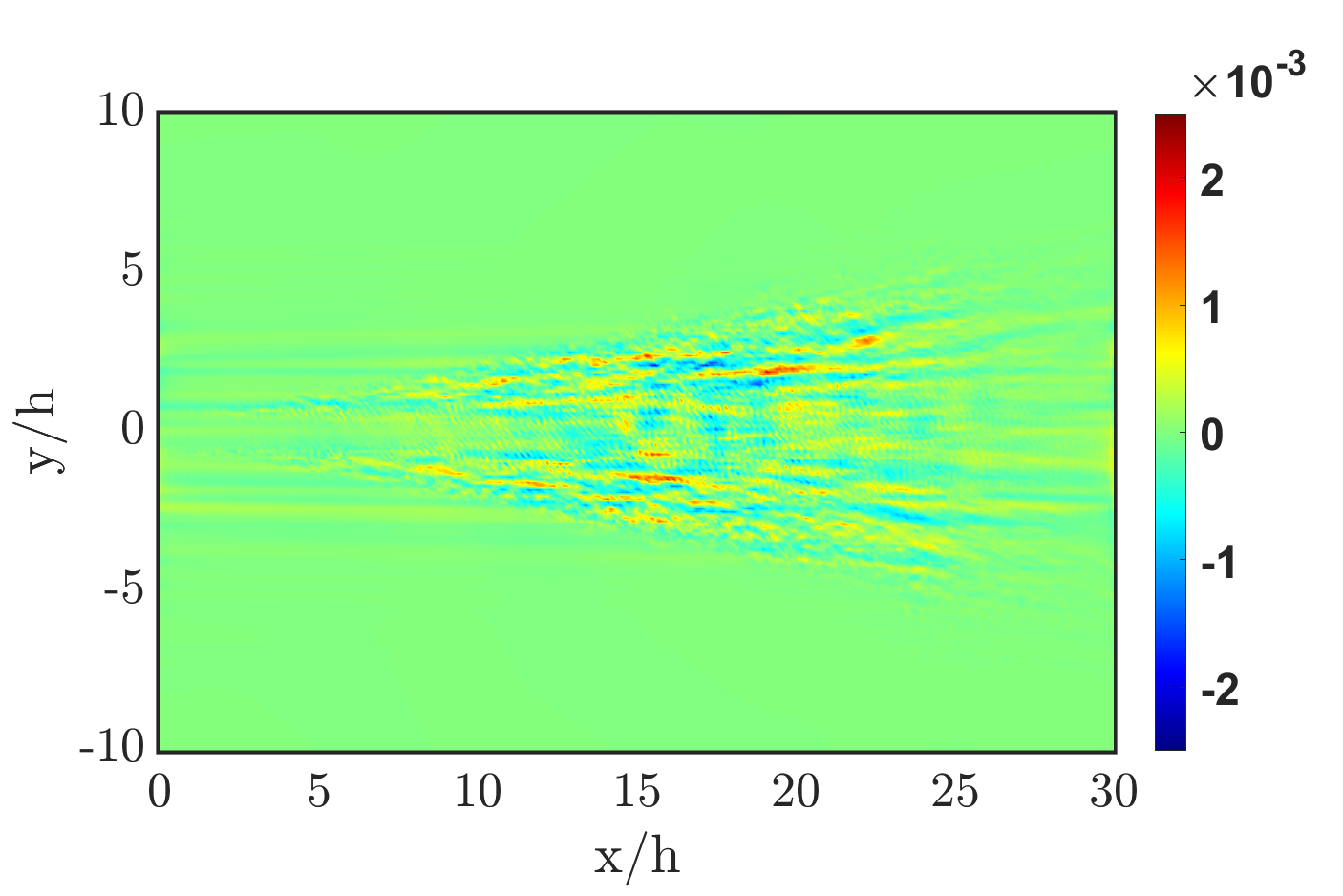} &
		\includegraphics[width=0.35\textwidth]{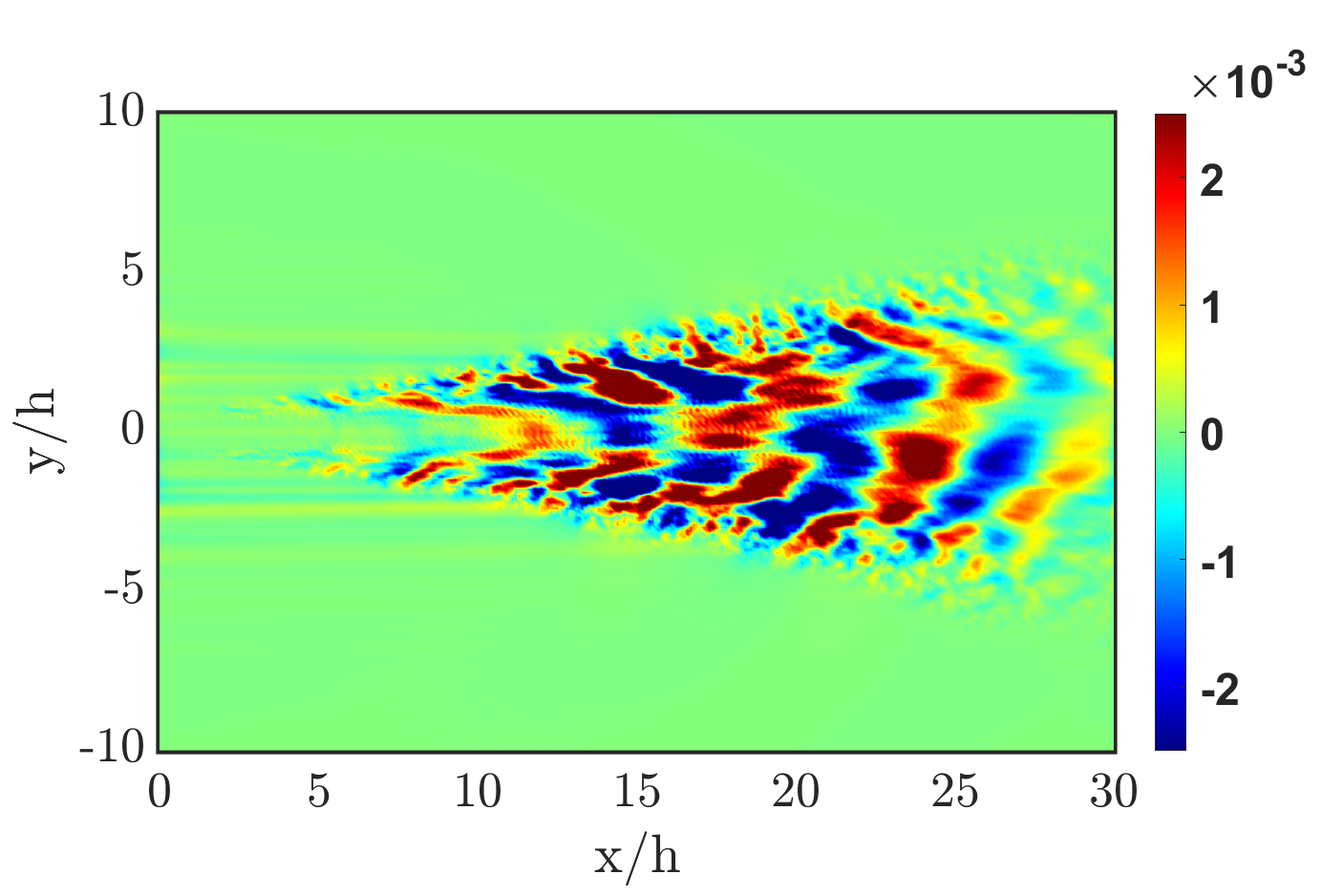} \\
		(c) & (d)
	\end{tabular}
	\caption{(a) Energy spectra of $w'$ SPOD modes, (b) modulus of the leading-order SPOD mode shape function for $w'$ at the screech fundamental frequency, (c) upstream-traveling waves in the leading-order $u'$ SPOD mode, (d) downstream-traveling waves in the leading-order $w'$ SPOD mode.}
	\label{fig:Spod_w}
\end{figure}

 This appendix presents the SPOD energy spectra and the leading-order SPOD mode shapes for pressure fluctuations $p'$, density fluctuations $\rho'$, and velocity fluctuations $u'$ and $w'$ from figures~\ref{fig:Spod_p} to~\ref{fig:Spod_w}. These complement the results for $v'$ SPOD modes in the paper. These modes are computed by optimizing the variance of the fluctuations of the corresponding variable. The key characteristics of the energy spectra and the mode shapes are similar to those of $v'$ presented in Section~\ref{sec:dominant coherence} for all the variables except for $w'$. The energy spectra for $w'$ SPOD modes do not show peaks at the screech frequency or its harmonics 
 in the minor-axis plane. There are no standing waves in the SPOD mode shapes of $w'$ at screech frequency and the coherent structures are mostly dominated by downstream-traveling K-H waves. This suggests that the out-of-plane velocity fluctuations (along the minor-axis plane) are not strongly correlated with the dominant coherence for the screech feedback in the rectangular jet.

\bibliographystyle{jfm}
\bibliography{references}

\begin{thebibliography}{40}
\expandafter\ifx\csname natexlab\endcsname\relax\def\natexlab#1{#1}\fi
\def\au#1{#1} \def\ed#1{#1} \def\yr#1{#1}\def\at#1{#1}\def\jt#1{\textit{#1}}
  \def\bt#1{#1}\def\bvol#1{\textbf{#1}} \def\vol#1{#1} \def\pg#1{#1}
  \def\publ#1{#1}\def\arxiv#1{#1}\def\org#1{#1}\def\st#1{\textit{#1}}

\bibitem[Alkislar {\em et~al.\/}(2005)Alkislar, Krothapalli, Choutapalli \&
  Lourenco]{alkislar2005structure}
{\sc \au{Alkislar, Mehmet~B}, \au{Krothapalli, Anjaneyulu}, \au{Choutapalli,
  Isaac} \& \au{Lourenco, Luiz~M}} \yr{2005}  \at{Structure of supersonic twin
  jets}.  \jt{AIAA journal}  \bvol{43}~(11),  \pg{2309--2318}.

\bibitem[Bogey \& Gojon(2017)]{Bogey_Gojon_JM2017_feedback}
{\sc \au{Bogey, Christophe} \& \au{Gojon, Romain}} \yr{2017}  \at{Feedback loop
  and upwind-propagating waves in ideally expanded supersonic impinging round
  jets}.  \jt{Journal of Fluid Mechanics}  \bvol{823},  \pg{562--591}.

\bibitem[Br{\`e}s {\em et~al.\/}(2018)Br{\`e}s, Bose, Emory, Ham, Schmidt,
  Rigas \& Colonius]{bres2018large}
{\sc \au{Br{\`e}s, Guillaume~A}, \au{Bose, Sanjeeb~T}, \au{Emory, Michael},
  \au{Ham, Frank~E}, \au{Schmidt, Oliver~T}, \au{Rigas, Georgios} \&
  \au{Colonius, Tim}} \yr{2018} Large-eddy simulations of co-annular turbulent
  jet using a voronoi-based mesh generation framework. AIAA Paper 2018-3302.

\bibitem[Br{\`e}s {\em et~al.\/}(2017)Br{\`e}s, Ham, Nichols \&
  Lele]{bres2017unstructured}
{\sc \au{Br{\`e}s, Guillaume~A}, \au{Ham, Frank~E}, \au{Nichols, Joseph~W} \&
  \au{Lele, Sanjiva~K}} \yr{2017}  \at{Unstructured large-eddy simulations of
  supersonic jets}.  \jt{AIAA journal}  \pg{pp. 1164--1184}.

\bibitem[Br{\`e}s {\em et~al.\/}(2019)Br{\`e}s, Towne \&
  Sanjiva]{bres2019investigating}
{\sc \au{Br{\`e}s, Guillaume~A}, \au{Towne, Aaron} \& \au{Sanjiva, K.~Lele}}
  \yr{2019} Investigating the effects of temperature non-uniformity on
  supersonic jet noise with large-eddy simulation. AIAA Paper 2019-2730.

\bibitem[Chandrashekar(2013)]{chandrashekar2013kinetic}
{\sc \au{Chandrashekar, Praveen}} \yr{2013}  \at{Kinetic energy preserving and
  entropy stable finite volume schemes for compressible euler and navier-stokes
  equations}.  \jt{Communications in Computational Physics}  \bvol{14}~(5),
  \pg{1252--1286}.

\bibitem[Edgington-Mitchell(2019)]{edgington2019Aeroacoustic}
{\sc \au{Edgington-Mitchell, Daniel}} \yr{2019}  \at{Aeroacoustic resonance and
  self-excitation in screeching and impinging supersonic jets – a review}.
  \jt{International Journal of Aeroacoustics}  \bvol{18}~(2-3),  \pg{118--188}.

\bibitem[Edgington-Mitchell {\em et~al.\/}(2018)Edgington-Mitchell, Jaunet,
  Jordan, Towne, Soria \& Honnery]{edgington2018upstream}
{\sc \au{Edgington-Mitchell, Daniel}, \au{Jaunet, Vincent}, \au{Jordan, Peter},
  \au{Towne, Aaron}, \au{Soria, Julio} \& \au{Honnery, Damon}} \yr{2018}
  \at{Upstream-travelling acoustic jet modes as a closure mechanism for
  screech}.  \jt{Journal of Fluid Mechanics}  \bvol{855}.

\bibitem[Edgington-Mitchell {\em et~al.\/}(2022)Edgington-Mitchell, Li, Liu,
  He, Wong, MacKenzie \& Nogueira]{edmitch_jfm2022_unifying}
{\sc \au{Edgington-Mitchell, Daniel}, \au{Li, Xiangru}, \au{Liu, Nianhua},
  \au{He, Feng}, \au{Wong, Tsz~Yeung}, \au{MacKenzie, Jacob} \& \au{Nogueira,
  Petronio}} \yr{2022}  \at{A unifying theory of jet screech}.  \jt{Journal of
  Fluid Mechanics}  \bvol{945}.

\bibitem[Edgington-Mitchell {\em et~al.\/}(2021)Edgington-Mitchell, Wang,
  Nogueira, Schmidt, Jaunet, Duke, Jordan \&
  Towne]{edmitch_2021JFM_waves_in_screech}
{\sc \au{Edgington-Mitchell, Daniel}, \au{Wang, Tianye}, \au{Nogueira,
  Petronio}, \au{Schmidt, Oliver}, \au{Jaunet, Vincent}, \au{Duke, Daniel},
  \au{Jordan, Peter} \& \au{Towne, Aaron}} \yr{2021}  \at{Waves in screeching
  jets}.  \jt{Journal of Fluid Mechanics}  \bvol{913}.

\bibitem[Edgington-Mitchell {\em et~al.\/}(2020)Edgington-Mitchell, Weightman,
  Lock, Kirby, Nair, Soria \& Honnery]{EdMitch_jfm2022_shockleakage}
{\sc \au{Edgington-Mitchell, Daniel}, \au{Weightman, Joel}, \au{Lock, Samuel},
  \au{Kirby, Rhiannon}, \au{Nair, Vineeth}, \au{Soria, Julio} \& \au{Honnery,
  Damon}} \yr{2020}  \at{The generation of screech tones by shock leakage}.
  \jt{Journal of Fluid Mechanics} .

\bibitem[Ffowcs-Williams \& Hawkings(1969)]{Ffowcs_etal_1969_Sound}
{\sc \au{Ffowcs-Williams, J.~E.} \& \au{Hawkings, D.~L.}} \yr{1969}  \at{Sound
  generation by turbulence and surfaces in arbitrary motion}.
  \jt{Philosophical Transactions of the Royal Society of London. Series A,
  Mathematical and Physical Sciences}  \bvol{264}~(1151),  \pg{321--342}.

\bibitem[Fisher \& Carpenter(2013)]{fisher2013high}
{\sc \au{Fisher, Travis~C.} \& \au{Carpenter, Mark~H.}} \yr{2013}
  \at{High-order entropy stable finite difference schemes for nonlinear
  conservation laws: finite domains}.  \jt{Journal of Computational Physics}
  \bvol{252},  \pg{518--557}.

\bibitem[Gojon {\em et~al.\/}(2019)Gojon, Gutmark \&
  Mihaescu]{gojon2019antisymmetric}
{\sc \au{Gojon, Romain}, \au{Gutmark, Ephraim} \& \au{Mihaescu, Mihai}}
  \yr{2019}  \at{Antisymmetric oscillation modes in rectangular screeching
  jets}.  \jt{AIAA Journal}  \bvol{57}~(8).

\bibitem[Lighthill(1952)]{Lighthill1952}
{\sc \au{Lighthill, M.~J.}} \yr{1952}  \at{On sound generated aerodynamically,
  part 1.}  \jt{Proceedings of the Royal Society of London. Series A.
  Mathematical and Physical Sciences}  \bvol{211},  \pg{564--587}.

\bibitem[Lockard(2000)]{lockard2000efficient}
{\sc \au{Lockard, David~P}} \yr{2000}  \at{An efficient, two-dimensional
  implementation of the ffowcs williams and hawkings equation}.  \jt{Journal of
  Sound and Vibration}  \bvol{229}~(4),  \pg{897--911}.

\bibitem[Lumley(1970)]{lumley1970stochastic}
{\sc \au{Lumley, J.~L.}}, ed. \yr{1970} {\em Stochastic Tools in Turbulence\/}.
   \publ{Academic Press.}

\bibitem[Majumdar(2014)]{majumdar2014}
{\sc \au{Majumdar, Dave}} \yr{2014}  \at{America’s \$400 billion stealth jet
  fleet is grounded}.  \jt{The Daily Beast News} .

\bibitem[Mancinelli {\em et~al.\/}(2021)Mancinelli, Jaunet, Jordan \&
  Towne]{Mancinelli2021}
{\sc \au{Mancinelli, Matteo}, \au{Jaunet, Vincent}, \au{Jordan, Peter} \&
  \au{Towne, Aaron}} \yr{2021}  \at{A complex-valued resonance model for
  axisymmetric screech tones in supersonic jets}.  \jt{Journal of Fluid
  Mechanics}  \bvol{928}.

\bibitem[Manning \& Lele(2000)]{manning2000numerical}
{\sc \au{Manning, Ted} \& \au{Lele, Sanjiva}} \yr{2000} A numerical
  investigation of sound generation in supersonic jet screech. AIAA Paper
  2000-2081.

\bibitem[Nekkanti \& Schmidt(2020)]{Nekkanti2020}
{\sc \au{Nekkanti, Akhil} \& \au{Schmidt, Oliver~T.}} \yr{2020} Modal analysis
  of the directivity of acoustic emissions from wavepackets in turbulent jets.
  AIAA Paper 2020-0745.

\bibitem[Nekkanti \& Schmidt(2021)]{Nekkanti2021}
{\sc \au{Nekkanti, Akhil} \& \au{Schmidt, Oliver~T.}} \yr{2021}  \at{Modal
  analysis of acoustic directivity in turbulent jets}.  \jt{AIAA Journal}
  \bvol{59},  \pg{228--239}.

\bibitem[Nogueira {\em et~al.\/}(2022{\natexlab{{\em a\/}}})Nogueira, Jaunet,
  Mancinelli, Jordan \& Edgington-Mitchell]{Nogueira2022_closure_mech_a1_a2}
{\sc \au{Nogueira, Petrônio~A.S.}, \au{Jaunet, Vincent}, \au{Mancinelli,
  Matteo}, \au{Jordan, Peter} \& \au{Edgington-Mitchell, Daniel}}
  \yr{2022{\natexlab{{\em a\/}}}}  \at{Closure mechanism of the a1 and a2 modes
  in jet screech}.  \jt{Journal of Fluid Mechanics}  \bvol{936}.

\bibitem[Nogueira {\em et~al.\/}(2022{\natexlab{{\em b\/}}})Nogueira, Jordan,
  Jaunet, Cavalieri, Towne \& Edgington-Mitchell]{Nogueira2022_absolute_instab}
{\sc \au{Nogueira, Petrônio~A.S.}, \au{Jordan, Peter}, \au{Jaunet, Vincent},
  \au{Cavalieri, André~V.G.}, \au{Towne, Aaron} \& \au{Edgington-Mitchell,
  Daniel}} \yr{2022{\natexlab{{\em b\/}}}}  \at{Absolute instability in
  shock-containing jets}.  \jt{Journal of Fluid Mechanics}  \bvol{930}.

\bibitem[Panda(1999)]{panda1999AnExp}
{\sc \au{Panda, J}} \yr{1999}  \at{An experimenal investigation of screech
  noise generation}.  \jt{Journal of Fluid Mechanics}  \bvol{378},
  \pg{71--96}.

\bibitem[Powell(1953)]{powell1953mechanism}
{\sc \au{Powell, Alan}} \yr{1953}  \at{On the mechanism of choked jet noise}.
  \jt{Proceedings of the Physical Society. Section B}  \bvol{66}~(12),
  \pg{1039}.

\bibitem[Raman(1997)]{raman1997Screech}
{\sc \au{Raman, Ganesh}} \yr{1997}  \at{Screech tones from rectangular jets
  with spanwise oblique shock-cell structures}.  \jt{Journal of Fluid
  Mechanics}  \bvol{330},  \pg{141--168}.

\bibitem[Raman(1999)]{raman1999Supersonic}
{\sc \au{Raman, Ganesh}} \yr{1999}  \at{Supersonic jet screech: half-century
  from powell to the present}.  \jt{Journal of Sound and Vibration}
  \bvol{225}~(3),  \pg{543--571}.

\bibitem[Raman {\em et~al.\/}(2012)Raman, Panickar \&
  Chelliah]{raman2012aeroacoustics}
{\sc \au{Raman, Ganesh}, \au{Panickar, Praveen} \& \au{Chelliah, Kanthasamy}}
  \yr{2012}  \at{Aeroacoustics of twin supersonic jets: a review}.
  \jt{International Journal of Aeroacoustics}  \bvol{11}~(7-8),  \pg{957--984}.

\bibitem[Schmidt \& Colonius(2020)]{Schmidt_Colonius_AIAA2020_Guide}
{\sc \au{Schmidt, Oliver~T.} \& \au{Colonius, Tim}} \yr{2020}  \at{Guide to
  spectral proper orthogonal decomposition}.  \jt{AIAA Journal}  \bvol{58}~(3),
   \pg{1023--1033}.

\bibitem[Schmidt {\em et~al.\/}(2018)Schmidt, Towne, Rigas, Colonius \&
  Brès]{Schmidt_etal_JFM_2018_Spectral_Jet_Turb}
{\sc \au{Schmidt, Oliver~T.}, \au{Towne, Aaron}, \au{Rigas, Georgios},
  \au{Colonius, Tim} \& \au{Brès, Guillaume~A.}} \yr{2018}  \at{Spectral
  analysis of jet turbulence}.  \jt{Journal of Fluid Mechanics}  \bvol{855},
  \pg{953--982}.

\bibitem[Suzuki \& Lele(2003)]{suzuki_lele_2003}
{\sc \au{Suzuki, Takao} \& \au{Lele, Sanjiva~K.}} \yr{2003}  \at{Shock leakage
  through an unsteady vortex-laden mixing layer: application to jet screech}.
  \jt{Journal of Fluid Mechanics}  \bvol{490},  \pg{139–167}.

\bibitem[Tadmor(2003)]{tadmor2003entropy}
{\sc \au{Tadmor, Eitan}} \yr{2003}  \at{Entropy stability theory for difference
  approximations of nonlinear conservation laws and related time-dependent
  problems}.  \jt{Acta Numerica}  \pg{pp. 451--512}.

\bibitem[Tam {\em et~al.\/}(1986)Tam, Seiner \& Yu]{tam1986proposed}
{\sc \au{Tam, Christopher~KW}, \au{Seiner, John~M} \& \au{Yu, JC}} \yr{1986}
  \at{Proposed relationship between broadband shock associated noise and
  screech tones}.  \jt{Journal of sound and vibration}  \bvol{110}~(2),
  \pg{309--321}.

\bibitem[Tam \& Ahuja(1990)]{Tam_Ahuja_JFM1990}
{\sc \au{Tam, C. K.~W.} \& \au{Ahuja, K.~K.}} \yr{1990}  \at{Theoretical-model
  of discrete tone generation by impinging jets}.  \jt{Journal of Fluid
  Mechanics}  \bvol{214},  \pg{67--87}.

\bibitem[Tam \& Hu(1989)]{tam1989onthethree}
{\sc \au{Tam, Christopher K~W} \& \au{Hu, F.~Q.}} \yr{1989}  \at{On the three
  families of instability waves of high-speed jets}.  \jt{Journal of Fluid
  Mechanics}  \bvol{201},  \pg{447--483}.

\bibitem[Towne {\em et~al.\/}(2018)Towne, Schmidt \&
  Colonius]{towne_JFM_2018_spectral_relation}
{\sc \au{Towne, Aaron}, \au{Schmidt, Oliver~T} \& \au{Colonius, Tim}} \yr{2018}
   \at{Spectral proper orthogonal decomposition and its relationship to dynamic
  mode decomposition and resolvent analysis}.  \jt{Journal of Fluid Mechanics}
  \bvol{847},  \pg{821--867}.

\bibitem[Valentich {\em et~al.\/}(2016)Valentich, Upadhyay \&
  Kumar]{valentich2016mixing}
{\sc \au{Valentich, Griffin}, \au{Upadhyay, Puja} \& \au{Kumar, Rajan}}
  \yr{2016}  \at{Mixing characteristics of a moderate aspect ratio screeching
  supersonic rectangular jet}.  \jt{Experiments in Fluids}  \bvol{57}~(5),
  \pg{71}.

\bibitem[Wu {\em et~al.\/}(2020)Wu, Lele \& Jeun]{wu2020_ctrab}
{\sc \au{Wu, Gao~Jun}, \au{Lele, Sanjiva~K.} \& \au{Jeun, Jinah}} \yr{2020}
  \at{Coherence and feedback in supersonic rectangular jet screech}.
  \jt{Center for Turbulence Research Annual Research Brief} .

\bibitem[Wu {\em et~al.\/}(2022)Wu, Lele, Jeun, Kumar \&
  Gustavsson]{wu2022_aiaaj_assessment}
{\sc \au{Wu, Gao~Jun}, \au{Lele, Sanjiva~K.}, \au{Jeun, Jinah}, \au{Kumar,
  Rajan} \& \au{Gustavsson, Jonas}} \yr{2022}  \at{Unstructured large-eddy
  simulations of rectangular jet screech: assessment and validation}.  \jt{AIAA
  Journal} .

\end{thebibliography}







\end{document}